\documentclass[twocolumn,aps,showpacs,preprintnumbers,nofootinbib,prd,
superscriptaddress,10pt]{revtex4-1}

\usepackage{lmodern}
\usepackage{amsmath,amssymb}
\usepackage[normalem]{ulem}
\usepackage{textcomp}
\usepackage{hyperref}
\usepackage{bm}
\usepackage{graphicx}
\usepackage{psfrag}
\usepackage{aas_macros}
\usepackage[usenames,dvipsnames]{xcolor}
\usepackage[utf8]{inputenc}
\usepackage{lipsum} 
\usepackage{multirow}
\usepackage{subfigure}
\usepackage{acronym}
\usepackage{cancel}
\usepackage{orcidlink}

\definecolor{darkgreen}{rgb}{0,0.5,0}

\hypersetup{
    unicode=false,          
    pdftoolbar=true,        
    pdfmenubar=true,        
    pdffitwindow=false,     
    pdfstartview={FitH},    
	pdftitle={Parametrizing the Unknown: },
    pdfauthor={Sumit K et al.},
    pdfsubject={Gravitational Waves},   
    pdfkeywords={gravitational waves, eparameter estimation},
    pdfnewwindow=true,      
    colorlinks=true,       
    linkcolor=red,          
    citecolor=Blue,        
    filecolor=magenta,      
    urlcolor=Blue,           
    linktocpage=true
}

\allowdisplaybreaks[4]

\DeclareFontFamily{OT1}{pzc}{}
\DeclareFontShape{OT1}{pzc}{m}{it}{<-> s * [1.10] pzcmi7t}{}
\DeclareMathAlphabet{\mathpzc}{OT1}{pzc}{m}{it}

\newcommand{\thetatr}{\Theta_\mathrm{tr}}
\newcommand{\thetabf}{\Theta_\mathrm{bf}}
\newcommand{\thetatropt}{\overline{\Theta}_\mathrm{tr}}
\usepackage{xspace}
\newcommand{\phenompvtwo}{\texttt{IMRPhenomPv2}\xspace}

\usepackage{standalone}  

\begin{document}
    \newacro{nr}[NR]{Numerical relativity}
    
    \newacro{gw}[GW]{gravitational wave}
    
    \newacro{pe}[PE]{parameter estimation}

    \newacro{pn}[PN]{Post-Newtonian}

\title{Accounting for the Known Unknowns: A Parametric Framework to Incorporate Systematic Waveform Errors in Gravitational-Wave Parameter Estimation}

\date{\today}

\author{Sumit Kumar \orcidlink{0000-0002-6404-0517}}
\email{sumit.kumar@aei.mpg.de}
\affiliation{Max-Planck-Institut f{\"u}r Gravitationsphysik (Albert-Einstein-Institut), D-30167 Hannover, Germany}
\affiliation{Leibniz Universit{\"a}t Hannover, D-30167 Hannover, Germany}
\affiliation{Institute for Gravitational and Subatomic Physics (GRASP), 
Utrecht University, Princetonplein 1, 3584 CC Utrecht, The Netherlands}
\affiliation{Nikhef -- National Institute for Subatomic Physics, 
Science Park 105, 1098 XG Amsterdam, The Netherlands}

\author{Max Melching \orcidlink{0009-0001-4899-9955}}
\email{max.melching@aei.mpg.de}
\affiliation{Max-Planck-Institut f{\"u}r Gravitationsphysik (Albert-Einstein-Institut), D-30167 Hannover, Germany}
\affiliation{Leibniz Universit{\"a}t Hannover, D-30167 Hannover, Germany}

\author{Frank Ohme \orcidlink{0000-0003-0493-5607}}
\email{frank.ohme@aei.mpg.de}
\affiliation{Max-Planck-Institut f{\"u}r Gravitationsphysik (Albert-Einstein-Institut), D-30167 Hannover, Germany}
\affiliation{Leibniz Universit{\"a}t Hannover, D-30167 Hannover, Germany}


\begin{abstract}
The \ac{pe} for \ac{gw} merger events relies on a waveform model calibrated using numerical simulations. Within the Bayesian framework, this waveform model represents the \ac{gw} signal produced during the merger and is crucial for estimating the likelihood function. However, these waveform models may possess systematic errors that can differ across the parameter space. Addressing these errors in the current data analysis pipeline is an active area of research.
We introduce parametrizations for the uncertainties in the amplitude and phase of the reference waveform model. When the error budget in the amplitude and phase of the waveform model, as a function of frequency, is known, it can be used as a prior distribution in the Bayesian framework. We also show that conservative priors can be used to quantify uncertainties in waveform modeling without any knowledge of waveform uncertainty error budgets.
Through zero-noise injections and \ac{pe} recoveries, we demonstrate that even 1\%-2\% of errors in relative phase to the actual waveform model, for a GW150914-like signal and advanced LIGO detector sensitivity, can introduce biases in the recovered parameters. These biases can be corrected when we account for waveform uncertainties within the \ac{pe} framework. By analyzing a series of simulated signals from mergers with precessing orbits and recovering them using a non-spinning waveform model, we demonstrate that we can reduce the ratio of systematic errors to statistical errors. This approach allows us to address scenarios where specific physical effects are missing in waveform modeling. The code that implements our parametrization for performing PE is available as a Python package \href{https://github.com/gwastro/pycbc_wferrors_plugin}{`pycbc\_wferrors\_plugin'}, compatible with the PyCBC open source GW analysis library.
\end{abstract}

\date{\today}

\maketitle
\section{Introduction}
\label{sec:intro}
\ac{gw} signals from compact binary merger events provide us with a unique opportunity to probe these highly energetic events. With the network of advanced \ac{gw} detectors, including the LIGO detectors in Hanford and Livingston, USA, the Virgo detector in Italy, and the KAGRA detector in Japan, the detection of \ac{gw} signals has become routine \citep{LIGOScientific:2014pky, aLIGO:2020wna, KAGRA:2013rdx, VIRGO:2014yos, Virgo:2019juy}. Recent catalogs of \ac{gw} mergers, compiled from the data analysis of the first three observation runs of the LIGO and Virgo detectors, compile over 90 merger events. These catalogs include the analysis from the LIGO-Virgo-KAGRA (LVK) collaboration \citep{KAGRA:2021vkt}, as well as analyses conducted by other independent groups \citep{Nitz:2021zwj,Wadekar:2023gea}. 
The emphasis of the \ac{gw} community has now expanded from detection to precision science, such as constraining the theory of general relativity \citep{LIGOScientific:2021sio},  inferring the astrophysical distribution of the population of compact binary mergers \citep{KAGRA:2021duu, Nitz:2021zwj}, inferring Hubble constant \citep{LIGOScientific:2021aug}, constraining the equation of state of a neutron star \citep{LIGOScientific:2018cki, LIGOScientific:2018hze, Capano:2019eae}, etc. 

The detection of GWs and the precision of measurements of source properties relies on the accurate modeling of the signals from the merger of compact binary systems such as binary black holes (BBH), binary neutron stars (BNS), and neutron star-black hole (NSBH). These waveform models have been developed over the years with the help of \ac{pn} analytical calculation for the early inspiral part of the signals \citep{Blanchet:2013haa}, and they rely on calibration from \ac{nr} simulations for merger and ringdown regimes \citep{Campanelli:2005dd, Pretorius:2005gq, Baker:2005vv}. The signal's loudness and detector sensitivity determine the width of statistical uncertainties in the measurement, while the inaccuracies in the waveform models could lead to systematic errors. The statistical uncertainties go down as the detectors become more and more sensitive while systematic errors remain the same.

In the coming years, the planned upgrades to existing detectors \citep{KAGRA:2013rdx, Abbott:2016xvh} and the addition of new ones, such as LIGO India \citep{Saleem:2021iwi}, are expected to enhance the sensitivity of the detector network. Efforts are currently underway to develop third-generation (3G) detectors, such as the Einstein Telescope (ET) \citep{Punturo:2010zza, Hild:2010id} and Cosmic Explorer (CE) \citep{Evans:2021gyd, Srivastava:2022slt, Evans:2023euw}, over the next decade. These 3G detectors are anticipated to be significantly more sensitive—by an order of magnitude—than the current generation of detectors, and they are expected to enhance low-frequency sensitivity. As the \ac{gw} detectors' sensitivities keep improving, the waveform modeling community tries to keep up with the required accuracy to provide unbiased estimates.

One important question is how accurate waveform models need to be. One may not want to burden the waveform modeling and \ac{nr} simulation community with generating waveform models much better than the analysis can resolve \citep{Lindblom:2008cm}. However, studies suggest that we are already reaching the point where we can not ignore the systematic bias in a fraction of loud events. 
In a recent study, \citep{Purrer:2019jcp} it is shown that the typical biases in the inferred source parameters will be roughly equivalent to the standard deviation for the design sensitivity of current-generation detectors, and these biases could increase significantly for third-generation (3G) detectors. While a small bias in the fraction of events may be irrelevant for some studies, it could be critical for those that depend on accurate estimates of source properties. This is particularly true for inferring population characteristics of BBH mergers \cite{Dhani:2024jja}, estimating the Hubble constant using localization volume, issuing pre-merger localization alerts \cite{Kumar:2022tto}, and conducting tests of general relativity \citep{Gupta:2024gun, Chandramouli:2024vhw}, among others.

The impacts of the waveform modeling accuracy on estimating the source parameters have been extensively studied \citep{Lindblom:2008cm, Purrer:2019jcp, Read:2023hkv, Dhani:2024jja}. The standard approach to account for waveform systematics is combining the posterior samples from different waveform models by assigning appropriate weights \citep{KAGRA:2021vkt, Ashton:2019leq, Jan:2020bdz} or using hyperparameters to sample over different waveform approximants at the likelihood evaluation level \citep{Ashton:2021cub, Hoy:2022tst, Puecher:2023rxw,PhysRevD.98.124030, PhysRevD.100.024046,  Hoy:2024vpc}. These approaches are expected to marginalize potential differences between different waveform approximants. However, if all the waveform models have similar systematic effects, their difference would not accurately represent the true systematics in each model. Another approach is to marginalize the uncertainties using prior distributions analytically. These priors can be constructed using the Gaussian processes regression technique from a training set of accurate templates \citep{PhysRevLett.113.251101}. Similar techniques can also be applied to address the waveform systematics arising from calibration with \ac{nr} simulations \citep{Pompili:2024yec, Bachhar:2024olc}. There are also approaches to provide probabilistic models for the waveform, which can be used in sampling with appropriate weights \citep{PhysRevD.109.104045}. The authors in \citep{Owen:2023mid} marginalize over higher order \ac{pn} terms to mitigate the systematic errors. 

This work presents a general framework that accounts for the uncertainties in waveform modeling through parametric models. The additional parameters in \ac{pe} analysis relate to the waveform's amplitude and phase errors. When we know the expected amplitude and phase error distribution in the waveform as a function of frequency, we can use them as priors in \ac{pe} analysis. These errors are expected to be a function of parameter space. In the absence of such knowledge, we can be agnostic and use wider priors to capture any possible deviation. We evaluate this framework by introducing arbitrary deviations into reference waveform models and recovering them through our data analysis pipeline. In addition to correcting for biases, we assess the method's ability to quantify the nature of deviations from the true signal, particularly for the higher signal-to-noise ratio (SNR) "golden" binary systems.

This paper is structured as follows: Section \ref{sec:wf_modeling} reviews current techniques used in waveform model developments and the potential sources of systematic biases they might contain. Section \ref{sec:pe} visits the state-of-the-art \ac{pe} techniques and also discusses current methods to deal with waveform systematics. Section \ref{sec:modeling_strain} discusses modeling \ac{gw} strain data from the detector and introduces the parametrizations that we use in this work to account for waveform errors in \ac{pe}. The following section (section \ref{sec:FM}) discusses the Fisher matrix approach to account for the systematic bias. We also discuss the limitations where the Fisher matrix approach is no longer valid, and we need to do a full \ac{pe} analysis. In section \ref{sec:simulations}, we present the detailed simulations followed by an injection-recovery campaign to test the validity of this method. We also discuss the scenario where we can use these parametric models to account for the missing physical effects not considered in the waveform model. In the end, we discuss the findings of the work and summarize them in section \ref{sec:summary}. If the reader is already familiar with the \ac{pe} techniques and the potential sources of errors, they may skip Section \ref{sec:pe}. Additionally, if the reader is acquainted with the Fisher matrix approach for correcting systematic biases and understands its limitations in specific scenarios, they can skip Section \ref{sec:FM}. The reader who is just interested in the methodology and results can read sections \ref{sec:modeling_strain}, \ref{sec:simulations}, and \ref{sec:summary} in that order.

\section{Technical Background}
\subsection{Waveform Modeling and Sources of Uncertainties}
\label{sec:wf_modeling}
The \ac{gw} signal of compact binary mergers can be predicted by solving Einstein's Equations for the two-body problem in the general theory of relativity. However, due to the complexity of these equations, some form of approximations need to be employed to model binary systems. The most sophisticated models combine several modeling techniques. A review of those techniques and models is beyond the scope of this paper, but we give a brief overview of aspects that are relevant to our discussion of systematic waveform errors. Here we focus on black-hole binaries on quasi-circular orbits, as they are the most frequent source of detected GWs. However, many considerations detailed below apply to more general binary sources as well.

A black hole binary and its emitted \ac{gw} signal are described by several parameters, such as the BH component masses ($m_1, m_2 $), their spins ($\vec{s}_1, \vec{s}_2$), luminosity distance ($D_L$), inclination of binary plane with the line of sight ($\iota$), polarization angle with respect to the detector ($\psi$), coalescence phase of the binary ($\phi$), right ascension (RA) and declination angle (dec) in the sky, and the coalescence time ($t_c$).

When the binary constituents are widely separated and their velocities are small (compared to the speed of light), the system's evolution can be approximately modeled by a power series expansion. The most commonly used approach is the \ac{pn} formalism that expands in the system's relative velocity (see \cite{Blanchet:2013haa} for a review). Alternative approaches exist, for example  the post-Minkowskian formalism expands in the gravitational constant G \cite{Detweiler:1996mq, Bern:2019nnu}. In the highly relativistic regime, where the spacetime curvature is large and velocities are high, these approaches become inaccurate. More concretely, \ac{pn} expansions are well suited to describe the early inspiral of a black-hole binary, but as the black holes approach each other and their velocities increase, purely \ac{pn} models become inaccurate and unable to describe the merger and post-merger stages.

While the post-merger ringdown of the remnant black hole can be well described by perturbation theory \cite{Vishveshwara:1970zz, Chandrasekhar:1975zza}, the merger itself can only be modeled by numerically solving Einstein's Equation. Since its breakthrough \cite{Pretorius:2005gq, Campanelli:2005dd, Baker:2005vv}, \ac{nr} has become an important, well established tool in \ac{gw} astronomy. However, because of its computational complexity, \ac{nr} simulation can only cover the late inspiral (typically a few tens of orbits), merger and ringdown of the binary coalescence.

Neither of the above mentioned approaches can model the entire \ac{gw} signal of interest. \ac{pn} models become inaccurate towards the merger and do not include any merger or ringdown portion. \ac{nr} is computationally too expensive and cannot cover hundreds or even thousands of orbits before merger. Therefore, several model approaches have been designed to bridge the gap between numerical and analytical methods \cite{Ohme:2011rm}. 

Three model families are regularly employed in the analysis of \ac{gw} observations \cite{KAGRA:2021vkt}. The phenomenological models \cite{Ajith:2007kx, Ajith:2009bn, Santamaria:2010yb, Khan:2015jqa, Khan:2018fmp, Khan:2019kot, Husa:2015iqa, London:2017bcn, Hannam:2013oca, Garcia-Quiros:2020qpx, Pratten:2020ceb, Pratten:2020fqn, Estelles:2020osj, Estelles:2021gvs, Ghosh:2023mhc, Thompson:2023ase} are based on a set of hybrid waveforms that smoothly connect analytically derived inspiral signals with \ac{nr} data. This data set is then described by \ac{pn} inspired phenomenological formulae with tune-able coefficients that fit across the binary's parameter space. The resulting model is a set of closed-form expressions that describe the amplitude and phase of the \ac{gw} signal.

Another approach is based on the effective-one-body (EOB) formalism \cite{Buonanno:1998gg}, in which the binary system is mapped to a Hamiltonian formulation of a particle orbiting in an effective potential. By introducing higher-order terms and tune-able coefficients that are fit to waveforms from \ac{nr} simulations, the resulting EOBNR waveforms can describe the binary coalescence accurately up to the merger. Completed by a suitable ringdown attachment, several models of the EOBNR family have been developed \cite{Pan:2011gk, Pan:2013rra, Taracchini:2013rva, Bohe:2016gbl, Cotesta:2018fcv, Ossokine:2020kjp, Pompili:2023tna, Gamboa:2024hli, Damour:2012ky, Gamba:2020ljo, Nagar:2024oyk} and applied to the analysis of \ac{gw} events.

The third model family is called surrogate models. These models are built by decomposing waveform data from \ac{nr} simulations into a suitable basis and interpolating between them \cite{Field:2013cfa}. \ac{nr} surrogate models  \cite{Blackman:2017dfb, Blackman:2017pcm, Varma:2018mmi, Varma:2019csw} are very faithful to the \ac{nr} simulations they were built from, but they are limited by the length and parameter coverage of \ac{nr} simulations.

All of the modeling approaches approximate the exact solutions of Einstein's Equation. Therefore, they carry systematic uncertainties that are difficult to quantify in their entirety. Numerical simulations start with approximate initial data and discretize the spacetime. The resulting resolution error is often estimated by providing simulation results of the same system for different resolutions. All modeling approaches described above rely on input from numerical simulations, but they also need to interpolate between different simulations to cover the entire parameter space of interest. This interpolation introduces additional inaccuracies that are often less well quantified. Part of the problem is not only the interpolation (or even extrapolation) of existing \ac{nr} data. The ansatz used to describe the data carries its own limitations, whether it is data driven (such as Gaussian process regression) or analytically motivated.

In summary, all modeling approaches have to balance representing known \ac{nr} simulation data as accurately as possible while ensuring a smooth and robust interpolation across the parameter space. The techniques employed to accomplish this have advanced over the years, but they cannot be perfect. Therefore, any model's prediction has to be treated as a ``best-guess'' waveform for each set of parameters. While the faithfulness to \ac{nr} data is often very impressive, one must not forget the uncertainties arising from each step of the modeling process.

Here we discuss and test a framework that moves away from the ``best-guess'' paradigm to accounting for uncertain signal predictions in parameter-estimation analyses.

\subsection{Parameter Estimation}
\label{sec:pe}
In this section, we review the \ac{pe} techniques for \ac{gw} data analysis and potential errors in estimating source properties. Readers who are familiar with this can skip to the next section. Time series strain data $s(t)$ in each detector consists of noise $n(t)$ and may contain a transient signal $h(t)$. It is assumed that when the signal is present, the noise and signal are additive to give the strain data, i.e.
\begin{equation}
s(t) = h(t)+n(t)
\end{equation} 
The noise from the detector is modeled as Gaussian and stationary. In the absence of the signal, the detectors are assumed to contain only Gaussian and stationary noise. The Gaussian and stationary time series noise $n(t)$ can be expressed as \citep{LIGOScientific:2019hgc}
\begin{equation}
    \mathbf{n} = \frac{1}{\sqrt{2\pi\Sigma}} \exp\{ \frac{1}{2}(n(t_i)-\hat{\mu})^T\Sigma^{-1}_{ij}(n(t_j)-\hat{\mu})\}, \label{eqn:gaussian_stationary_noise}
\end{equation}
\noindent
where $\mathbf{n}$ is a noise realization which is represented as a vector with discrete time samples $n(t_i)$, $\Sigma$ is a covariance matrix, and $\Sigma_{ij}$ represent covariance between $i$ and $j$ time bins, and $\mu$ is the expectation value of noise $\mathbf{n}$. 
The \ac{gw} signal buried in strain data are modeled in terms of the parameters $\vec{\Theta}$ described in section \ref{sec:wf_modeling} as $h(t;\Theta)$.  Under these assumptions, search pipelines use matched filter techniques to estimate the SNR as,
\begin{equation}\label{eq:snr_def}
\rho^2 = 4\int_{f_\mathrm{low}}^{f_\mathrm{high}} \frac{|\tilde{s}(f)\tilde{h}(f, \Theta)|}{S_n(f)} df,
\end{equation}
\noindent
where $\tilde{s}(f)$ and $\tilde{h}(f)$ are the functions of frequency and are the Fourier transforms of the time domain quantities $s(t)$ and $h(t)$, respectively. $S_n(f)$ is the power spectral density, which quantifies the noise properties of the \ac{gw} detector. $f_\mathrm{low}$ and $f_\mathrm{high}$ are the low and high-frequency cutoffs for the integration limit. A template bank-based approach is employed in data analysis pipelines for \ac{gw} searches \citep{Prix:2007ks, Usman:2015kfa, Allen:2021yuy}. The SNR time series is evaluated for each point in the template bank. When the SNR surpasses a certain threshold, it is deemed a trigger and is saved for further analysis. Triggers from individual detectors that occur within the light travel time between the detectors are known as coincident triggers. In the next stage of \ac{gw} searches, these coincident triggers are further examined, and their significance is estimated by comparing them with the expected distribution of background triggers. Statistically significant triggers are classified as \ac{gw} merger events \citep{Usman:2015kfa, LIGOScientific:2019hgc}. A more detailed analysis is done using \ac{pe} techniques to determine the waveform model's parameters ($\vec{\Theta}$) \citep{Veitch:2014wba,Biwer:2018osg}.

In order to estimate the parameters of a \ac{gw} merger signal, the standard method is to use a Bayesian framework. For the given data $d$, the likelihood function $\mathcal{L}(d | \vec{\Theta}, I)$ describes the probability of obtaining the data for a given model with any other prior information $I$. The posterior probability distribution $p(\vec{\Theta} | d, I)$ is then calculated using Bayes' theorem,
\begin{equation}
    P(\vec{\Theta} | d, I) = \frac{\mathcal{L}(d | \vec{\Theta}, I) \pi(\vec{\Theta} | I)}{p(d|I)} \label{eqn:bayes_theorem}
\end{equation}
\noindent
where $\pi(\vec{\Theta} | I)$ represents the prior probability distribution for parameters $\vec{\Theta}$ and $p(d|I)$ describes the marginalised likelihood. 

\subsection{Sources of errors in the parameter estimation}
There can be systematic errors present in the \ac{pe} analysis due to any of the following reasons:
\begin{itemize}
\item \textbf{Data analysis artifacts due to mis-modeling the noise:} For certain periods of strain data, the noise in the \ac{gw} detector may not adhere to the assumptions of Gaussianity and stationarity. This deviation can occur due to environmental factors or issues with the instrument itself. In a study, Mozzon et al. \citep{Mozzon2020} found that approximately three percent of the \ac{gw} detector data does not conform to the assumptions of Gaussianity and stationarity. This represents a situation where the noise model fails. If left unaddressed, this could result in biased estimates of the properties of the binary source (systematic errors) or inaccurate statistical uncertainties (statistical errors) \citep{Edy2021, Kumar:2022tto}. 
\item \textbf{Waveform systematics:} With the improvements in the sensitivity of the \ac{gw} detectors and with the addition of more physics in the description of waveform from the \ac{gw} merger, such as the inclusion of higher modes, eccentricity, precision, the requirements for the accuracies of the waveform models are becoming stricter. We have already reached an era where the systematics of the waveform models can not be ignored. If the waveform model does not describe the underlying reality, and if we do not account for the possible errors, the estimates of source properties are expected to be biased.
\item \textbf{Other unaccounted effects:} Even if our models for noise and the waveform describe the strain data perfectly, there can be rare cases such as modification of \ac{gw} waveform by strong lensing by intermediate-mass black hole object or presence of sub-threshold overlapping signals, etc. It can give rise to unmodeled effects, which can again bias the \ac{pe} results.
\end{itemize}
It is crucial to consider any systematic errors that may arise from the above mentioned effects. This work focuses on the systematic and statistical errors arising from inaccuracies in waveform models. In \ac{pe}, the systematic errors can be represented by $|\Delta\mathbf{\Theta}|=|\mathbf{\Theta}_\mathrm{rec}-\mathbf{\Theta}_\mathrm{true}|$, where $\mathbf{\Theta}_\mathrm{true}$ represents the true value of waveform parameters and $\mathbf{\Theta}_\mathrm{rec}$ represents the recovered value of the parameters. For individual waveform parameters, we define $\Delta\Theta^i=|\Theta^i_{rec} - \Theta^i_{true}|$ where $\Theta^i_{rec}$ is the median value of one dimensional marginalized posterior samples for $i$-th waveform parameter. Similarly, the statistical error $\sigma_i$ represents the standard deviation of marginalized posterior distribution of $i$-th waveform parameter. For a given waveform model, the statistical errors depend on the signal loudness and detector sensitivity. We can ignore systematic errors when,
\begin{equation}
|\Delta\Theta^i| << \sigma_i ~~\forall ~~i.
\end{equation}  
Conversely, the systematic errors become important when,
\begin{equation}
 \frac{|\Delta\Theta^i|}{\sigma_i} \sim \mathcal{O}(1) ~~~~\text{for any  i }\label{eqn:systematic_errors_validity_regime}
\end{equation}

\section{Modeling uncertainties in strain}
\label{sec:modeling_strain}
The description of transient signal $h(t, \vec{\Theta})$ can be affected by following:
\begin{itemize}
    \item {\textbf{Calibration uncertainties:}} The \ac{gw} detectors are calibrated periodically to map between input strain and time series output. \ac{gw} detectors are not perfectly calibrated and the calibration uncertainties are provided with detector characterization \citep{ligoLIGOT2100313v3LIGO}.
    \item {\textbf{Inaccuracies in waveform modeling:}} These inaccuracies might arise due to one of many reasons, such as i) errors in calibrating to \ac{nr} waveforms, ii) not accounting for all the physics in the waveform modeling, iii) inherent errors in \ac{nr} simulations.
\end{itemize}
To model general uncertainties in the observed strain data, we go to the frequency domain where the signal, $\Tilde{h}_{\mathrm{obs}}(f)$, can be modeled as,
\begin{equation}
\Tilde{h}_{\mathrm{\mathrm{obs}}}(f) = \Tilde{A}_{\mathrm{obs}}(f)\exp\{\iota\Tilde{\phi}_{\mathrm{obs}}(f)\},
\end{equation}
\noindent
where $A_{\mathrm{obs}}(f)$ and $\phi_{\mathrm{obs}}(f)$ are observed amplitude and phase of the signal. From now on, we will drop writing the explicit frequency dependence. Whenever we present a quantity in the frequency domain, with overhead $\Tilde{ }$, it is assumed that it is generally an explicit function of frequency unless stated otherwise. Now we assume that the observed signal $\Tilde{h}_{\mathrm{obs}}$ can be expressed in terms of `true' amplitude and phase with linear perturbation,
\begin{eqnarray}
    \Tilde{h}_{\mathrm{obs}} &=& \Tilde{A}_{\mathrm{obs}}\exp\{\iota\Tilde{\phi}_{\mathrm{obs}}\} \nonumber \\
    &=& (\Tilde{A}_{\mathrm{true}} + \delta\Tilde{A}_{\mathrm{abs}})\exp{\{\iota (\Tilde{\phi}_{\mathrm{true}} + \delta \Tilde{\phi}_{\mathrm{abs}}}) \}\\
    &=& \Tilde{A}_{\mathrm{true}}(1+\delta\Tilde{A}_{\mathrm{rel}})\exp\{\iota \Tilde{\phi}_{\mathrm{true}}(1+\delta\Tilde{\phi}_{\mathrm{rel}})\}  
\end{eqnarray}
\noindent
where $\Tilde{A}_{\mathrm{true}}$ and $\Tilde{\phi}_{\mathrm{true}}$ are true amplitude and phase of the signal. $\delta\Tilde{A}_{\mathrm{abs}}$ and $\delta\Tilde{\phi}_{\mathrm{abs}}$ are absolute errors in the true amplitude and phase, respectively, while $\delta\Tilde{A}_{\mathrm{rel}}$ and $\delta\Tilde{\phi}_{\mathrm{rel}}$ are relative deviation in true amplitude and phase respectively. The relation between absolute and relative deviation (or errors) are, $\delta\Tilde{A}_{\mathrm{rel}} = \frac{\delta\Tilde{A}_{\mathrm{abs}}}{\Tilde{A}_{\mathrm{true}}}$ and $\delta\Tilde{\phi}_{\mathrm{rel}} =\frac{\delta\Tilde{\phi}_\mathrm{abs}}{\Tilde{\phi}_{\mathrm{true}}}$. 

\subsection{Detector calibration uncertainties}
\begin{figure*}
    \centering
    \includegraphics[scale=0.6]{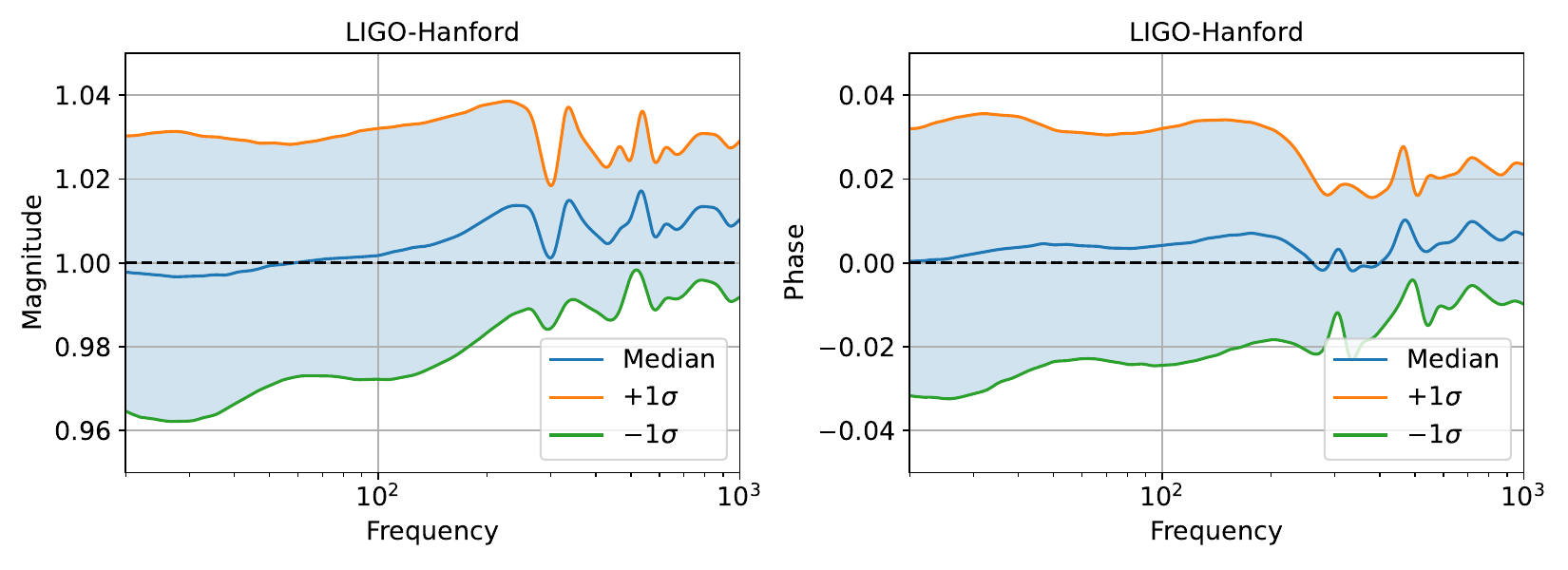}
    \caption{The Figure shows the calibration uncertainties as a function of frequency for the LIGO Hanford detectors around the time of the first \ac{gw} detection: GW150914. The left panel shows the systematic error in the magnitude, while the right panel shows the systematic error in the phase angle (in radians). The shaded region represents the $\pm 1\sigma$ band around the median systematic error. The systematic error is described by the ratio of the true response function to the modeled response function of the detector. The LIGO and Virgo calibration uncertainty files for O1, O2, and O3 observation runs are  available at \citep{ligoLIGOT2100313v3LIGO}.}
    \label{fig:calibration_envelopes}
\end{figure*}

When \ac{gw} passes through the detector, it produces a differential arm displacement between two L-shaped arms relative to the original arm lengths. This differential arm displacement generates laser power fluctuations in the photodetector, reproduced as the time series data. The conversion to the digital strain data regarding time series from the differential arm displacement happens through the control loop and calibration pipeline \citep{2017PhRvD..96j2001C, Sun:2020wke}. The calibration of the strain data are modeled with systematic and statistical errors, which are then provided as calibration uncertainties as function of frequency \citep{Sun:2020wke, Sun:2021qcg, 2017PhRvD..96j2001C, VIRGO:2021kfv, Virgo:2018gxa}. Figure \ref{fig:calibration_envelopes} shows the calibration envelopes for the LIGO-Hanford detector around the event GW150914. 

 The calibration uncertainties are modeled as frequency dependent errors in phase and amplitude as follows \citep{SplineCalMarg-T1400682, LIGOScientific:2017aaj}:
\begin{equation}
    \Tilde{h}_{\mathrm{obs}} = \Tilde{h}_{\mathrm{true}}(1+\delta\Tilde{A}_\mathrm{cal}) \exp{(\iota \delta\Tilde{\phi}_\mathrm{cal})}, \label{eqn:calibration_model}
\end{equation} 
In terms of the general parametrization we discussed above, $\delta\Tilde{A}_\mathrm{cal}$ can be identified as relative error in amplitude while $\delta\Tilde{\phi}_\mathrm{cal}$ is absolute error in the phase. In the calibration uncertainty model, equation \eqref{eqn:calibration_model},  the exponential term can be expanded as a Taylor series and is rewritten as,
\begin{equation}
    \Tilde{h}_{\mathrm{obs}} = \Tilde{h}_{\mathrm{true}} (1+\delta\Tilde{A}_\mathrm{cal}) \frac{2 + \iota \delta \Tilde{\psi}_\mathrm{cal}}{2 - \iota \delta \Tilde{\psi}_\mathrm{cal}},
\end{equation}
where the ratio involving $\delta\Tilde{\psi}$ is designed to have a modulus of unity, thus being a pure phase \citep{SplineCalMarg-T1400682}. In existing \ac{pe} pipelines such as \textsc{PyCBC}-Inference \citep{Biwer:2018osg} and \textsc{BilBy} \citep{Ashton:2018jfp}, the projected strain in each detector is modeled with phase and amplitude correction using cubic-spline curves drawn from the distribution provided by calibration envelopes, as shown in Figure \ref{fig:calibration_envelopes}.

\begin{figure*}
    \centering
    \includegraphics[width=1\textwidth]{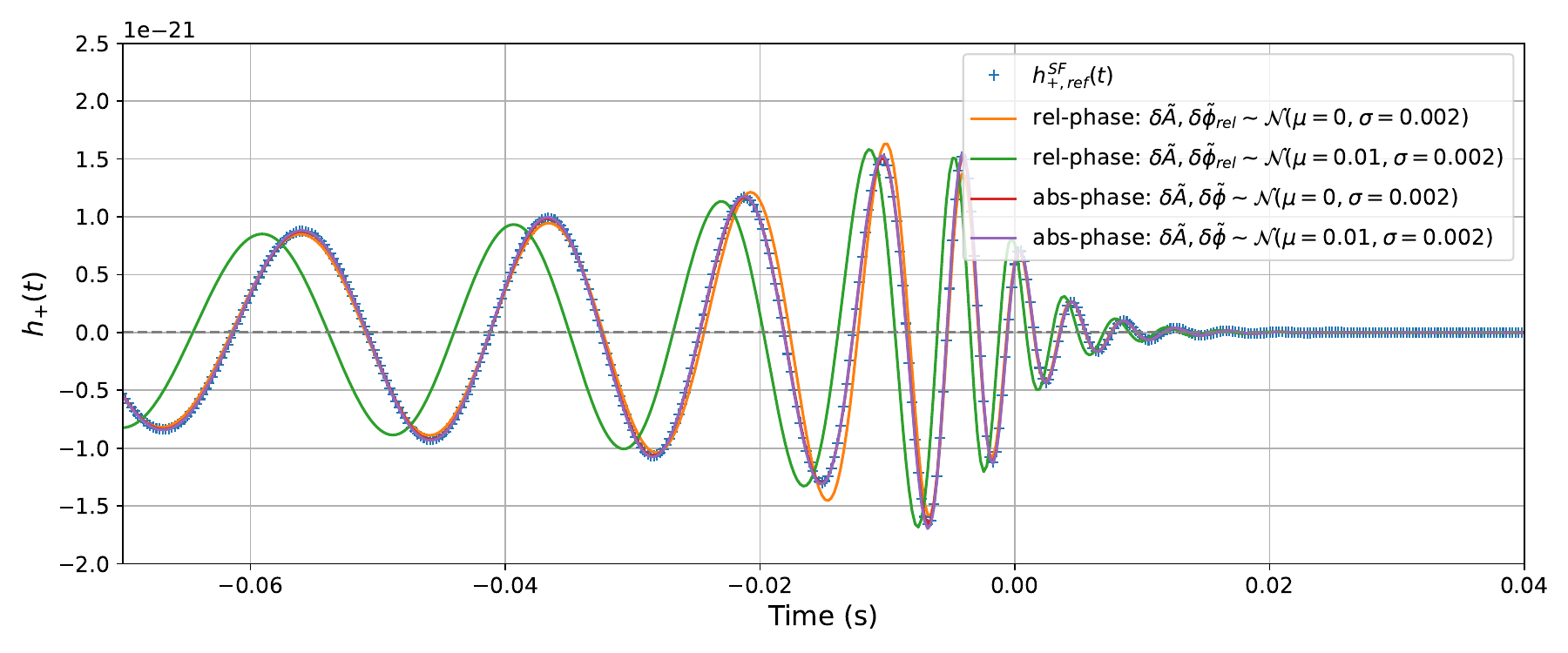}
    \caption{We show the time domain waveform in a signal frame (before projecting it in the detector) for one of the polarization $(+)$ for a GW150914-type signal. Apart from the reference waveform model \phenompvtwo $h_{+, ref}^{SF}(t)$ (blue, denoted by `+' markers), we also show the modification to the reference signal by one of the parametrizations described in the text. Red and violet waveforms represent the abs-phase modification given by \eqref{eqn:wferror_modeling_abs}. We use cubic splines for the parameters $(\delta\Tilde{A}, \delta\Tilde{\phi})$ with a realization from the normal distribution as shown in corresponding plot legends. Orange and green waveforms represent the modification with rel-phase parametrization given by equation \eqref{eqn:wferror_modeling_rel}.}
    \label{fig:wf_modification_general}
\end{figure*}
\subsection{\ac{gw} waveform  model uncertainties}
Motivated by how detector calibration uncertainties are handled, we suggest that uncertainties in waveform models can be treated similarly. Specifically, we propose parametrizing the uncertainties in the description of \ac{gw} waveform in the frequency domain. Furthermore, we suggest that waveform developers provide uncertainty bands with their model. A similar approach was proposed in \citep{Read:2023hkv}. Here, we examine the technical details, feasibility, choices of parametrization, and degeneracy with the detector calibration-correction framework. We use the following parametrizations to account for waveform modeling errors:
\begin{itemize}
    \item {\bf abs-phase}: Relative errors in amplitude and absolute errors in phase:
    \begin{eqnarray}
        \Tilde{h}_{model} &=& \Tilde{A}_{0}(1+\delta\Tilde{A})\exp\{\iota (\Tilde{\phi}_{0} + \delta\Tilde{\phi})\}, \nonumber\\
        &=& \Tilde{h}_{0}(1+\delta\Tilde{A})\exp\{\iota\delta\Tilde{\phi}\}\label{eqn:wferror_modeling_abs},
    \end{eqnarray}
    \item {\bf rel-phase}: Relative errors in amplitude and phase:
    \begin{eqnarray}
        \Tilde{h}_{model} &=& \Tilde{A}_0(1+\delta\Tilde{A})\exp\{\iota \Tilde{\phi}_{0}(1+\delta\Tilde{\phi}_{\mathrm{rel}})\}, \nonumber\\
        &=& \Tilde{h}_{0}(1+\delta\Tilde{A})\exp\{\iota\Tilde{\phi}_{0}\delta\Tilde{\phi}_{\mathrm{rel}}\}\label{eqn:wferror_modeling_rel},
    \end{eqnarray}
\end{itemize}
\noindent
where $\Tilde{h}_0=\Tilde{A}_0\exp{(\iota\Tilde{\phi_0})}$ represent the baseline (or reference) model, which aspires to be a `true' model. As no waveform model is perfect to arbitrary accuracy, we introduce additional terms to account for potential systematics. $\delta\Tilde{A}$ represent the fractional or relative error in the amplitude of the signal, and $\delta\Tilde{\phi}$ ($\delta\Tilde{\phi}_{\mathrm{rel}}$) represent absolute (relative/fractional) errors in phase. 

The abs-phase parametrization described by Eq.~\eqref{eqn:wferror_modeling_abs} is also used in the correction for the detector calibration. Other studies also use this parametrization to quantify the waveform uncertainties \citep{Edelman:2020aqj, Read:2023hkv}. Since $\delta \Tilde{A}$ is the relative change in amplitude, it is a dimensionless parameter in these parametrizations. The units of $\delta\Tilde{\phi}$ is radian in abs-phase parametrization while $\delta\Tilde{\phi}_\mathrm{rel}$ is a dimensionless parameter. 

A point of caution must be mentioned here for description of the relative phase error. The baseline phase $\Tilde \phi_0$ is only uniquely defined up to integer multiples of $2 \pi$. However, the multiplication with $\delta\Tilde{\phi}_{\mathrm{rel}}$ would give the phase error a different physical meaning depending on the chosen branch of the baseline phase. Therefore, a convention must be to introduced to remove this ambiguity. We implement a simple method that ensures the baseline phase to be between $- \pi$ and $\pi$ at the starting frequency of the waveform generation. The rest of the phase evolution follows a standard unwrapping algorithm as implemented in \textsc{Numpy} \citep{harris2020array}. 

We note that alternative choices for removing this ambiguity exist. For example, one could parameterize the phase by an exact value $\Tilde \phi_{\rm ref}$ at a given reference frequency $f_{\rm ref}$. By this convention, the error would vanish at the reference frequency, and one could explicitly set
\begin{eqnarray}
      \Tilde{\phi}_{0}(f) \delta\Tilde{\phi}_{\rm rel}(f) \mapsto \left[ \Tilde{\phi}_{0}(f) - \Tilde{\phi}_{0}(f_{\rm ref}) \right] \delta\Tilde{\phi}_{\rm rel}(f) +  \Tilde{\phi}_{\rm ref}.
\end{eqnarray}
In this formulation, phase errors would accumulate as one moves away (both in the positive and negative $f$-direction) from $f_{\rm ref}$.

From now on, we will drop the subscript `rel' from the $\delta\Tilde{\phi}$ parameter and we will refer these parameters as the pair $(\delta\Tilde{A}, \delta\Tilde{\phi})$, which in general are functions of frequency. These parameters will represent either the abs-phase or rel-phase model, depending on the context of which parametrization is being discussed. We refer to these models as waveform error parametrization models or WF-Error parametrization for short.

Following the calibration uncertainty framework, we use N nodal points (or knots) between the frequency interval $f \in [f_\mathrm{min}, f_\mathrm{max}]$. Here, $ f_{\mathrm{min}} $ serves as the low-frequency cutoff, or the frequency for the first nodal point, while $f_{\mathrm{max}}$ represents the maximum frequency of our analysis which can be a frequency immediately following the merger. In this way, we can account for waveform errors throughout the signal in the detector band. The values of the WF-Error parametrization parameters at the waveform nodal points can be utilized for cubic spline interpolation. This method employs piecewise polynomial curves that are cubic in order and have continuous second derivatives. We utilize the CubicSpline implementation from the publicly available \textsc{SciPy} package \citep{2020SciPy-NMeth} for our WF-Error parametrizations. Using $f_i \in [f_1, f_2, .... f_N]$, which are the nodes of the polynomial in the frequency, we construct $\delta\Tilde{A}(f)$ and $\delta\Tilde{\phi}(f)$ curves using:
\begin{eqnarray}
    \delta\Tilde{A}(f) = P^{3}_s(f; \{ f_i, \delta \Tilde{A}_i\}) \\
    \delta\Tilde{\phi}(f) = P^{3}_s(f; \{ f_i, \delta \Tilde{\phi}_i\}) 
\end{eqnarray}
\noindent
where $P_s^3$ is the cubic spline polynomial. In the \ac{pe} analysis, we provide the priors for $\delta \Tilde{A}_i$ and $\delta \Tilde{\phi}_i$ at each frequency node. Each realization of ($\delta \Tilde{A}_i, \delta \Tilde{\phi}_i$) points corresponds to the curves $\delta\Tilde{A}(f)$, and $\delta\Tilde{\phi}(f)$ which is used to modify the reference waveform model $\Tilde{h}_{ref}(f)$ in accordance to equations \eqref{eqn:wferror_modeling_abs} and \eqref{eqn:wferror_modeling_rel}.

We now turn to the question why we introduce two different phase parametrizations. The SNR (\ref{eq:snr_def}) and the inner product (\ref{eq:inner_prod}) used in \ac{pe} obey a symmetry between the observed data and the \ac{gw} model. Therefore, if one accounts for calibration uncertainties in the data, this should have the same effect as accounting for waveform systematics. Since the calibration uncertainties are modeled with absolute errors in phase, marginalizing over calibration uncertainties can account for waveform systematics if they are of the same order. However, it is worth pointing out that all \ac{gw} Transient Catalogs (GWTCs) included the effect of uncertain detector calibration in the analysis of \ac{gw} events. Despite marginalizing over amplitude and (absolute) phase uncertainty for each detector, the properties of some binaries showed significant systematic differences between the waveform models that have been employed for the analysis (e.g., GW190412 \cite{LIGOScientific:2020stg}, GW191109\_010717, GW191219\_163120, GW200129\_065458; see Sec.~III~E in \cite{KAGRA:2021vkt}). Evidently, marginalizing over calibration uncertainties did not incorporate all waveform systematics.

This is not unexpected. Systematic uncertainties in waveform models can often appear as secular phase drifts caused by numerical errors, (unavoidably) incomplete series expansions, interpolation inaccuracies, or even missing physics. The resulting waveform errors can, therefore, become much more severe than random small phase variations. If one wants to allow for possibly significant waveform uncertainties, large absolute phase errors must be considered, which may lead to an overly pessimistic error estimate.  Alternatively, a small relative phase error is likely more appropriate to model systematic waveform uncertainties at the leading order.

We now turn to incorporating WF-Error parametrizations in \ac{pe} analysis. Ideally, a waveform model shall consist of the best-estimated reference model $\Tilde{h}_{\mathrm{ref}}$, and priors for the parameters $\delta\Tilde{A}$ and $\delta\Tilde{\phi} $ for rel-phase or abs-phase parametrization to account for the waveform systematics. We call them WF-Error priors or error budget. The WF-Error priors should generally depend on the parameter space. We expect these priors to be narrower in the regime where many \ac{nr} simulations are available, and waveform models are calibrated to greater accuracy. For the regions where \ac{nr} simulations are sparse and theoretical waveform description is not so accurate, we expect them to be broader. For a given signal in parameter space, we also expect these priors to be broader for the frequencies close to the merger, where the analytical models extending from the inspiral regime to the merger are calibrated with \ac{nr} simulations.

In the absence of such priors from the waveform modeling side, we can use a conservative approach and use wide enough priors so that we expect to capture any systematic and use data to constrain the parameters related to waveform errors. It might lead to broader posterior distribution resulting in larger variance on signal parameters (especially for weak signals), but it is expected to be unbiased. We will test these assumptions in the section \ref{sec:simulations} with simulated signals and recovery.

\begin{figure}
    \includegraphics[scale=0.42]{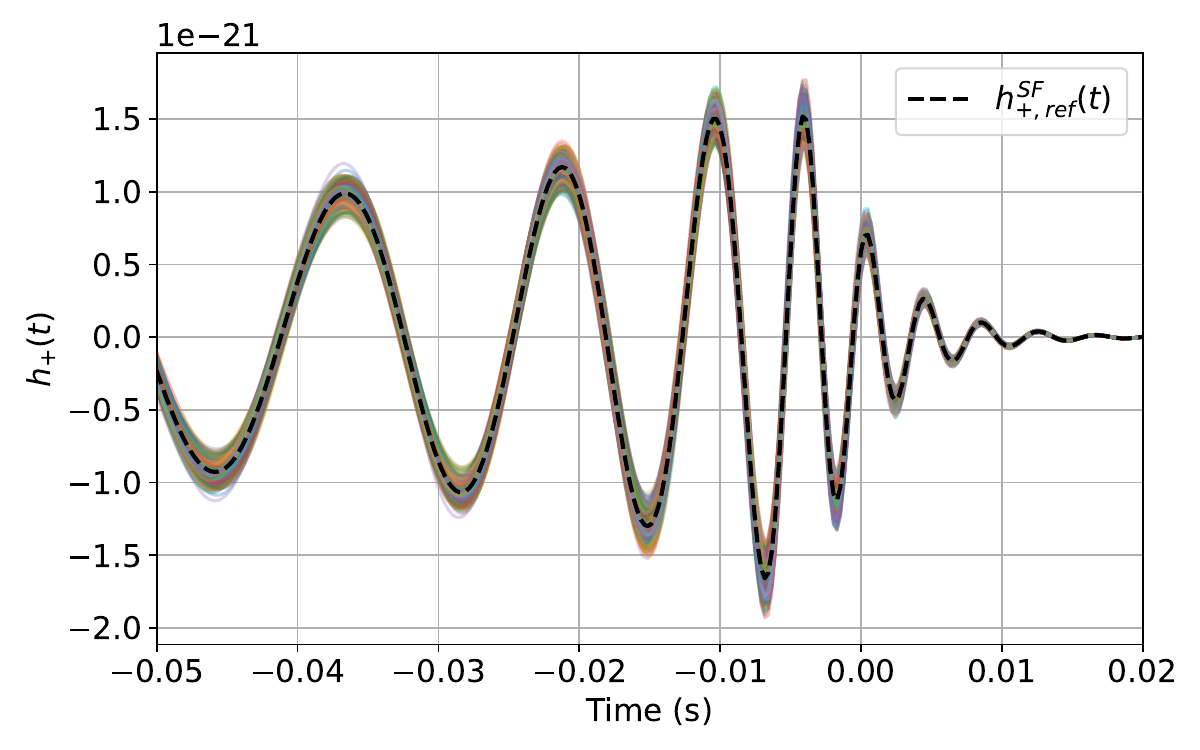}
    \caption{We show the time domain waveform for $(+)$ polarization for the GW150914 type signal. The reference waveform model \phenompvtwo: $\Tilde{h}_{+, ref}^{SF}(t)$ is shown as dashed black curve. Other curves are modified waveform with abs-phase modification, described by equation \eqref{eqn:wferror_modeling_abs}, with WF-Error parameters sampled from the distribution $\delta\Tilde{A}, \delta\Tilde{\phi} \sim \mathcal{N}(\mu=0, \sigma=0.05)$.}
    \label{fig:wf_modification_abs}
\end{figure}

Figure \ref{fig:wf_modification_general} shows an example of waveform with GW150914 type signals and the modifications applied in the frequency domain using cubic splines. To generate the reference signal, we use the \phenompvtwo waveform model with source frame masses $(m^\mathrm{src}_1, m^\mathrm{src}_2)=(36 M_\odot, 29M_\odot)$ at the luminosity distance of 500 Mpc. We used non-spinning injections for this example. For the modifications, we use both the parametrizations described previously with WF-Error parameters chosen from a realization of the Normal distribution described in the figure. For the same values of $\delta\Tilde{\phi}$, the rel-phase parametrization significantly modifies the reference model compared to abs-phase parametrization. In Figure \ref{fig:wf_modification_abs}, we show the reference waveform model for the GW150914 type signal described above along with the modified waveform curves with abs-phase modification. We use cubic splines with ten waveform nodal points in frequencies with values $(\delta\Tilde{A}, \delta\Tilde{\phi})$ drawn from a normal distribution $\mathcal{N}(\mu=0, \sigma=0.05)$.\\

It is important to note that the WF-Error parametrizations (abs-phase and rel-phase) can be transformed into one another using simple mathematical relations by evaluating the phase $\tilde{\phi}_0$ of the reference waveform model $\Tilde{h}_\mathrm{ref}$. We can use one of the parametrizations and mathematical transformations to convert it into the other according to the situation. However, using the correct parametrization is desirable for practical purposes in \ac{pe} analysis. As an example, if the rel-phase parametrization is used for waveform systematics, and if one chooses abs-phase parametrization in \ac{pe}, the priors for $\delta\Tilde{\phi}$ need to become wider for larger frequencies to account for the growing reference phase $\Tilde{\phi}_0$.

\begin{figure}
    \includegraphics[scale=0.42]{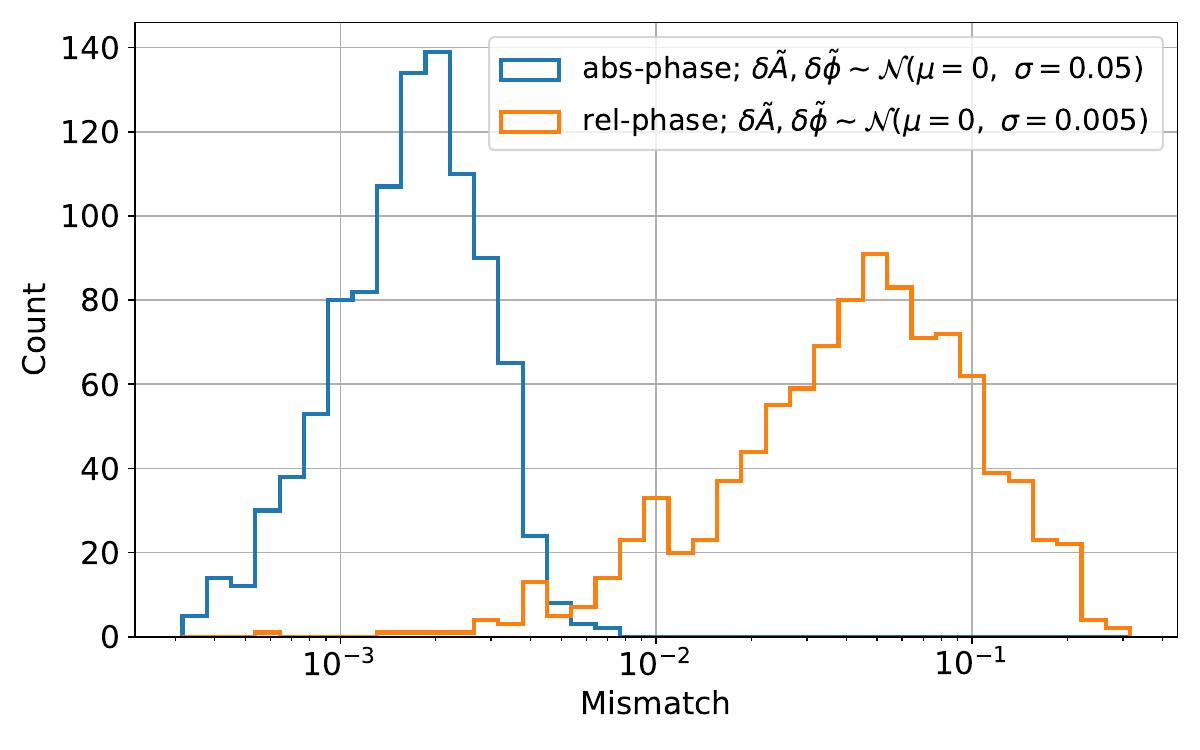}
    \caption{On the x-axis, we show the mismatch between the reference waveform model $h_\mathrm{ref}$ and the modified waveform model $h_\mathrm{mod}$. The modified signal, $h_\mathrm{mod}$, is generated by either abs-phase or rel-phase WF-Error parametrization. The WF-Error parameters are taken from the normal distribution, and we generate cubic spline curves to modify the reference model. The rel-phase model is generally well suited to model comparatively more significant mismatches.}
    \label{fig:mismatch_abs_and_rel_parametrization}
\end{figure}

We can use hints from the mismatch (see Appendix \ref{appendix:FM} for definition) studies to decide which WF-Error parametrization is better suited for practical purposes. In Figure \ref{fig:mismatch_abs_and_rel_parametrization}, we show the distribution of mismatches between the reference waveform model and the modified one using both parametrizations. We use the Gaussian prior with zero mean and a fixed standard deviation across frequency range to generate the modified signal $h_\mathrm{mod}$.
As expected, even with the narrower Gaussian prior, the rel-phase parametrization produces significant deviation to the $h_\mathrm{ref}$ compared to the abs-phase model, hence more significant mismatches. When we are in the parameter space where larger mismatches are expected, we can set our \ac{pe} analysis framework to use rel-phase parametrization.

\subsection{Degeneracy between detector calibration and waveform uncertainties}
A comprehensive \ac{pe} data analysis pipeline for \ac{pe} should include i) methods to incorporate uncertainties in waveform modeling, ii) corrections due to detector calibration, and iii) adjustments for possible deviations from Gaussian and stationary noise, whenever applicable. If we choose cubic-spline curves to mitigate waveform uncertainties and detector calibration uncertainties, a natural question arises: can these systematics be decoupled? We will explore the specifics of this query in a future study. However, we would like to point out that there are fundamental differences in how these corrections are applied: While the calibration corrections are applied in the detector frame and are different for each detector, the WF-Error parametrization corrections are applied before projecting the waveform into each detector. Therefore, the modification due to the calibration uncertainties and abs-phase waveform errors can be decoupled for a sufficiently loud signal.

\section{Can Fisher Matrix Estimates be used for Accounting Systematic Biases?}
\label{sec:FM}

A standard tool in the literature to quantify systematic biases between two waveform models is the Fisher matrix formalism \cite{Dhani:2024jja, dupletsa2024}. In this section, we will apply it to see how significantly errors modeled as \eqref{eqn:wferror_modeling_rel} change the recovery of source parameters $\Theta^\mu$ and verify whether the formalism can account for the induced biases.  If it can, this will help us predict the systematic parameter errors induced by externally provided estimates of waveform errors.  Then we can assess the significance of these errors and determine if corrections are necessary. For the science case for next-generation \ac{gw} detectors, many investigations focus on forecast studies using Fisher matrix formalism \citep{Iacovelli:2022bbs, Iacovelli:2022mbg, Li:2021mbo}. This section specifies the conditions under which corrections based on Fisher matrix analysis are applicable or not. Appendix \ref{sec:fm_formalism} introduces the Fisher matrix formalism. 

\subsection{Setup of the analysis}

In this part, we introduce methods to test how well the Fisher estimates work in trying to recover biases induced by certain choices of $\delta\Tilde{A}, \delta\Tilde{\phi}$. We do that by injecting data generated from the signal model \eqref{eqn:wferror_modeling_rel} for chosen values $\delta\Tilde{A}, \delta\Tilde{\phi}$, and recovering using the baseline waveform model. The injection parameters chosen for these analyses are quoted in table \ref{tab:inj_params}. 
We focus on non-spinning binaries and keep the extrinsic parameters $\psi, \alpha, \delta$ fixed, so in the end only $\mathcal{M}^z, q, D_L, \iota$ and time of coalescence $t_0$, global phase $\phi_0$ are varied. Accordingly, the set of parameters constituting the Fisher matrices we calculate is
\begin{equation}\label{eq:fisher_params}
    \{\mathcal{M}^z, q, D_L, \iota, t_c, \phi_0\} \, .
\end{equation}
Please refer to Appendix \ref{appendix:FM} for details on the calculation.

\begin{table*}
    \centering

\renewcommand{\arraystretch}{1.5}  

\begin{ruledtabular}
\begin{tabular}{cccccccccccc}

     & $m_1^z \, [M_\odot]$ & $m_2^z \, [M_\odot]$ & $D_L \, [\mathrm{Mpc}]$ & $\iota \, [\mathrm{rad}]$ & $f_\mathrm{ref} \, [\mathrm{Hz}]$ & $\mathrm{ra} \, [\mathrm{rad}]$ & $\mathrm{dec} \, [\mathrm{rad}]$ & $\psi \, [\mathrm{rad}]$ & Detector & $t_\mathrm{gps} \, [\mathrm{s}]$ & SNR \\
 \hline
    Base Signal & $20$ & $10$ & $500$ & $0.2$ & $20$ & $4.67$ & $0.65$ & $2.34$ & Hanford & $1192529720$ & $16.78$
    
\end{tabular}
\end{ruledtabular}

    \caption{Injection Parameters for the base signal. In the high-SNR case, the luminosity distance is changed, now taking a value of $100 \, \mathrm{Mpc}$ instead and accordingly, the SNR changes to $83.91$. Note that the SNR values quoted here refer to the non-modified injection, for the modified one the SNR will be slightly different (with deviations being on the order of $1\%$).\\
    We use priors that are uniform in component mass for $\mathcal{M}^\mathrm{z}, q$, uniform in $\cos(\iota)$ for the inclination (i.e.~$p(\iota) \propto \sin(\iota)$), uniform in comoving volume for the luminosity distance, and uniform in coalescence time $t_c$ (no prior for $\phi_0$ because we use a marginalized-phase likelihood).}
    \label{tab:inj_params}
\end{table*}

\begin{table*}
    \centering

\renewcommand{\arraystretch}{1.5}  

\begin{ruledtabular}
\begin{tabular}{cccccccccc}
     \multicolumn{2}{c}{\multirow{2}*{Configuration}} &
     \multicolumn{2}{c}{$\Delta \mathcal{M}^z \, [M_\odot]$} & \multicolumn{2}{c}{$\sigma_{\mathcal{M}^z} \, [M_\odot]$} & \multicolumn{2}{c}{$\Delta q$} & \multicolumn{2}{c}{$\sigma_q$}
    \\
     & & PE & Fisher & PE & Fisher & PE & Fisher & PE & Fisher
     \\
    \hline
    \multirow{2}*{$\delta \Tilde{A} = \delta \Tilde{\phi} = 0.004$}
    
     & Base Signal & $-0.023$ & $-0.026$ & $0.040$ & $0.039$ & $-0.012$ & $0.001$ & $0.143$ & $0.137$ \\
    
     & high-SNR & $-0.028$ & $-0.026$ & $0.0077$ & $0.0078$ & $0.007$ & $0.001$ & $0.0271$ & $0.0273$ %
    \\
    \hline
    \multirow{2}*{$\delta \Tilde{A} = \delta \Tilde{\phi} = 0.01$}
     & Base Signal & $-0.064$ & $-0.04$ & $0.040$ & $0.039$ & $-0.0048$ & $-0.057$ & $0.143$ & $0.137$ 
    \\
     & high-SNR & $-0.069$ & $-0.039$ & $0.0078$ & $0.0078$ & $0.013$ & $-0.058$ & $0.0275$ & $0.0273$ %
\end{tabular}
\end{ruledtabular}

    \caption{Summary of results from full \ac{pe} runs and Fisher matrix estimates. The numbers one should primarily compare the neighboring columns labeled "\ac{pe}" and "Fisher", for each of the four quantities that appear in the upper row. Moreover, when comparing them one should keep in mind that neither number is expected to be perfectly accurate, explaining certain small deviations; \ac{pe} results are always subject to sampling uncertainties (which are noticeable for instance for $\Delta q$), while limitations of the Fisher approach have been laid out at the beginning of this section and Appendix \ref{appendix:FM}.}
    \label{tab:fisher_results}
\end{table*}

There are two properties of the posteriors that we wish to compare to the respective Fisher matrix estimate:
\begin{enumerate}
    \item First of all, the systematic error $\Delta \Theta^\mu$. From \eqref{eq:sys_bias} we see that it is formally defined as the difference between maximum a posteriori estimate $\thetabf$ and injected parameters $\thetatr$.
    
    In practice, we follow the method chosen in the gravitational-wave transient catalogs \cite{gwtc1, gwtc2, LIGOScientific:2021usb, KAGRA:2021vkt}, and use median values as proxies for the maximum a posteriori points.

    \item Secondly, the posterior width. Ref.~\cite{Vallisneri_2008} shows that, in the high-SNR limit, it reads
    \begin{equation}\label{eq:post_width}
        \sigma_{\Theta^\mu} \simeq \sqrt{(\Gamma^{-1})^{\mu \mu}}
    \end{equation}
    and this number is compared to the standard deviation of the marginalized posterior for $\Theta^\mu$.
\end{enumerate}

\subsection{Results of comparing Fisher matrix estimates with PE runs}
In total, we have analyzed four scenarios: two different sets of error values of $\delta\Tilde{A} = 0.004, \delta\Tilde{\phi} = 0.004$ and $\delta\Tilde{A} = 0.01, \delta\Tilde{\phi} = 0.01$, and for each of them one moderate- and one high-SNR case (cf.~table \ref{tab:inj_params}). All of the signals have been injected as a zero-noise injection into the Hanford detector with an O4 PSD. The results are summarized in table \ref{tab:fisher_results}.

Since the chosen signal lies in the non-spinning part of the parameter space, the remaining intrinsic parameters are the masses parametrized by the tuple $\mathcal{M}^\mathrm{z}, q$. We will direct our attention toward these two, not discussing biases in the luminosity distance $D_L$ and inclination $\iota$ in detail. This is because the uncertainty in $D_L$ is typically very large. Moreover, there is a known correlation between these two parameters, which makes their recovery and the assessment of their bias results more complicated. We have verified that this correlation has no significant impact on results for other parameters.

The first important finding is that the \ac{pe} results are biased if a modified signal is recovered using the baseline, non-modified model. This finding implies that the error model \eqref{eqn:wferror_modeling_rel} is capable of parametrizing systematic errors in the way we intend it to. Moreover, we find that these biases can be recovered in the Fisher matrix framework, validating that the biases have a physical origin rather than being an artifact of flawed application of \ac{pe} or an error in the implementation. The statistical bias is recovered very well in all four scenarios, with differences usually appearing in the second significant digit. This agreement is an interesting feature of the results because it indicates that the width estimate \eqref{eq:stat_bias_avg} seems to be more robust to increasing waveform differences than the systematic bias estimate \eqref{eq:sys_bias} (at least for our applications), where only the first configuration $\delta\Tilde{A} = \delta\Tilde{\phi} = 0.004$ shows consistency between Fisher matrix estimates and PE results.

Those results motivate the following interpretation: in case of $\delta\Tilde{A} = \delta\Tilde{\phi} = 0.004$, the waveform differences introduced into the waveform are still well approximated by the linear expansion \eqref{eq:lsa}, the LSA is justified; this clearly changes for $\delta\Tilde{A} = \delta\Tilde{\phi} = 0.01$, where the systematic bias estimates begin to diverge from the corresponding \ac{pe} results -- the LSA is not a faithful approximation anymore. This divergence happens in various ways, with both magnitude and sign being sources of inconsistency.

We have seen that Fisher estimates can work, but not always. First of all, it was expected that the LSA will not hold up to arbitrarily large values of $\delta\Tilde{A}, \delta\Tilde{\phi}$, simply because we use the relative error model where errors accumulate from low to high frequencies and thus will quickly result in large waveform differences. However, this being the case for a $1\%$ error and the simplest setup we can think of (constant shifts, same deviations in both polarizations) has serious implications for our objectives of a practical application of the Fisher estimates: for realistic scenarios, where errors of this magnitude are expected, they are not a suitable prospect to predict systematic errors in intrinsic parameters induced by $\delta \tilde{A}, \delta \tilde{\phi}$. Of course, in other parts of the parameter space, other thresholds for the LSA validity might exist, but the existence of such a counterexample in a part of the parameter that is certainly relevant for detected signals is already sufficient to recognize that we should not rely on a faithful estimation of biases. The numbers are particularly concerning for the high-SNR signal with $\delta\Tilde{A} = \delta\Tilde{\phi} = 0.01$ since this corresponds to the to data from next-generation \ac{gw} detectors, where a tenfold increase in sensitivity is aimed for. Here, Fisher estimate and true recovery are basically fully separate. All of this plays into why we choose to not use Fisher matrix estimates for the rest of this paper. A suitable replacement to still account for these systematic errors by using waveform with an error model like \eqref{eqn:wferror_modeling_rel} on top of them is discussed in the next section.

\section{Simulations and Tests of Parameter Estimation Framework}
\label{sec:simulations}

This section uses simulations to test the WF-Error parametrizations described in section \ref{sec:wf_modeling}. Though we do not have access to the `true' waveform model, we can still check the validity of the \ac{pe} framework we describe by considering a reference waveform model, $h_{\mathrm{ref}}$, modifying the $h_{\mathrm{ref}}$, and checking if our parametrization can pick up the introduced modification. We take a reference signal $h_{\mathrm{ref}}$ described by a waveform approximant and introduce errors in the model described by parametrizations in equations \eqref{eqn:wferror_modeling_abs} and \eqref{eqn:wferror_modeling_rel}. Now we have a modified signal $h_{\mathrm{mod}}$, which shall have a systematic error compared to the $h_{\mathrm{ref}}$. We consider $h_{\mathrm{mod}}$ as the `true' signal and inject it in the strain data of LIGO detectors at Hanford (H1), Livingston (L1), and Virgo detector (V1). Then, we use the Bayesian \ac{pe} data analysis pipeline to recover the source properties using $h_{\mathrm{ref}}$ as our waveform model with and without incorporating waveform uncertainties in \ac{pe}. We use non-spinning signals (spin parameters in waveform approximant set to zero) with zero noise, except the last subsection where we deal with missing physics. In the subsequent subsections, we describe each injection-recovery campaign's injected signals and \ac{pe} scheme.

\begin{table*}[]
\centering
\begin{ruledtabular}
\begin{tabular}{llll}
Recovery Model &
  Sampling Parameters &
  Description &
  Prior \\ \hline
\multirow{7}{*}{IMRPhenomPv2} & $\mathcal{M}^\mathrm{src}$ & Source frame chirp mass &\multirow{2}{*}{Uniform in component masses} \\
 &
  $q = \frac{m_2}{m_1}~(m_1 \geq m_2)$ &
  Mass ratio &
  \\
 &
  $V_c$ &
  Comoving volume &
  Uniform: $\pi(V_c) = \mathcal{U}(5\times10^3, 9\times10^9)$ \\
 &
  $\iota$ &
  {Inclination angle} &
  $\pi(\iota) \propto \sin(\iota)$   \\
 &
  $t_c$ &
  Time of arrival &
  Uniform: $\pi(t_c) = \mathcal{U}(t_c-0.1, t_c+0.1)$ \\
 &
  RA, dec &
  \begin{tabular}[c]{@{}l@{}}Right ascension \\ and declination\end{tabular} &
  Uniform Sky \\
 &
  $\psi$ &
  Polarization Angle &
  Uniform angle: $\pi(\psi) = \mathcal{U}(0,2\pi)$ \\ \hline
\begin{tabular}[c]{@{}l@{}}IMRPhenomPv2 + \\ WF Error\end{tabular} &
  $\delta A_i, \delta \phi_i$ & WF-Error parameters
   &
  $\mathcal{N}(\mu=0, \sigma=0.032)$
\end{tabular}
\end{ruledtabular}

\caption{This table describes the priors used in the \ac{pe} of simulated signals. The distribution $\mathcal{U}(a,b)$ refers to a one-dimensional uniform distribution defined within the interval $(a,b)$. For the WF-Error parameters, when we employ cubic spline curves with ten nodal points across the frequency range of $(20, 500)$ Hz. The prior distributions for each $\delta\Tilde{A}_i$ and $\delta\Tilde{\phi}_i$ are specified accordingly. The same prior is used for the special test cases, when we do not use cubic spline but a constant shift in amplitude and phase with parameters $\delta \Tilde{A}, \delta\Tilde{\phi}$. $\pi(\theta)$ denotes the prior distribution used for a parameter $\theta$.}
\label{tab:pe_priors}
\end{table*}
For \ac{pe}, we use the publicly available toolkit \textsc{PyCBC} Inference~\citep{Biwer:2018osg}, and for sampling the likelihood function, we use the nested sampling algorithm implemented in the publicly available code \textsc{Dynesty}~\citep{speagle:2019}. We also implement WF-Error parametrizations in code that serves as a plugin \citep{GitHubPlugin} for \textsc{PyCBC}. We use the \textsc{aLIGOZeroDetHighPower} PSD for the LIGO Hanford and Livingston detectors and the \textsc{AdvVirgo} PSD for the Virgo detector. These PSD functions are implemented in publicly available modules \textsc{LALSimulation}, which is a part of the \textsc{LALSuite} code~\citep{lalsuite}. We use two sets of runs for each injection: 
\begin{itemize}
\item In first set of run, we sample the parameter space with the following parameters: source frame chirp mass ($\mathcal{M}^\mathrm{src}$), mass ratio $q (q=\frac{m_1}{m_2}, m_1 \geq m_2)$, comoving volume, inclination angle, and geocentric time of arrival of the signal ($t_c$). We fix right ascension (RA), declination (dec), and polarization angle ($\psi$) for these runs.
\item In the second sets of runs, in addition to the parameters described above, we also vary ra, dec, and $\psi$.
\end{itemize}

We transform the sampling parameters to the waveform model parameters, such as detector frame masses ($m_1^\mathrm{z}, m_2^\mathrm{z}$ ), and luminosity distance ($D_L$) through standard transformations. We fix the spins to be zero for the study in this manuscript, except for in sub-section \ref{subsec:missing_physics}. For the \ac{pe} runs described by WF-Error parametrizations, we use additional parameters. We use two additional WF-Error parameters for the constant shift models: $\delta\Tilde{A}$ and $\delta\Tilde{\phi}$. For the cubic-spline model, we use ten frequency nodal points with log spacing between a minimum frequency of 20 Hz and a maximum frequency of 500 Hz.

In Table \ref{tab:pe_priors}, we describe the prior distribution for each of the abovementioned parameters. All the injections described in this section are zero-noise injections.

\subsection{Simplest modification: A constant shift in amplitude and phase with rel-phase parametrization}
We consider a GW1500915-like signal with source frame masses $m^\mathrm{src}_1, m^\mathrm{src}_2=(36~M_\odot,~29~M_\odot)$ at a distance of 500 Mpc. We generate the corresponding reference signal using the \phenompvtwo waveform model and introduce the errors in reference signal with rel-phase parametrization corresponding to: 
\begin{equation}
    (\delta\Tilde{A}, \delta\Tilde{\phi}_{\mathrm{rel}}) = (0.01, 0.01)
\end{equation}

\begin{figure*}
    \centering
    \subfigure{\includegraphics[width=0.49\textwidth]{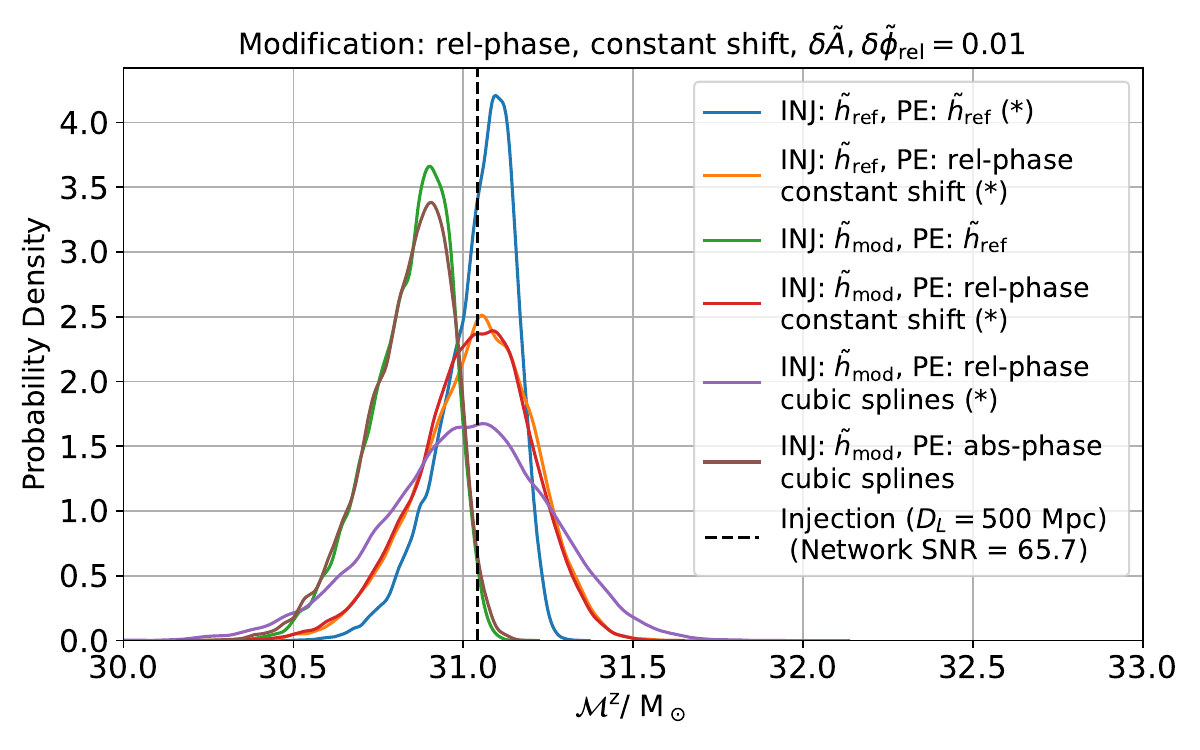}} 
    \subfigure{\includegraphics[width=0.49\textwidth]{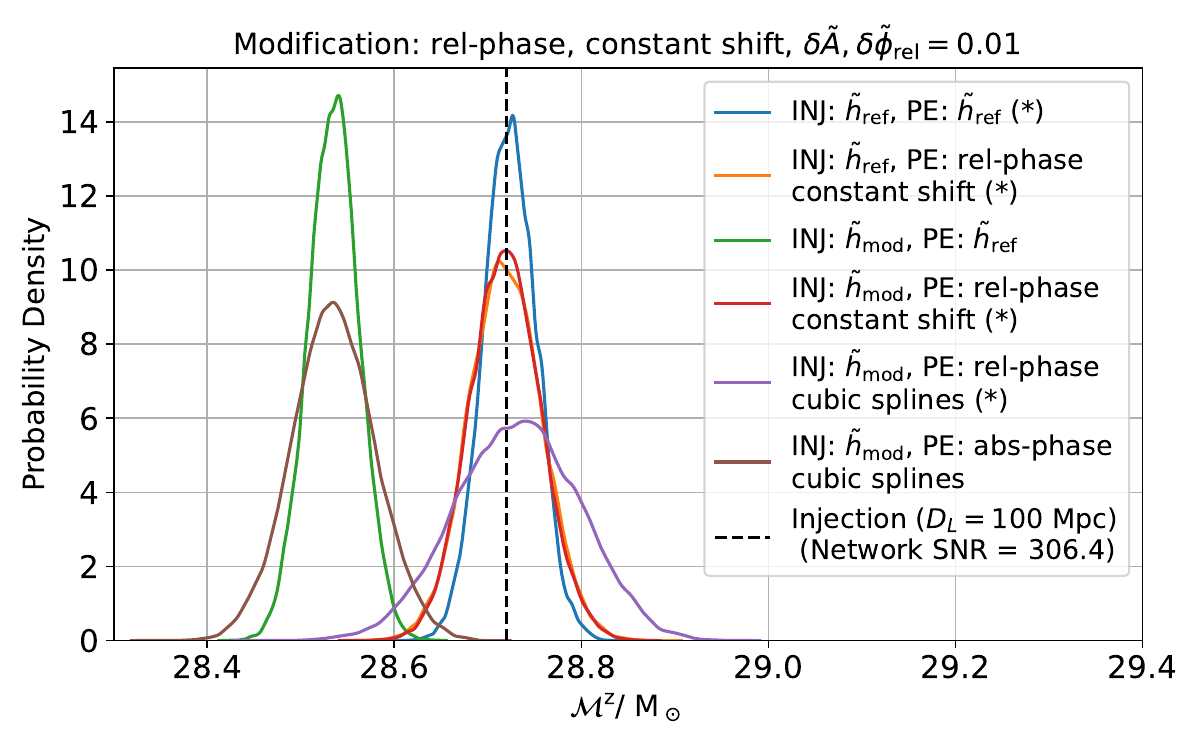}} 
    \caption{Injection and recovery are shown for a non-spinning GW150914-like signal with source frame masses $(m^{src}_1, m^{src}_2)=(36,29)~M_{\odot}$, at a luminosity distance of 500 Mpc (left panel) and a distance of 100 Mpc (right panel). We also inject a modified signal with rel-phase modification with parameters $(\delta\Tilde{A}, \delta\Tilde{\phi})=(0.01,0.01)$. When we inject a modified signal and use the incorrect model to recover, we see a bias in the recovery of detector frame chirp mass $\mathcal{M}^\mathrm{z}$. We also show that using the correct parametrization (in this case, rel-phase) can correct the bias and get a broader marginalized posterior sample. The bias is more prominent in a louder signal (right panel) than in a comparatively weaker signal (left panel). We mark that combination of injection and recovery by an asterisk ($\star$) where we do not expect a bias. The vertical dashed line represents the injected value of $\mathcal{M}^\mathrm{z}$. For these \ac{pe} runs, we fix the parameters RA, dec, and polarization to the injected values. The network SNR for reference injection is shown in each panel.}
    \label{fig:mchirp_bias_mod_rel_cs}
\end{figure*}

\begin{figure*}
    \centering
    \subfigure{\includegraphics[width=0.49\textwidth]{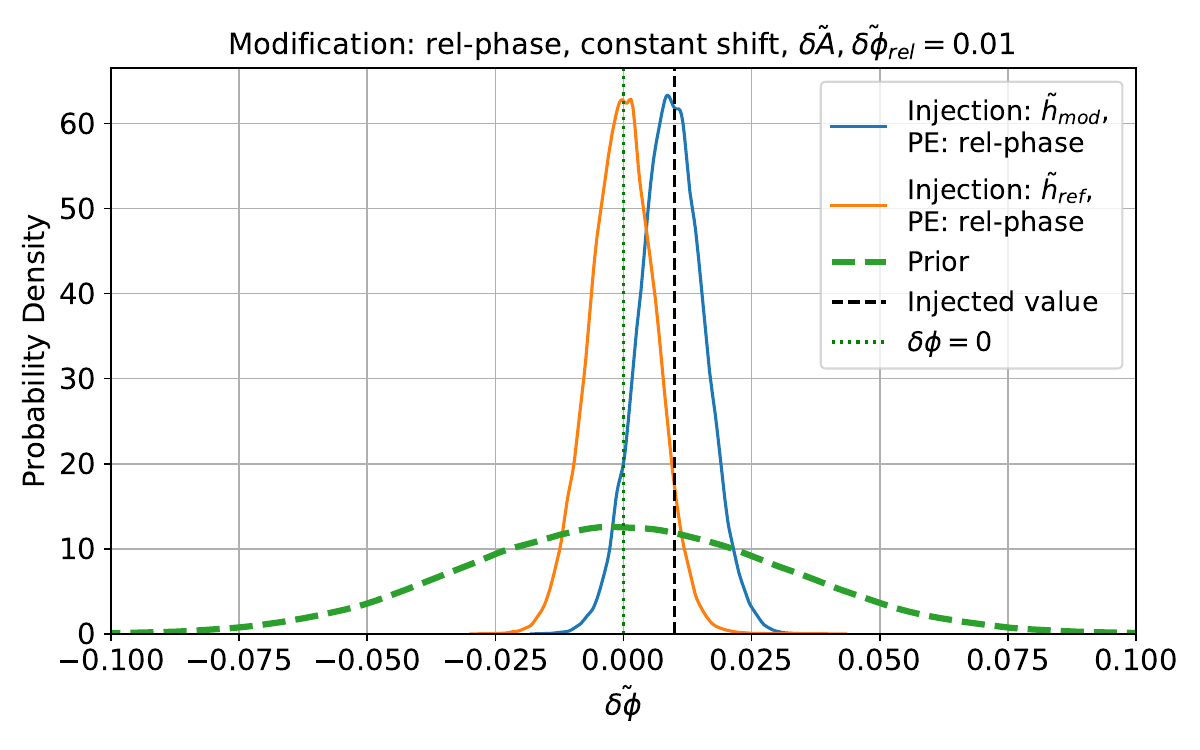}} 
    \subfigure{\includegraphics[width=0.49\textwidth]{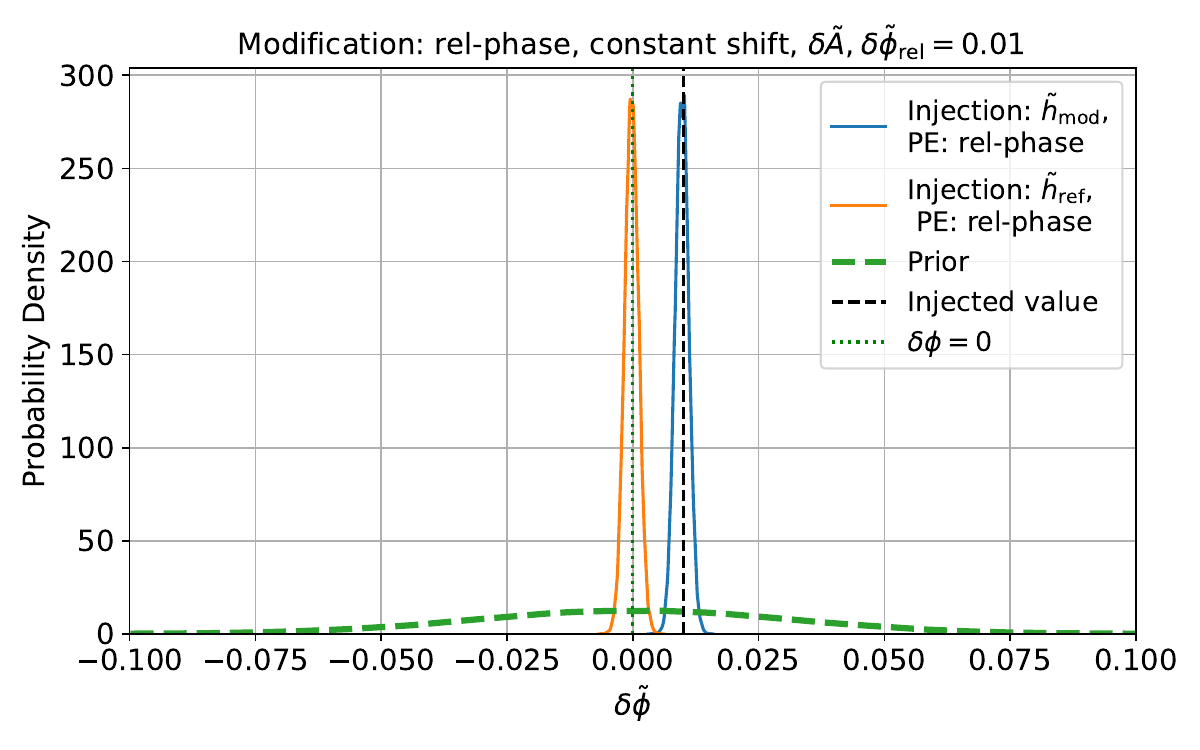}} 
    \caption{Recovery of the parameter $\delta\Tilde{\phi}$ for rel-phase parametrization for various scenarios of injected parameters. The signal injected here is a non-spinning GW150914-like, as in Figure~\ref{fig:mchirp_bias_mod_rel_cs}. The left panel has an injection at 500 Mpc, while the right panel shows the case with an injection at 100 Mpc. The vertical dashed line shows the injected value for $\delta\Tilde{\phi}$ for the modified signal. The green dashed curve shows the prior distribution for the $\delta\Tilde{\phi}$ parameter. We can recover the injected value of $\delta\Tilde{\phi}$ with the correct parametrization for \ac{pe}. We recover $\delta\Tilde{\phi}=0$ for the reference injection. For these \ac{pe} runs, we fix the parameters RA, dec, and polarization to the injected values.}
    \label{fig:dphi_mod_rel_cs}
\end{figure*}

For this most straightforward case, a constant phase shift is expected to be significant for rel-phase parametrization. For abs-phase parametrization, a constant phase shift will not introduce any physical effect except a constant shift in time. We will consider the general modification with abs-phase parametrization in the following subsections. We perform \ac{pe} with the injected signal in the detector network, HLV, with the following combination:
\begin{itemize}
\item \textbf{Injection: standard, \ac{pe}: standard} In this scenario, we inject the reference signal as described above and perform the \ac{pe} with the same waveform model without introducing the WF-Error parametrization.
\item \textbf{Injection: modified, \ac{pe}: standard} In this scenario, we inject the modified signal and perform the \ac{pe} with the reference waveform model without introducing the waveform errors.
\item \textbf{Injection: modified, \ac{pe}: WF-Errors} In this scenario, we inject the modified signal, perform the \ac{pe} with the reference waveform model + WF-Error parametrization. 
\item \textbf{Injection: standard, \ac{pe}: WF-Errors} Here, we inject the reference signal, perform the \ac{pe} with the reference waveform model+WF-Error.
\end{itemize}

In Figure \ref{fig:mchirp_bias_mod_rel_cs}, we show the injection and recovery of various combinations. We show that when we have a modified injection (equivalent to a systematic error with rel-phase parametrization), and we perform \ac{pe} with reference waveform model, we get a bias in the detector frame chirp mass $\mathcal{M}^\mathrm{z}$. This bias is corrected when we use the right parametrization to account for waveform systematics. We also note that we can correct the bias when we use cubic splines for constant shift error in the reference signal. However, it is to be noted that when we use abs-phase parametrization to correct for the bias introduced by rel-phase parametrization, it fails to correct for the bias. These effects are more visible in a louder signal, which is simulated by injection at a closer luminosity distance (100) Mpc compared to one at 500 Mpc. We also show the network SNR, $\rho^\mathrm{net}=\sqrt{\sum_i\rho_i^2}$, for the reference waveform injection and recovery, where $\rho_i$ is the SNR for the $i$-th detector. Figure \ref{fig:dphi_mod_rel_cs} shows the recovery of parameter $\delta\Tilde{\phi}$ with rel-phase parametrization. We show that we can recover the injected value of $\delta\Tilde{\phi}$ with the correct parametrization. When there is no modification to the reference signal, the rel-phase parametrization recovers $\delta\phi=0$. We fix the RA, dec, and $\psi$ for the runs shown in figures \ref{fig:mchirp_bias_mod_rel_cs} and \ref{fig:dphi_mod_rel_cs}. We find that the $\delta\Tilde{A}$ parameter returns the prior distribution and does not have a constraining power for the \ac{pe} runs described in this subsection. For simplistic cases, we expect $\delta\Tilde{A}$ to be degenerate with distance, i.e., a small constant shift in the amplitude can be modeled in the luminosity distance. At this stage, it can be argued that we can get rid of $\delta\Tilde{A}$ parameter in the WF-Error parametrization and focus on $\delta\Tilde{\phi}$ parameter. However, we shall keep this parameter in our model for more general scenarios where errors in amplitude can be a function of the frequency. We do not find any bias in the other parameters considered in the simulations in this subsection, except for the detector frame chirp mass.
\begin{figure}
    \includegraphics[scale=0.42]{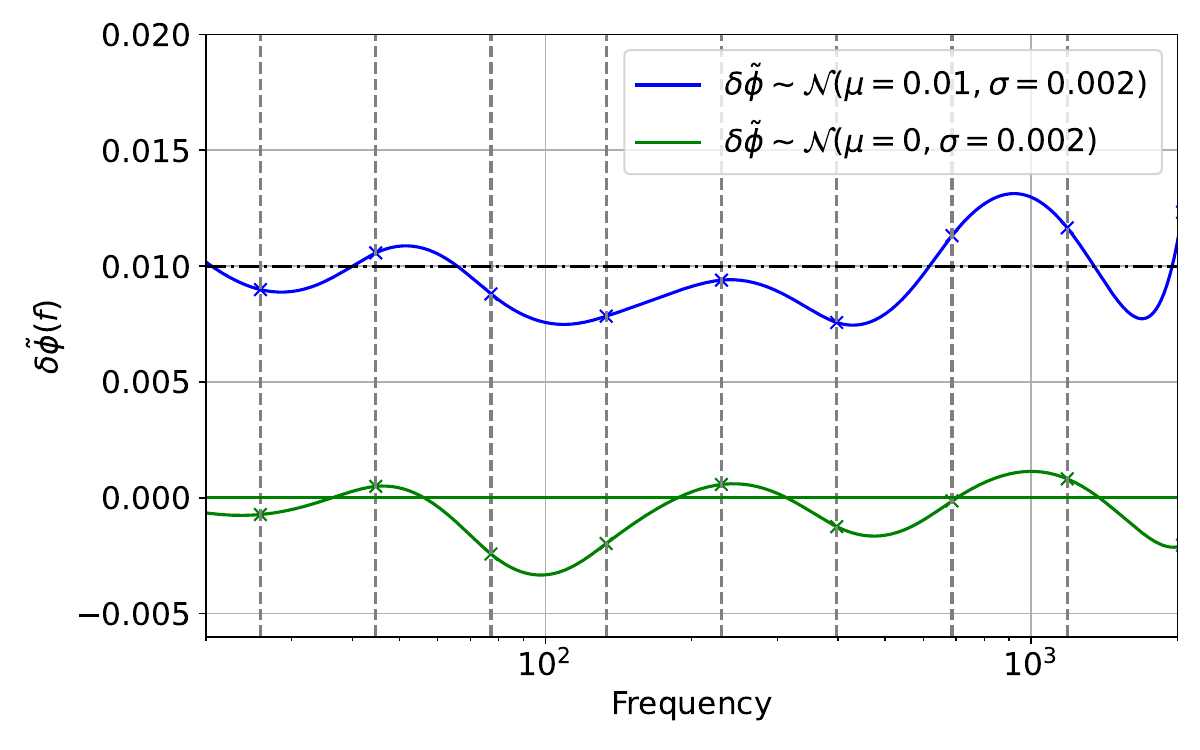}
    \caption{We show a realization of cubic spline curves with ten frequency nodal points (vertical dashed curves) for $\delta\Tilde{\phi}_i$ taken from two normal distributions: one with zero mean (green curve) and the another with non-zero mean $\mu=0.01$ (blue curve). The points marked with a cross (x) are the realization of the normal distribution. In reality, these curves will represent a general function of frequency that captures the waveform systematics or deviation from the `true' waveform model.}
    \label{fig:cubicspline_envelop}
\end{figure}
\begin{figure*}
    \centering
    \subfigure[]{\includegraphics[width=1.\textwidth]{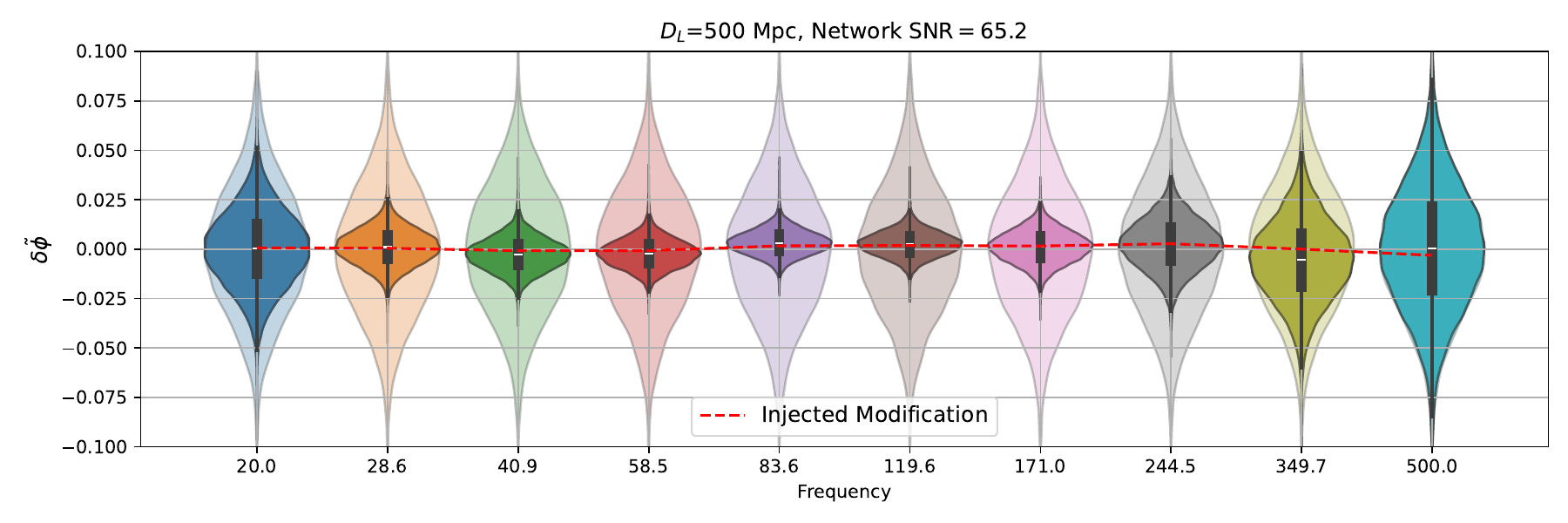}}\\
    \subfigure[]{\includegraphics[width=1.\textwidth]{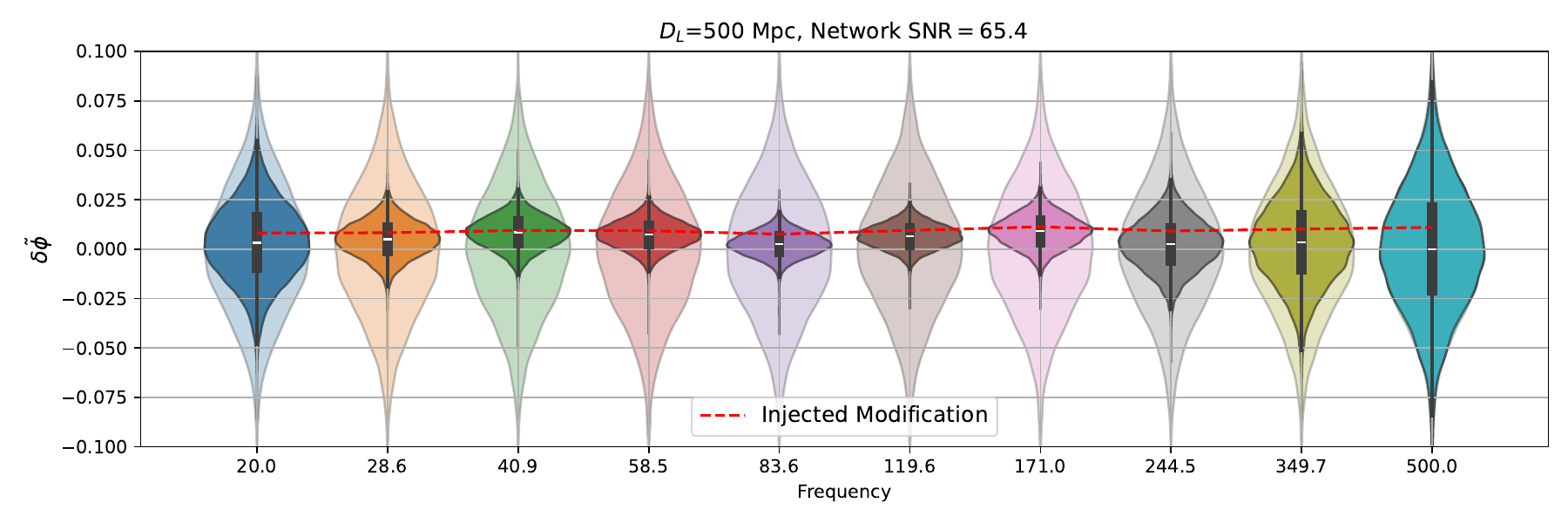}}\\
    \subfigure[]{\includegraphics[width=1.\textwidth]{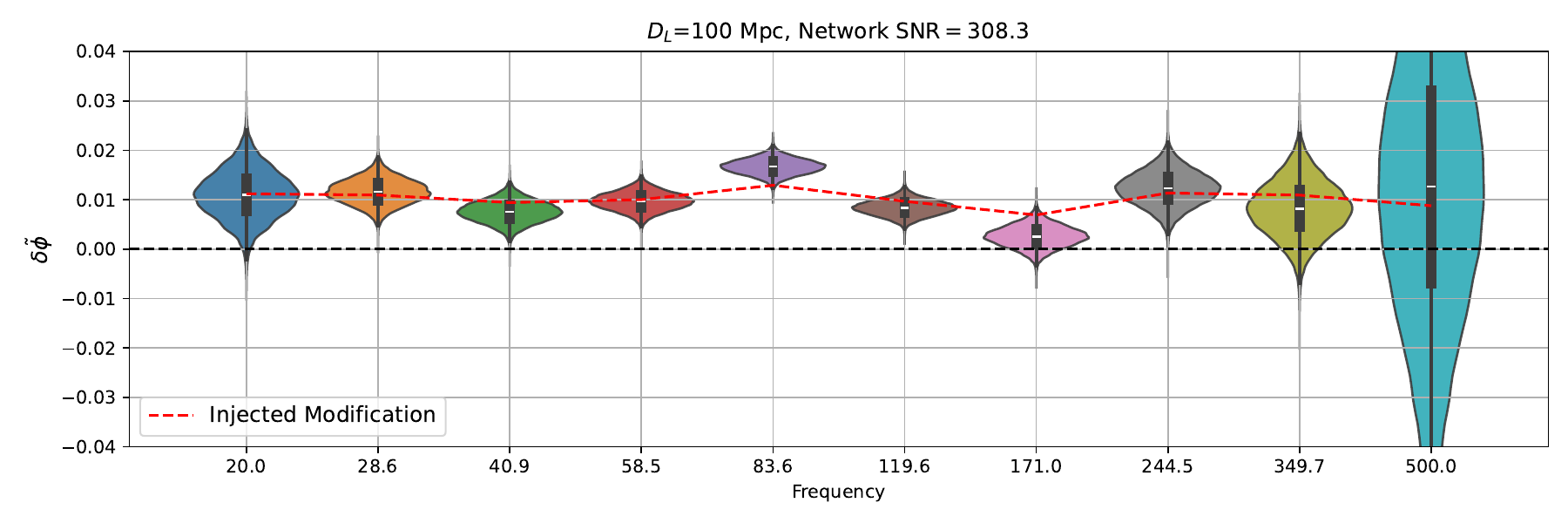}}
    \caption{The recovery of WF-Error parameter $\delta\Tilde{\phi}_i$ for rel-phase parametrization is shown for each frequency nodal point used in \ac{pe}. The light violin plot represents the prior distribution for each $\delta\Tilde{\phi}_i$, and the dark-shaded region represents the one-dimensional marginalized posterior samples. The last frequency nodal point at f=500 Hz returns the prior because the signal merges before this frequency. The constraints are also weaker at $f_i=20$ Hz because this frequency bin's signal-to-noise ratio is comparatively low. a) The modification to the reference waveform is applied using the rel-phase parametrization with a cubic spline realization generated from the distribution $\delta\Tilde{A}_i, \delta\Tilde{\phi}_i \sim \mathcal{N}(\mu=0, \sigma=0.002)$. Panel b) \textbf{and c)} represent the modification from the distribution $\delta\Tilde{A}_i, \delta\Tilde{\phi}_i \sim \mathcal{N}(\mu=0.01, \sigma=0.002)$. The network SNR and luminosity distance of the injected (GW150914-like) signal are shown at the top of each panel. For these \ac{pe} runs, we fix the parameters RA, dec, and polarization to the injected values.}
    \label{fig:wferror_cubicspline_recovery}
\end{figure*}

\begin{figure*}
    \centering
    {\includegraphics[width=0.98\textwidth]{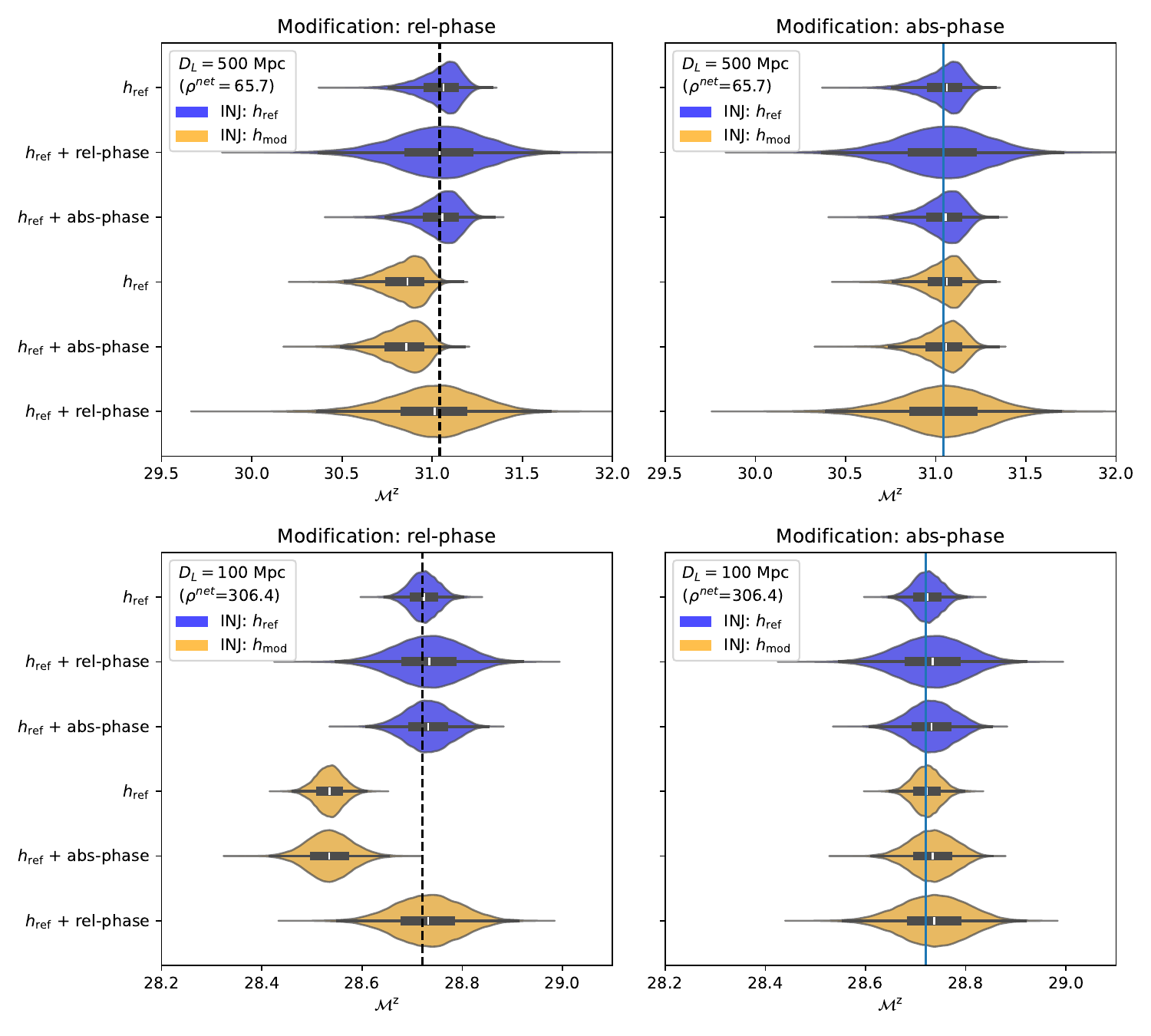}}\\
    \caption{A GW150914-like signal is injected and recovered using two WF-Error formalisms. The blue violin plots show the cases with reference injection $h_{\mathrm{ref}}$ while orange violin plots show the modified injections. The left panels represent the cases where the modification is done with rel-phase modification, while the right panel shows abs-phase modification. The top rows represent the injections with the source placed at a luminosity distance of 500 Mpc, while the bottom panels show the injection corresponds to the same source placed at 100 Mpc. Labels on the y-axis show the waveform models used in \ac{pe} runs. For these \ac{pe} runs, we fix the parameters RA, dec, and polarization to the injected values. The network SNR, $\rho^\mathrm{net}$, for reference injection is shown in each panel.}
    \label{fig:mc_bias}
\end{figure*}
Through these simulations, we have established that for a simplistic model,  a systematic error corresponding to $\delta\Tilde{\phi}=0.01$ can lead to significant bias in the detector frame chirp mass for GW150914-type signal, with the detector sensitivity of advanced LIGO design sensitivity. We have also shown that, by using WF-Error parametrization, we can account for the systematic errors introduced in the reference waveform. We can also recover the injected modification in the $\delta\phi$ parameter with correct parametrization. However, when we modify the reference waveform with rel-phase parametrization, the \ac{pe} recovery with abs-phase parametrization fails to correct the observed bias in $\mathcal{M}^\mathrm{z}$. This is because the relative-phase parametrization can accommodate more significant deviations in the reference waveform than the absolute-phase parametrization for the same range of priors in $\delta\Tilde{\phi}$, as previously discussed.

\subsection{A more realistic modification using cubic splines}\label{subsec:cubic_splines}
Within the framework of WF-Error parametrizations, a waveform model $h_{\mathrm{ref}}$ will deviate from the `true' model described by general functions of frequency: $\delta\Tilde{A}(f), \delta\Tilde{\phi}(f)$. We use cubic splines to generate them at specific frequency nodal points. In this subsection, we introduce a more general type of modification to our reference GW150914-like signal within the framework of both the parametrizations:
\begin{eqnarray}
\delta\Tilde{A}, \delta\Tilde{\phi} &\sim& \mathcal{N}(\mu=0,\sigma=0.002 ) \label{eqn:cubicspline_distribution_zm}\\
\delta\Tilde{A}, \delta\Tilde{\phi} &\sim& \mathcal{N}(\mu=0.01,\sigma=0.002) \label{eqn:cubicspline_distribution_nzm}
\end{eqnarray}

Figure \ref{fig:cubicspline_envelop} shows examples of cubic spline curves generated from the abovementioned distributions. We use a realization of $(\delta\Tilde{A}_i, \delta\Tilde{\phi}_i)$ at nodal frequency points generated from the distributions described by the equations \eqref{eqn:cubicspline_distribution_zm} and \eqref{eqn:cubicspline_distribution_nzm}.

In Figure \ref{fig:wferror_cubicspline_recovery}, we show the recovery of $\delta\Tilde{\phi}$ parameter at each frequency nodal point for the modified injection generated using rel-phase parametrization. We demonstrate that the WF-Error parametrization framework can recover the $\delta\Tilde{\phi}$ curves used to modify the reference waveform model. For a relatively louder signal, corresponding to the source located at $D_L=100$ Mpc, we observe that the posterior samples at nodal points align well with the injected values of $\delta\Tilde{\phi}$. These examples demonstrate that if the reference waveform model deviates from reality, which is described by rel-phase parametrization, the framework presented here has the potential to account for that. In other words, in the absence of true knowledge of WF-Error priors, we can use a data-driven approach to determine if there is a deviation from the reference waveform model, especially for loud signals.

\begin{figure}
    \includegraphics[scale=0.68]{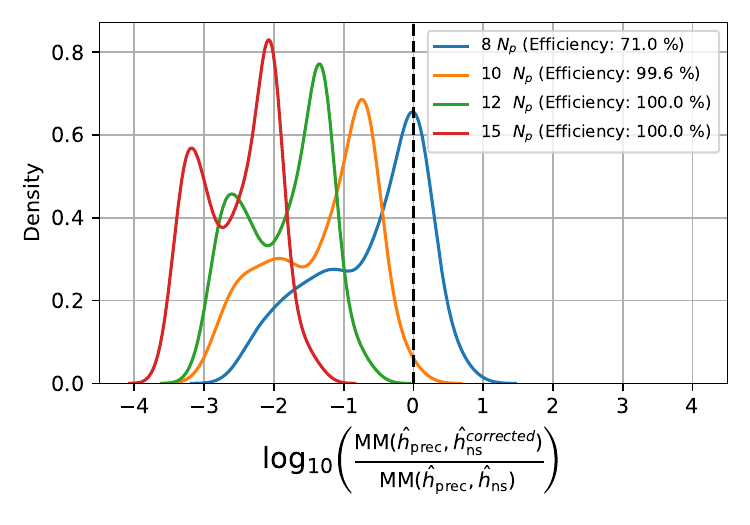}
    \caption{We demonstrate the efficiency of the cubic spline method based on the number of nodal points $(N_p)$ used to correct the baseline model. The x-axis represents the logarithm of the ratio of mismatches: the mismatch after correction compared to the mismatch before correction. The function $\mathrm{MM}(h_1,h_2)$ is defined in equation \ref{eq:inner_prod}. Here, $h_\mathrm{prec}$ represents the injected waveform strain with precessing signals, while $h_\mathrm{ns}$ refers to the strain generated with the same parameters except the spin parameters set to zero. $h^\mathrm{corrected}_\mathrm{ns}$ indicates the corrected strain using waveform-error parametrization with $N_p$ frequency nodal points. A logarithmic mismatch ratio less than one indicates an improvement, meaning that the corrected model is closer to the injected waveform $h_\mathrm{prec}$. The vertical dashed line serves as a reference. The efficiency is defined as the percentage of simulation points where an improvement is achieved (i.e., where the logarithmic mismatch ratio is less than 0).}
    \label{fig:mismatch_ratio}
\end{figure}
We now focus on the other parameters, such as chirp mass $\mathcal{M}$. In Figure \ref{fig:mc_bias}, we present various combinations of injections and \ac{pe} recovery. Our findings can be summarized as follows:

\begin{itemize}
    \item When the injection uses the reference waveform model \(h_{\mathrm{ref}}\) and \ac{pe} recovery is performed with the same model, we achieve the expected recovery with no bias.
    \item If we enable the waveform error parametrizations (rel-phase or abs-phase) while using the correct recovery model, we observe a broadening of the posterior samples as anticipated. The broadening of the posterior distribution is more pronounced with the rel-phase parametrization than the abs-phase parametrization.
    \item In the scenario when the injection is performed by modifying the reference signal and we use the reference signal to recover the source parameters, we get bias in the chirp mass. As expected, the bias is more prominent for a large SNR signal. 
    \item In the case of modified injections, using WF-Error parametrization in conjunction with the \( h_{\mathrm{ref}} \) recovery model broadens the posteriors. Additionally, employing the correct parametrization leads to a correction in bias. The bias introduced by rel-phase parametrization can only be corrected using rel-phase parametrization. The abs-phase parametrization does not correct the biases introduced by rel-phase parametrization.
    \item For the abs-phase parametrization, the moderate deviation introduced by the cubic spline curves generated from the abovementioned distribution does not introduce significant bias for a GW150914-like signal with advanced LIGO detector sensitivity. In order to see any significant bias for abs-phase parametrization, we either need a comparatively large deviation in phase or a very high SNR, such as in the case of 3G detectors. 
\end{itemize}
Through these simulations, we conclude that if the `true' waveform model deviates from the reference waveform model in a manner described by the parametrizations in equations \eqref{eqn:wferror_modeling_abs} and \eqref{eqn:wferror_modeling_rel}, our framework can correct any bias in source parameters, if present. Additionally, a sufficiently loud signal can recover the frequency dependence of WF-Error parameters. We still need to ensure that the priors are sufficiently broad to capture the deviations. Additionally, the selection of knots or nodal points for cubic splines should accurately reflect the true nature of $\delta\Tilde{ \phi}$ as a function of frequency. In these examples, we use binning in log frequency and use a total of ten nodal points between 20 Hz and 500 Hz.
\begin{figure*}
    \centering
    {\includegraphics[width=1\textwidth]{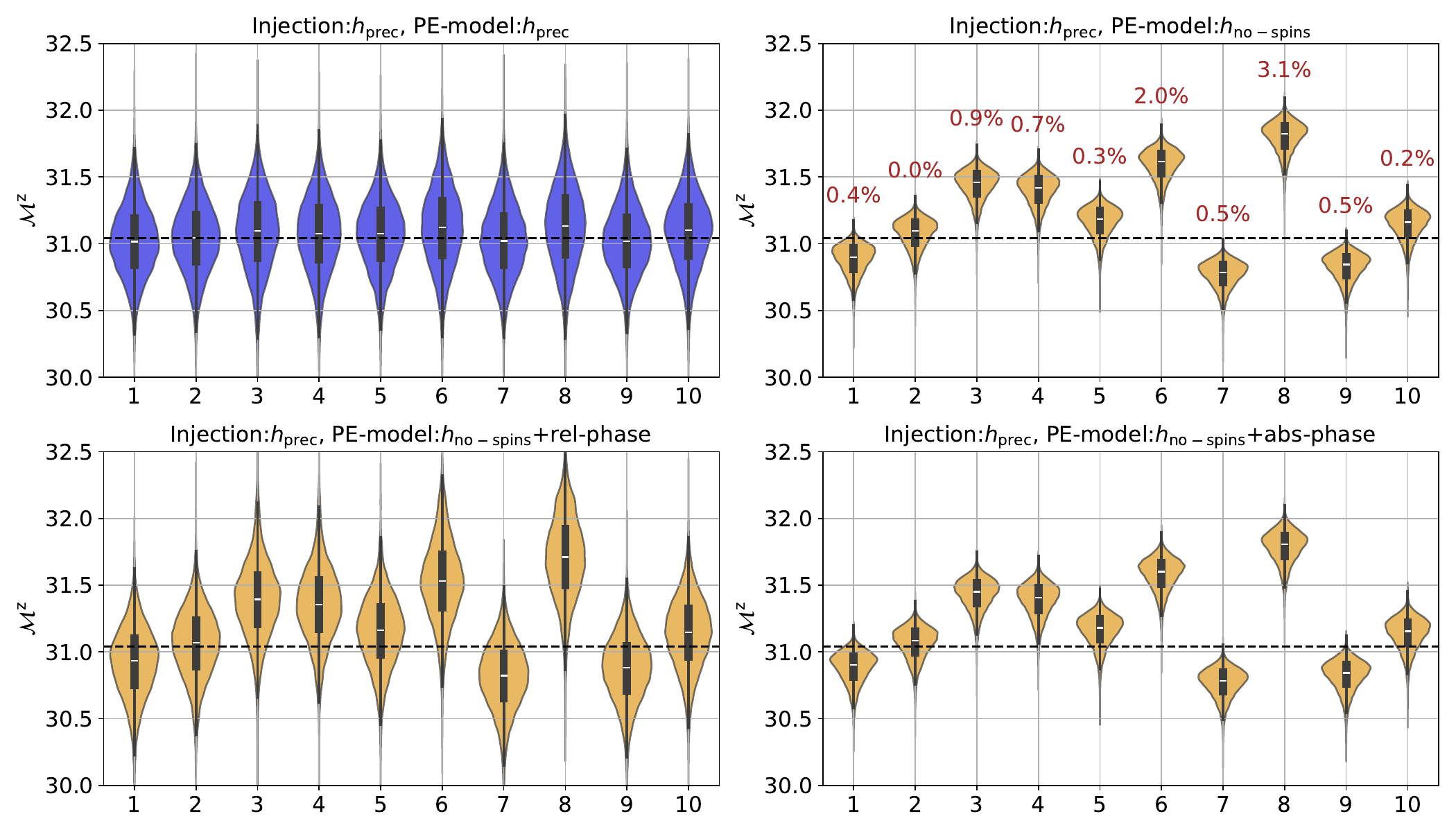}}\\
    \caption{We use ten precessing injections in the HLV network. The top left panel shows the posterior samples from each \ac{pe} run where we use the precessing recovery model: $h_{\mathrm{prec}}$. We use the nonspinning model $h_{\mathrm{no-spins}}$ for \ac{pe} in the top right panel. The bottom panels use a nonspinning model along with the WF-Error framework. The bottom left plot shows recovery with rel-phase parametrization, and the bottom right model uses abs-phase parametrization. The horizontal dashed line represents the injected chirp mass ($\mathcal{M}^\mathrm{z}$) value in the detector frame. The x-axes represent the injection simulation IDs. We use all the parameters listed in Table \ref{tab:pe_priors} for these \ac{pe} runs. We also use component spin parameters $s_{1,2}^{x,y,z}$ for the precessing recovery. We use the same priors for $(\delta\tilde{A}_i, \delta\tilde{\phi}_i)$ parameters for both the WF-Error parametrizations in bottom panels. The numbers above the data points in the top right panel represent mismatch ($\%$) between the injected template waveform $h_\mathrm{prec}$ and the corresponding non-spinning template $h_\mathrm{no-spin}$.}
    \label{fig:missing_physics}
\end{figure*}

\subsection{Incorporating missing physics}
\label{subsec:missing_physics}
One use of WF-Error parametrization can be to account for the missing physics. For BBH mergers, the most up-to-date waveform model includes effects such as higher-order harmonics and system precession, and the waveform developers are working on systems with eccentricity \citep{Paul:2022xfy, Henry:2023tka, Ramos-Buades:2021adz, Paul:2024ujx}. The most general description of a BBH merger can include other environmental effects such as matter accretion, another binary or heavy object nearby, or a merger near a supermassive BH near the host galaxy's center. 

However, the waveform model used in \ac{pe} might not fully describe the reality. We want to see if the WF-Error parametrization can capture the deviation from the actual model because the missing effects are not considered in the waveform model. In order to test this scenario, we injected a set of precessing signals with components coming from the uniform distribution $s_{1,2}^{x,y,z}\in \mathcal{U}(-0.1,0.1)$. We use the \phenompvtwo model and other parameters described in the previous section. We call the injected model $h_{\mathrm{prec}}$. We expect bias in the measured parameters for a precessing injection if we use a recovery model that does not include spins in the waveform description: $h_{\mathrm{no-spins}}$, where we set spin parameters to zero in the \textsc{IMRPhenomPv2} waveform model. For the \ac{pe} analysis, we use the priors described in table \ref{tab:pe_priors}. For the precessing \ac{pe} runs, we use isotropic spin priors for all the spin components $s^{x,y,z}_{1,2}$. Our selection of spin parameters aims to modify the reference waveform model, which includes the missing physics. Our goal is not to model deviations based on an actual astrophysical distribution, but rather to investigate whether we can incorporate such deviations for individual signals, if they are present.

In order to test if cubic-spline corrections to the $h_\mathrm{no-spin}$ work, we simulate 500 realizations of the \ac{gw} strain $h_\mathrm{prec}$. We then estimate the amplitude and phase differences ($\delta\tilde{A}_\mathrm{rel}(f), \delta\tilde{\phi}_\mathrm{rel}(f)$) with respect to the non-spinning waveform template $\tilde{h}_\mathrm{no-spin}$. We then choose a different number of nodal points, $N_p$, and apply the correction to the non-spinning template, using cubic splines generated using these nodal points. We define the mismatch between two normalized templates $\hat{h}_1$ and $\hat{h}_1$ as,
\begin{equation}
\text{MM}(\hat{h}_1 , \hat{h}_1) = 1-\langle\hat{h}_1|\hat{h}_2\rangle,
\end{equation}
where $\hat{h}$ is a normalized template such that the inner product $\langle\hat{h}|\hat{h}\rangle=1$. We estimate the mismatch between the waveform strain before and after the correction, i.e., MM$_1$=MM$(\hat{h}_\mathrm{prec},\hat{h}_\mathrm{no-spin})$, and MM$_2$=MM$(\hat{h}_\mathrm{prec},\hat{h}^\mathrm{corrected}_\mathrm{no-spin})$. When the cubic spline corrections to $h_\mathrm{no-spin}$ bring the model closer to the injected signal $h_\mathrm{prec}$, the ratio $\frac{\mathrm{MM}_2}{\mathrm{MM}_1}$ is less than unity, or the logarithm of this ratio is less than 0. In Figure \ref{fig:mismatch_ratio}, we show the distribution of the ratio $\frac{\mathrm{MM}_2}{\mathrm{MM}_1}$ for different number of nodal points used for cubic spline curves. For $N_p=8$, we get this ratio less than one for $>70\%$ simulated precessing signals. For $N_p=10$, more than $99\%$ of cases show the improvement. This number was $100\%$ for $N_p\geq12$. This indicate that, in principle, using $\geq10$ nodal points to apply corrections to baseline non-spinning waveform model should be enough to incorporate missing physics considered in this specific study.

We now focus on \ac{pe} runs done in this study with ten realizations of precessing signals generated from the distribution described earlier in this subsection. Figure \ref{fig:missing_physics} shows that when we use a non-spinning waveform model $h_{\mathrm{no-spins}}$ to recover precessing injections, we get bias in the recovered chirp mass parameter $\mathcal{M}^\mathrm{z}$ in detector frame. However, when we use WF-Error parametrizations along with $h_{\mathrm{no-spins}}$ waveform model, we notice that, for rel-phase parametrization, it results in broader marginalized posterior samples. However, abs-phase parametrization cannot make their posterior samples broad enough to correct the biases for this specific example. We observe that for injections with a significant mismatch $(\geq 1\%)$ to the reference model $h_\mathrm{ref}$, the correction remains inadequate to account for the missing physics fully. We will revisit this issue later in this subsection.

\begin{figure*}
    \centering
    \subfigure{\includegraphics[width=0.49\textwidth]{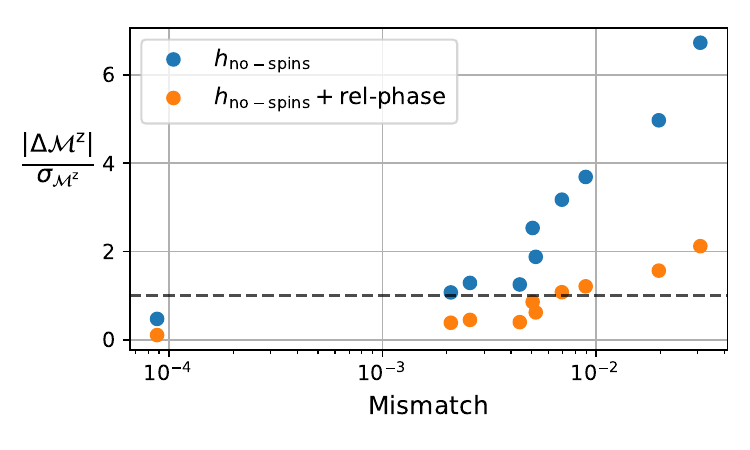}} 
    \subfigure{\includegraphics[width=0.49\textwidth]{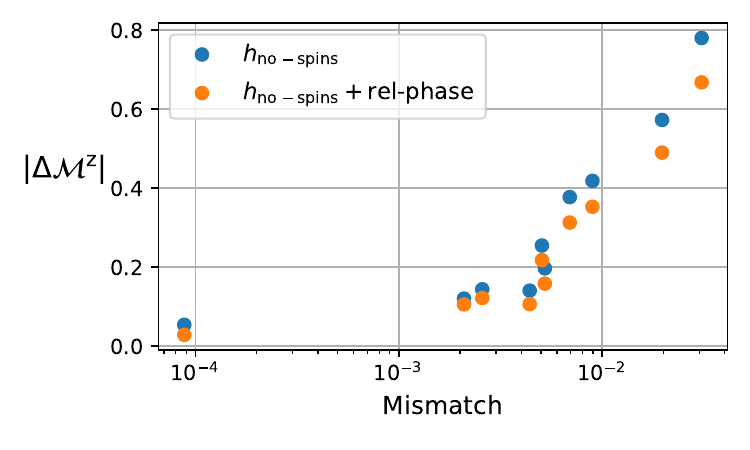}} 
    \caption{For the injections and recovery shown in Figure \ref{fig:missing_physics}, on the x-axis, we show the mismatch between the waveform generated with non-spinning and precessing parameter values. We kept all other parameters the same (see text). \textbf{Left panel}: The y-axis shows the ratio of the absolute value of the bias ($\Delta\mathcal{M}^\mathrm{z} = \mathcal{M}^\mathrm{z}_{median} - \mathcal{M}^\mathrm{z}_{inj}$) to the standard deviation of the 1D marginalized posterior samples of $\mathcal{M}^\mathrm{z}$. Horizontal dashed line represent the the line where $\Delta \mathcal{M}^\mathrm{z} = \sigma_{\mathcal{M}^\mathrm{z}}$. Right panel: The y-axis shows the absolute value of the bias $\Delta\mathcal{M}^\mathrm{z}$.}
    \label{fig:bias_by_sigma}
\end{figure*}

We assess the performance of the WF-Error parametrization using the ratio of systematic errors to statistical errors, expressed as \(|\Delta \mathcal{M}^\mathrm{z}|\textfractionsolidus\sigma_{\mathcal{M}^\mathrm{z}}\), with two runs consist of one that employs the \(h_\mathrm{no-spin}\) waveform model and another that applies the same model with the rel-phase parametrization. As demonstrated in equation \eqref{eqn:systematic_errors_validity_regime}, systematic errors become important when this ratio is of the order of unity. In Figure \ref{fig:bias_by_sigma}, on the x-axis, we show the mismatch between two templates: $h_\mathrm{prec}$ and $h_\mathrm{no-spins}$, keeping the values of non-spinning parameters the same. It quantifies how much overlap there is between the waveform templates. On the y-axis, we show the quantity $|\Delta \mathcal{M}^\mathrm{z}|\textfractionsolidus\sigma_{\mathcal{M}^\mathrm{z}}$ (left panel), and the bias $\mathcal{M}^\mathrm{z}$ (right panel). The bias is estimated as the difference between the injected value and the median of the 1D marginalized posterior samples. The ratio $|\Delta \mathcal{M}^\mathrm{z}|\textfractionsolidus\sigma_{\mathcal{M}^\mathrm{z}}$ increases when there is a larger mismatch between the two templates. It shows a significant drop in the ratio when the rel-phase WF-Error parametrization is used with the $h_\mathrm{no-spins}$ waveform model. This drop is a combination of both the effects: broadening of posterior samples (larger $\sigma$), as well as a decrease in bias $\Delta\mathcal{M}^\mathrm{z}$ (See right panel of Figure \ref{fig:bias_by_sigma}). In this set of \ac{pe} runs, we do not find significant biases in any other parameter except $\mathcal{M}^z$. 

The functional form of the parameters $(\delta\tilde{A},\delta\tilde{\phi})$ for the simulation set considered in this section is shown in the appendix \ref{appendix:Additional}. In our \ac{pe} runs, though we notice consistent decrease in the bias value $|\Delta\mathcal{M}^z|$ and the ratio $|\Delta \mathcal{M}^\mathrm{z}|\textfractionsolidus\sigma_{\mathcal{M}^\mathrm{z}}$, we focus on two injections where the original mismatch was more than $1\%$. The current framework fails to fully mitigate the systematic bias (ideally $|\Delta\mathcal{M}^z| \rightarrow0$). Upon further investigation, we find that this could be because of a combination of i) A most generalized waveform modification should be applied on both polarizations ($h^+, h^\times$) independently, and ii) a large prior ranges for $\delta\tilde{A}_i, \delta\tilde{\phi}_i$ are needed to incorporate significant deviations for large mismatch. We leave the details of such an investigation for a follow-up study.

We want to emphasize that this is an extreme example of the proof of concept. However, it indicates that we can switch on appropriate WF-Error parametrization in \ac{pe} analysis whenever we are unsure if the waveform model captures the realistic description. If the `true' model differs from the reference model in such a way that the WF-Error parametrizations can capture it, we should be able to account for potential biases in the source parameters of the binary system.

\section{Comparison with other approaches in the literature}
This section compares our framework with existing approaches in the literature regarding waveform systematics. We categorize the types of systematic errors that waveform modeling may encounter, focusing on at least two primary categories:
\begin{itemize}
\item \textbf{Type-A errors}: These errors are realized when \ac{gw} waveform morphology of a true signal $h_\mathrm{true}$ can not be achieved by the waveform model $h_\mathrm{model}(\mathbf{\Theta})$. Mathematically,
\begin{equation}
\left<\hat{h}_\mathrm{true}|\hat{h}(\mathbf{\Theta})\right> ~~< ~~1~~\forall ~~\mathbf{\Theta}, \label{eqn:type_1_err}
\end{equation} 
where $\hat{h}$ is the normalized strain such that $\langle\hat{h}|\hat{h}\rangle=1$.
\item \textbf{Type-B errors}: The waveform model can generate the true signal's morphology, but for the wrong parameters $\mathbf{\Theta}$. These errors are challenging to address, particularly if we rely on only one waveform model or if all available waveform models exhibit similar systematic errors. 
\end{itemize}
Current approaches to mitigate the waveform systematics can be categorized in several ways, with some methods fitting into multiple categories. However, we will list them under their primary category:
\begin{itemize}
\item \textbf{Account for differences between waveform models}: In some regions of parameter space, such as high spins or highly unequal mass ratios, waveform models fall short of the accuracy requirements \citep{Hu:2022rjq}, and different waveform models may give inconsistent results in \ac{pe} \citep{KAGRA:2021vkt}. Several approaches try to account for these differences, e.g., by combining the posterior samples with appropriate weighting \citep{KAGRA:2021vkt, Ashton:2019leq, Jan:2020bdz} or using hyperparameters \citep{Ashton:2021cub, Hoy:2022tst, Puecher:2023rxw,PhysRevD.98.124030, PhysRevD.100.024046,  Hoy:2024vpc} to sample across different waveform models. It is important to understand that if all waveform models share common systematics, particularly type-A errors, then these methods will not be able to address those systematics. However, if one of the waveform models exhibits type-B errors, this approach can help mitigate the impact of waveform systematics by diluting the contribution from that model.  
\item \textbf{Numerical relativity calibration errors}: Other approaches include using Gaussian process regression techniques to model deviation from the reference waveform \citep{PhysRevLett.113.251101}. The deviations can be modeled by interpolating from differences found with accurate templates, e.g., numerical relativity simulations, or by modeling the numerical relativity calibration parameters used in waveform models based on effective one-body (EOB) approaches \citep{Pompili:2024yec, Bachhar:2024olc} These approaches may fail to mitigate type-A errors if the \ac{nr} simulations used for the waveform model calibration themselves contain the systematics e.g. thermal systematics in binary neutron star waveforms \citep{Hammond:2021vtv, Gittins:2024jui}.
\item \textbf{Probabilistic waveform models}: Another suggested approach is that waveform developers provide the probabilistic models across the parameter space. For each waveform call, a bunch of curves with different weights associated with the model or an error band as a function of time or frequency can be provided \citep{PhysRevD.109.104045}. 
\end{itemize}

Our framework provides a generic, data-driven approach in which the cubic splines can account for various systematic errors, including waveform systematics. While providing the prior ranges, we expect them to contain error budgets from waveform systematics, missing physics, and other data analysis artifacts. Examples of the data analysis artifacts include residual power from glitch removal near the signal or non-stationary noise. We can integrate other approaches in the literature with our framework to provide error budgets for waveform systematics, thus serving as a complementary method. Another unique feature of our framework is the ability to mitigate type-A systematic errors. If these types of systematic errors are present, they can show up in the posterior distribution of the nodal points as deviating away from $(\delta\tilde{A}_i, \delta\tilde{\phi}_i) = 0$. It can guide determining that type-A systematic errors are present in the data, as shown in panel (c) of Figure \ref{fig:wferror_cubicspline_recovery}. For type-B errors mitigation, we can combine our framework with the methods that account for differences between waveform models.

\section{Discussion and Summary}
\label{sec:summary}
The systematic errors in the \ac{pe} analysis of a \ac{gw} merger can arise due to waveform modeling errors, data analysis artifacts, or missing physics. As the \ac{gw} detectors become more sensitive, the statistical errors become smaller, and we are reaching an era where systematic errors can not be ignored. Several approaches in the literature deal with accounting potential systematic errors in \ac{pe}. To account for potential differences between existing waveform models, posterior samples are combined \citep{KAGRA:2021vkt, Ashton:2019leq, Jan:2020bdz}, or hyperparameter models are used to sample likelihood samples \citep{Ashton:2021cub, Hoy:2022tst, Puecher:2023rxw, PhysRevD.98.124030, Hoy:2024vpc}. Other approaches include marginalizing the potential differences using prior distribution analytically \citep{PhysRevLett.113.251101}. These priors can be constructed by estimating the \ac{nr} calibration errors \citep{Pompili:2024yec, Bachhar:2024olc} or using \ac{pn}-based methods to mitigate the errors \citep{Owen:2023mid}.

In this work, we present a general \ac{pe} framework to account for the errors in the waveform modeling of \ac{gw} sources. We assume that the reference signal, $h_{\mathrm{ref}}$, used in the recovery of the \ac{gw} source parameters, is different from the `true' signal $h_{\mathrm{true}}$; the parameters can model the differences $(\delta\Tilde{A}, \delta\Tilde{ \phi})$, as differences in the amplitude and the phase of the signal. We present two WF-Error parametrizations: one corresponds to the relative phase, while the other corresponds to the absolute phase difference from the so-called true phase.

We use the cubic-spline method to modify the reference waveform model. We developed a Python code as a plugin that can be easily integrated with the publicly available \ac{gw} analysis tool \textsc{PyCBC}. The source code can be downloaded from \cite{GitHubPlugin}. We also release the frame files for the simulated injections and configuration files of all the \ac{pe} runs done in this analysis \cite{kumar_2025_14975178}.

If the WF-Error budgets for the parameters $(\delta\Tilde{A}, \delta\Tilde{ \phi})$ are available for a given waveform model, these error budgets can be used as a prior in WF-Error parametrization. Without such error budgets, we can use wider priors, potentially making our errors in estimating other parameters broader. This code and analysis framework can be utilized for the following purposes: i) to account for systematic errors in waveform models by incorporating appropriate priors for $(\delta\Tilde{A}, \delta\Tilde{\phi})$, such as calibration errors from numerical relativity (NR), or ii) to address the potential bias that may arise from missing physical effects in the waveform description.

We use Fisher matrix formalism to study its abilities and limitations in accounting for the bias induced by $(\delta\Tilde{A}, \delta\Tilde{ \phi})$. In the LSA regime, where the modified model is very close to the reference model, we can use the Fisher matrix formalism to account for systematic biases. However, in the non-LSA regime, where the overlap between the waveform models is smaller, we can no longer trust the Fisher matrix approach to correctly predict the systematic bias.

Our findings indicate that even a one-percent phase error in the rel-phase parametrization can introduce bias in the observed chirp mass. The bias in the chirp mass could affect the conclusion of follow-up studies that rely on the correct distribution of source parameters, such as inferring the properties of an intrinsic population of BBH mergers. For the abs-phase parametrization, the moderate deviation in $\delta\phi$ of the order of $\sim\mathcal{O}(0.01)$ radians, with cubic splines, do not introduce significant bias, at least in current generation detectors. We need a more significant deviation or a very high SNR signal to notice any bias. However, even in the absence of significant bias, we observe a broadening of the posterior samples by up to 20 percent ($\frac{\sigma}{\sigma_{\mathrm{ref}}} \approx 0.2$), where $\sigma$ is the standard deviation of 1D marginalized posterior of parameter $\mathcal{M}$ using the abs-phase parametrization and $\sigma_{\mathrm{ref}}$ is the standard deviation with the reference waveform model.

We make a case for waveform developers to provide error budgets alongside the waveform models. It is necessary to accurately account for the WF-Error parametrization in the \ac{pe} analysis. These error budgets can be frequency dependent $(\delta\Tilde{A}, \delta\Tilde{\phi})$ regions of $1\sigma$ uncertainties. We call them WF-Error envelopes. These WF-Error envelopes are expected to be a function of parameter space. These envelopes can be used as priors for $(\delta\Tilde{A}, \delta\Tilde{\phi})$ parameters in WF-Error parametrization. However, if such envelopes are unavailable, we can use wider priors to account for potential waveform systematics, allowing the signal's loudness and data to inform us about the constraints on the WF-Error parameters. In this scenario, we may face a penalty in the form of broader posterior samples due to the significantly larger prior volume.

One of the use cases of WF-Error parametrization is to see if the waveform model in use is missing a description of reality. The next generation of waveform models are expected to include combined effects such as eccentric orbits, precessing systems, and higher-order modes. If any of the above-mentioned effects are missing from the description of the waveform model, we might introduce bias in inferred parameters such as chirp mass. Assuming the true model differs from the reference waveform model used for the \ac{pe}, we can use WF-Error parametrization to determine if the data favors this hypothesis. At the very least, we expect the posterior samples to be broad enough to bring down the bias and standard deviation ratio. In order to test this, we used ten random realizations of mildly processing systems and performed \ac{pe} analysis with the non-spinning model. We find expected bias in this scenario. However, when we switch on rel-phase parametrization, the ratio $|\Delta \mathcal{M}^\mathrm{z}|\textfractionsolidus\sigma$ comes down, up to,  by a factor of three. In this case, the abs-phase parametrization could not account for correcting the bias with given prior range.

The method we present here is based on certain assumptions that need to be tested in broader scenarios. We assume that the reference waveform  differs from the true  waveform in a manner that can be parameterized using one of the WF-Error parametric models described by equations \eqref{eqn:wferror_modeling_abs} or \eqref{eqn:wferror_modeling_rel}. Additionally, we assume that the number of knots or nodal points selected within the frequency bins are sufficient to produce cubic spline curves that accurately model the deviation. 

As we approach an era where loud signals are more common, we expect systematic biases to become amplified. It is important to account for the WF-Error parametrization or any other scheme that can address potential systematics in the reference waveform model. The pipeline we present can work in scenarios where the types of systematics are unknown. We can select our preferred waveform model, and the parametrization best describes the deviations, allowing the data to indicate any possible systematic errors. At the same time, this approach results in broader posterior samples, and such an outcome is anticipated. 

The authors plan to explore the use of WF-Error parametrization to correct for potential biases in source parameters due to missing physics in more detail in the follow-up studies. For example, the next generation waveform models are expected to include the effects such as spin-precession, eccentric orbits, as well as higher modes. Until this is achieved, there might always be a question of whether there is inherent bias in estimating source properties due to a missing description of one of the above effects. In our follow-up studies, we will use existing \ac{nr} simulation and state-of-the-art waveform models to study if any potential biases can be accounted for, and corrected if possible.

We make certain assumptions for the method presented here, and these assumptions need to be tested in broader scenarios. We assume that the reference waveform differs from the true waveform in a manner that can be parameterized using one of the WF-Error parametric models described by equations \eqref{eqn:wferror_modeling_abs} or \eqref{eqn:wferror_modeling_rel}. Additionally, we assume that the number of knots or nodal points selected within the frequency bins are sufficient to produce cubic spline curves that accurately model the deviation. In our future studies, we will thoroughly test these assumptions for more general scenarios such as the systems with precession and/or eccentricity. Moreover, if the amplitude and phase corrections are highly oscillating functions of frequency, which can not be captured by the cubic-spline curves and a small number of fixed nodal points, we need to test other schemes, such as flexible nodal points and/or more complicated interpolation schemes.

\acknowledgments
We acknowledge the Max Planck Gesellschaft and the Max Planck Independent Research Group Program, through which this work was supported. SK also acknowledges support by the research programme of the Netherlands Organisation for Scientific Research~(NWO). We thank the computing team from AEI Hannover for their significant technical support. We also thank Binary Merger Observations and \ac{nr} group members at AEI for their feedback and valuable comments. We further thank Francesco Jimenez Forteza for in depth discussions and feedback. We are also grateful to Krishnendu NV, Prayush Kumar, and Chris Van Den Broeck for discussions and valuable inputs. We thank Michael Pürrer for valuable comments on an earlier version of this manuscript. We thank the anonymous referee for their valuable feedback. We used the Holodeck computing cluster and the Atlas cluster at AEI Hannover for all the computation.

\appendix

\section{Fisher Matrix Formalism}\label{sec:fm_formalism}
\label{appendix:FM}
In section \ref{sec:FM}, we have applied the Fisher matrix formalism to quantify systematic biases, but we have assumed a familiarity with the topic. This section serves as an introduction to the formalism itself.

The framework in which Fisher matrix estimates are derived is based on a geometric perspective on \ac{pe}. Using a metric on the parameter space, the Fisher matrix
\begin{equation}\label{eq:fisher_matrix}
    \Gamma_{\mu \nu} = \langle \partial_\mu h, \partial_\nu h \rangle \, ,
\end{equation}
one can derive estimates about various errors in the process.\footnote{As far as we can see, Ref.~\cite{Cutler_2007} was the first one to use these estimates. The related geometric interpretation was first applied to \ac{gw} data analysis by Ref.~\cite{Balasub_1996}.} Here, $h$ is some fiducial waveform model and $\partial_\mu$ refers to the derivative with respect to the parameter $\Theta^\mu$. Eq.~\eqref{eq:fisher_matrix} is defined in terms of the noise-weighted inner product
\begin{equation}\label{eq:inner_prod}
    \langle a, b \rangle = 4 \Re \int_{f_\mathrm{low}}^{f_\mathrm{high}} \frac{\tilde{a}(f) \, \tilde{b}^*(f)}{S_n(f)} \, df \, ,
\end{equation}
which has already been used in Eq.~\eqref{eq:snr_def}. It is this inner product that several important notions in \ac{gw} data analysis are based on. Specifically, we will use `match' to refer to the inner product optimized over relative time and phase shifts $t_0, \phi_0$ between $a, b$ and `overlap' to refer to the normalized match, where the result is divided by the product $||a|| \cdot ||b||$. A complementary notion is the `mismatch', which is defined as $1 - \text{match}$.

The Fisher matrix formalism is a very popular tool because it allows to estimate biases in the source parameters $\Theta$, which we will denote as $\Delta \Theta = \thetabf - \thetatr$. The two parameters $\thetatr, \thetabf$ are to be understood in the following context: say we have data that contains a signal produced by a model $h_1$ (and potentially some noise on top of that), and we wish to run \ac{pe} on this data using another model $h_2$; then we call the parameters that $h_1$ is evaluated in the `true parameters' $\thetatr$ and the parameters obtained from \ac{pe} `best-fitting' parameters $\thetabf$. What the Fisher matrix formalism does for us is that it allows to obtain an estimate of the parameter difference between true and recovered parameters without having to perform a full \ac{pe} run. This difference will have two contributions: the systematic bias due to $h_1 \neq h_2$ can be estimated as \cite{Cutler_2007}
\begin{equation}\label{eq:sys_bias}
    \Delta \Theta^\mu_\mathrm{sys} = \sum_\nu (\Gamma^{-1})^{\mu \nu} \, \langle h_1 - h_2, \partial_\nu h_2 \rangle
\end{equation}
and the corresponding measurement uncertainty caused by noise (which is also called statistical bias) as \cite{Cutler_2007}
\begin{equation}\label{eq:stat_bias}
    \Delta \Theta^\mu_\mathrm{stat} = \sum_\nu (\Gamma^{-1})^{\mu \nu} \langle n, \partial_\nu h_2 \rangle \, .
\end{equation}
In reality, where $n$ is not known, it is more common to work with the corresponding standard deviation \cite{Cutler_2007}

\begin{equation}\label{eq:stat_bias_avg}
	\sigma_{\Delta \Theta^\mu_\mathrm{stat}} = \sqrt{(\Gamma^{-1})^{\mu \mu}} \, .
\end{equation}
These estimates are very valuable because they can be calculated without having to do an extra \ac{pe} run. The right hand side of both equations can be evaluated either in the true, injected parameters $\thetatr$ or in the best-fitting parameters $\thetabf$ that would be the result of a \ac{pe} run using $h_2$ on the signal $h_1(\thetatr)$. To calculate the Fisher matrix in this context, $h_2$ must be used.\\

However, there is also a caveat. The whole approach depends crucially on the validity of the linear signal approximation (LSA)
\begin{equation}\label{eq:lsa}
    h_2(\thetatr) \simeq h_2(\thetabf) + \partial_\mu h_2(\thetabf) (\thetatr - \thetabf)^\mu \, .
\end{equation}
Besides numerical issues with condition numbers of the Fisher matrix, which are also common problems, this is the major bottleneck of Fisher matrix estimates. If $h_1$ and $h_2$ exhibit large differences, they might produce maximum-a-posteriori estimates $\thetatr, \thetabf$ for which \eqref{eq:lsa} is not a good approximation. In this case, the accuracy of later applied estimates such as \eqref{eq:sys_bias} may be compromised.

Recently, the authors of \cite{Dhani:2024jja} introduced a scheme to improve this issue, which we have adopted in this paper as well. The idea, which we only briefly review, is to replace the waveform difference $h_1(\thetatr) - h_2(\thetatr)$ by another difference $h_1(\thetatr) - h_2(\thetatropt)$ in a procedure called alignment.\footnote{Here we assume estimate \eqref{eq:sys_bias} to be evaluated in $\thetatr$, without loss of generality (the discussion also applies when using $\thetabf$).} The new point $\thetatropt$ is defined in such a way that most of its components still coincide with the one of $\thetatr$; only the ones that belong to parameters from a set $S$ are changed in order to minimize the normalized inner product
\begin{equation}\label{eq:opt_overlap}
    \frac{\langle h_1(\thetatr), h_2(\thetatropt) \rangle}{\sqrt{\langle h_1(\thetatr), h_1(\thetatr) \rangle \langle h_2(\thetatropt), h_2(\thetatropt) \rangle}} \, .
\end{equation}
\\
\\
In principle, $S$ can contain an arbitrary combination of parameters because there is no restriction as to which parameters can be optimized over in \eqref{eq:opt_overlap}. From the motivation, it is natural to select parameters for which conventional differences might occur and as a basic (yet effective) version of $S$, we choose to include relative time and phase shifts $t_0, \phi_0$ (so that \eqref{eq:opt_overlap} coincides with our definition of the `overlap'). This also allows a computationally very efficient optimization, exploiting that \eqref{eq:inner_prod} can be written as an inverse Fourier transform \cite{Allen_2005}. Of course, the estimate \eqref{eq:sys_bias} has to be changed accordingly, and the details of how to do this are presented in section III.~B.~of Ref.~\cite{Dhani:2024jja} (the basic structure remains the same, which is why we do not cover it explicitly here).

\begin{figure*}
    \centering
    \includegraphics[width=0.98\textwidth]{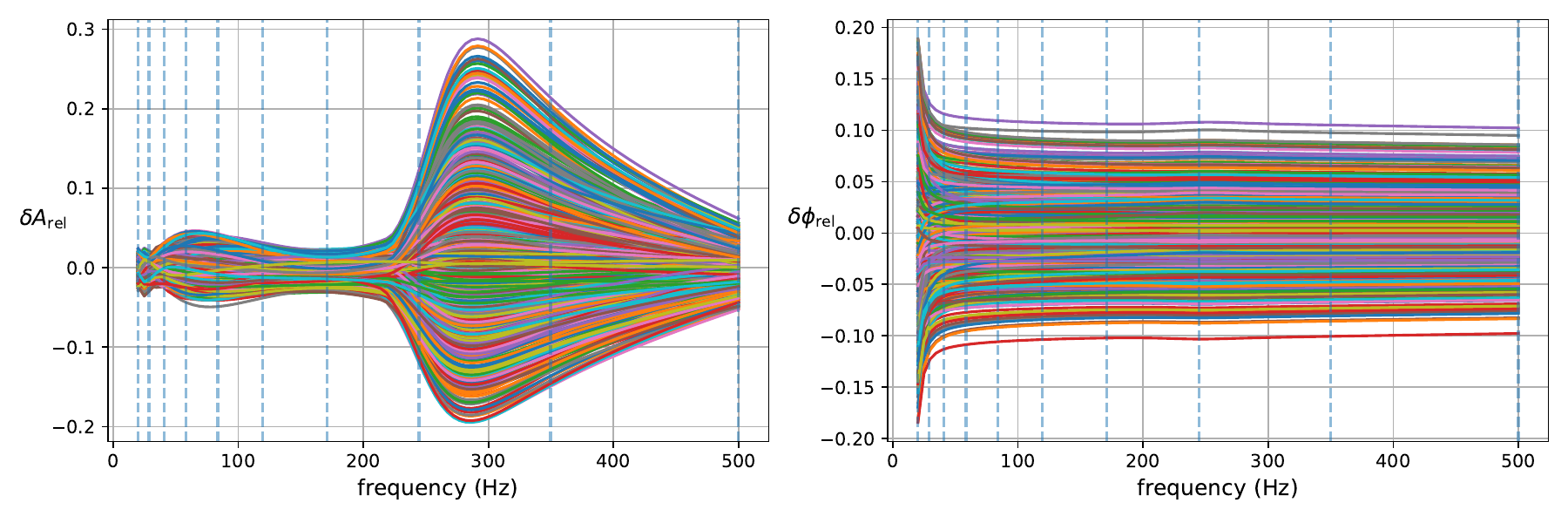} 
    \caption{For the simulations done in section \ref{subsec:missing_physics}, we show the relative errors in amplitude and phase (between precessing and non-spinning waveform template) as function of frequency. For reference, we also show 10 waveform nodal points as dashed vertical lines.}
    \label{fig:prior_range_estimation}
\end{figure*}
\section{Additional details of simulations and \ac{pe} results}
\label{appendix:Additional}
 In the examples we considered in section \ref{sec:simulations}, we find that it is primarily the $\delta\phi$ parameter (for both the parametrizations) that is constrained by the data. We obtain the prior distribution only for the $\delta\Tilde{A}$. It can be understood because, in our simulations, we introduce $1\%-2\%$ relative errors in amplitude. It seems too low for the simulations to pick up variation of that order in the amplitude parameter. Moreover, a constant shift of $\delta\Tilde{A}$ is expected to be absorbed in distance posteriors. For the 2G \ac{pe} analysis, we can use WF-Error parametrizations \eqref{eqn:wferror_modeling_abs}, and \eqref{eqn:wferror_modeling_rel} which uses only $\delta\Tilde{\phi}$ parameter. We choose to keep the $\delta\Tilde{A}$ in our parametrizations to account for most general cases of deviation.

In Figure 13, we show the relative amplitude and phase errors between $h_\mathrm{no-spin}$ and 500 simulated $h_\mathrm{prec}$ \ac{gw} strain for the GW150914-like system. The details of the waveform parameters are given in section \ref{subsec:missing_physics}. In \ac{pe}, we use ten frequency nodal points with logarithmic spacing in the range $f \in [20,500]$. Distribution of $\delta\tilde{A}(f_i), \delta\tilde{\phi}(f_i)$ at given frequency nodal point $f_i$ can be used as prior distribution.

\bibliographystyle{apsrev4-1}
\bibliography{references}

\begin{thebibliography}{126}%
\makeatletter
\providecommand \@ifxundefined [1]{%
 \@ifx{#1\undefined}
}%
\providecommand \@ifnum [1]{%
 \ifnum #1\expandafter \@firstoftwo
 \else \expandafter \@secondoftwo
 \fi
}%
\providecommand \@ifx [1]{%
 \ifx #1\expandafter \@firstoftwo
 \else \expandafter \@secondoftwo
 \fi
}%
\providecommand \natexlab [1]{#1}%
\providecommand \enquote  [1]{``#1''}%
\providecommand \bibnamefont  [1]{#1}%
\providecommand \bibfnamefont [1]{#1}%
\providecommand \citenamefont [1]{#1}%
\providecommand \href@noop [0]{\@secondoftwo}%
\providecommand \href [0]{\begingroup \@sanitize@url \@href}%
\providecommand \@href[1]{\@@startlink{#1}\@@href}%
\providecommand \@@href[1]{\endgroup#1\@@endlink}%
\providecommand \@sanitize@url [0]{\catcode `\\12\catcode `\$12\catcode
  `\&12\catcode `\#12\catcode `\^12\catcode `\_12\catcode `\%12\relax}%
\providecommand \@@startlink[1]{}%
\providecommand \@@endlink[0]{}%
\providecommand \url  [0]{\begingroup\@sanitize@url \@url }%
\providecommand \@url [1]{\endgroup\@href {#1}{\urlprefix }}%
\providecommand \urlprefix  [0]{URL }%
\providecommand \Eprint [0]{\href }%
\providecommand \doibase [0]{http://dx.doi.org/}%
\providecommand \selectlanguage [0]{\@gobble}%
\providecommand \bibinfo  [0]{\@secondoftwo}%
\providecommand \bibfield  [0]{\@secondoftwo}%
\providecommand \translation [1]{[#1]}%
\providecommand \BibitemOpen [0]{}%
\providecommand \bibitemStop [0]{}%
\providecommand \bibitemNoStop [0]{.\EOS\space}%
\providecommand \EOS [0]{\spacefactor3000\relax}%
\providecommand \BibitemShut  [1]{\csname bibitem#1\endcsname}%
\let\auto@bib@innerbib\@empty
\bibitem [{\citenamefont {Aasi}\ \emph {et~al.}(2015)\citenamefont {Aasi} \emph
  {et~al.}}]{LIGOScientific:2014pky}%
  \BibitemOpen
  \bibfield  {author} {\bibinfo {author} {\bibfnamefont {J.}~\bibnamefont
  {Aasi}} \emph {et~al.} (\bibinfo {collaboration} {LIGO Scientific}),\ }\href
  {\doibase 10.1088/0264-9381/32/7/074001} {\bibfield  {journal} {\bibinfo
  {journal} {Class. Quant. Grav.}\ }\textbf {\bibinfo {volume} {32}},\ \bibinfo
  {pages} {074001} (\bibinfo {year} {2015})},\ \Eprint
  {http://arxiv.org/abs/1411.4547} {arXiv:1411.4547 [gr-qc]} \BibitemShut
  {NoStop}%
\bibitem [{\citenamefont {Buikema}\ \emph {et~al.}(2020)\citenamefont {Buikema}
  \emph {et~al.}}]{aLIGO:2020wna}%
  \BibitemOpen
  \bibfield  {author} {\bibinfo {author} {\bibfnamefont {A.}~\bibnamefont
  {Buikema}} \emph {et~al.} (\bibinfo {collaboration} {aLIGO}),\ }\href
  {\doibase 10.1103/PhysRevD.102.062003} {\bibfield  {journal} {\bibinfo
  {journal} {Phys. Rev. D}\ }\textbf {\bibinfo {volume} {102}},\ \bibinfo
  {pages} {062003} (\bibinfo {year} {2020})},\ \Eprint
  {http://arxiv.org/abs/2008.01301} {arXiv:2008.01301 [astro-ph.IM]}
  \BibitemShut {NoStop}%
\bibitem [{\citenamefont {Abbott}\ \emph
  {et~al.}(2016{\natexlab{a}})\citenamefont {Abbott} \emph
  {et~al.}}]{KAGRA:2013rdx}%
  \BibitemOpen
  \bibfield  {author} {\bibinfo {author} {\bibfnamefont {B.~P.}\ \bibnamefont
  {Abbott}} \emph {et~al.} (\bibinfo {collaboration} {KAGRA, LIGO Scientific,
  Virgo}),\ }\href {\doibase 10.1007/s41114-020-00026-9} {\bibfield  {journal}
  {\bibinfo  {journal} {Living Rev. Rel.}\ }\textbf {\bibinfo {volume} {19}},\
  \bibinfo {pages} {1} (\bibinfo {year} {2016}{\natexlab{a}})},\ \Eprint
  {http://arxiv.org/abs/1304.0670} {arXiv:1304.0670 [gr-qc]} \BibitemShut
  {NoStop}%
\bibitem [{\citenamefont {Acernese}\ \emph {et~al.}(2015)\citenamefont
  {Acernese} \emph {et~al.}}]{VIRGO:2014yos}%
  \BibitemOpen
  \bibfield  {author} {\bibinfo {author} {\bibfnamefont {F.}~\bibnamefont
  {Acernese}} \emph {et~al.} (\bibinfo {collaboration} {VIRGO}),\ }\href
  {\doibase 10.1088/0264-9381/32/2/024001} {\bibfield  {journal} {\bibinfo
  {journal} {Class. Quant. Grav.}\ }\textbf {\bibinfo {volume} {32}},\ \bibinfo
  {pages} {024001} (\bibinfo {year} {2015})},\ \Eprint
  {http://arxiv.org/abs/1408.3978} {arXiv:1408.3978 [gr-qc]} \BibitemShut
  {NoStop}%
\bibitem [{\citenamefont {Acernese}\ \emph {et~al.}(2019)\citenamefont
  {Acernese} \emph {et~al.}}]{Virgo:2019juy}%
  \BibitemOpen
  \bibfield  {author} {\bibinfo {author} {\bibfnamefont {F.}~\bibnamefont
  {Acernese}} \emph {et~al.} (\bibinfo {collaboration} {Virgo}),\ }\href
  {\doibase 10.1103/PhysRevLett.123.231108} {\bibfield  {journal} {\bibinfo
  {journal} {Phys. Rev. Lett.}\ }\textbf {\bibinfo {volume} {123}},\ \bibinfo
  {pages} {231108} (\bibinfo {year} {2019})}\BibitemShut {NoStop}%
\bibitem [{\citenamefont {Abbott}\ \emph
  {et~al.}(2023{\natexlab{a}})\citenamefont {Abbott} \emph
  {et~al.}}]{KAGRA:2021vkt}%
  \BibitemOpen
  \bibfield  {author} {\bibinfo {author} {\bibfnamefont {R.}~\bibnamefont
  {Abbott}} \emph {et~al.} (\bibinfo {collaboration} {KAGRA, VIRGO, LIGO
  Scientific}),\ }\href {\doibase 10.1103/PhysRevX.13.041039} {\bibfield
  {journal} {\bibinfo  {journal} {Phys. Rev. X}\ }\textbf {\bibinfo {volume}
  {13}},\ \bibinfo {pages} {041039} (\bibinfo {year} {2023}{\natexlab{a}})},\
  \Eprint {http://arxiv.org/abs/2111.03606} {arXiv:2111.03606 [gr-qc]}
  \BibitemShut {NoStop}%
\bibitem [{\citenamefont {Nitz}\ \emph {et~al.}(2023)\citenamefont {Nitz},
  \citenamefont {Kumar}, \citenamefont {Wang}, \citenamefont {Kastha},
  \citenamefont {Wu}, \citenamefont {Sch\"afer}, \citenamefont {Dhurkunde},\
  and\ \citenamefont {Capano}}]{Nitz:2021zwj}%
  \BibitemOpen
  \bibfield  {author} {\bibinfo {author} {\bibfnamefont {A.~H.}\ \bibnamefont
  {Nitz}}, \bibinfo {author} {\bibfnamefont {S.}~\bibnamefont {Kumar}},
  \bibinfo {author} {\bibfnamefont {Y.-F.}\ \bibnamefont {Wang}}, \bibinfo
  {author} {\bibfnamefont {S.}~\bibnamefont {Kastha}}, \bibinfo {author}
  {\bibfnamefont {S.}~\bibnamefont {Wu}}, \bibinfo {author} {\bibfnamefont
  {M.}~\bibnamefont {Sch\"afer}}, \bibinfo {author} {\bibfnamefont
  {R.}~\bibnamefont {Dhurkunde}}, \ and\ \bibinfo {author} {\bibfnamefont
  {C.~D.}\ \bibnamefont {Capano}},\ }\href {\doibase 10.3847/1538-4357/aca591}
  {\bibfield  {journal} {\bibinfo  {journal} {Astrophys. J.}\ }\textbf
  {\bibinfo {volume} {946}},\ \bibinfo {pages} {59} (\bibinfo {year} {2023})},\
  \Eprint {http://arxiv.org/abs/2112.06878} {arXiv:2112.06878 [astro-ph.HE]}
  \BibitemShut {NoStop}%
\bibitem [{\citenamefont {Wadekar}\ \emph {et~al.}(2023)\citenamefont
  {Wadekar}, \citenamefont {Roulet}, \citenamefont {Venumadhav}, \citenamefont
  {Mehta}, \citenamefont {Zackay}, \citenamefont {Mushkin}, \citenamefont
  {Olsen},\ and\ \citenamefont {Zaldarriaga}}]{Wadekar:2023gea}%
  \BibitemOpen
  \bibfield  {author} {\bibinfo {author} {\bibfnamefont {D.}~\bibnamefont
  {Wadekar}}, \bibinfo {author} {\bibfnamefont {J.}~\bibnamefont {Roulet}},
  \bibinfo {author} {\bibfnamefont {T.}~\bibnamefont {Venumadhav}}, \bibinfo
  {author} {\bibfnamefont {A.~K.}\ \bibnamefont {Mehta}}, \bibinfo {author}
  {\bibfnamefont {B.}~\bibnamefont {Zackay}}, \bibinfo {author} {\bibfnamefont
  {J.}~\bibnamefont {Mushkin}}, \bibinfo {author} {\bibfnamefont
  {S.}~\bibnamefont {Olsen}}, \ and\ \bibinfo {author} {\bibfnamefont
  {M.}~\bibnamefont {Zaldarriaga}},\ }\href@noop {} {\  (\bibinfo {year}
  {2023})},\ \Eprint {http://arxiv.org/abs/2312.06631} {arXiv:2312.06631
  [gr-qc]} \BibitemShut {NoStop}%
\bibitem [{\citenamefont {Abbott}\ \emph
  {et~al.}(2021{\natexlab{a}})\citenamefont {Abbott} \emph
  {et~al.}}]{LIGOScientific:2021sio}%
  \BibitemOpen
  \bibfield  {author} {\bibinfo {author} {\bibfnamefont {R.}~\bibnamefont
  {Abbott}} \emph {et~al.} (\bibinfo {collaboration} {LIGO Scientific, VIRGO,
  KAGRA}),\ }\href@noop {} {\  (\bibinfo {year} {2021}{\natexlab{a}})},\
  \Eprint {http://arxiv.org/abs/2112.06861} {arXiv:2112.06861 [gr-qc]}
  \BibitemShut {NoStop}%
\bibitem [{\citenamefont {Abbott}\ \emph
  {et~al.}(2023{\natexlab{b}})\citenamefont {Abbott} \emph
  {et~al.}}]{KAGRA:2021duu}%
  \BibitemOpen
  \bibfield  {author} {\bibinfo {author} {\bibfnamefont {R.}~\bibnamefont
  {Abbott}} \emph {et~al.} (\bibinfo {collaboration} {KAGRA, VIRGO, LIGO
  Scientific}),\ }\href {\doibase 10.1103/PhysRevX.13.011048} {\bibfield
  {journal} {\bibinfo  {journal} {Phys. Rev. X}\ }\textbf {\bibinfo {volume}
  {13}},\ \bibinfo {pages} {011048} (\bibinfo {year} {2023}{\natexlab{b}})},\
  \Eprint {http://arxiv.org/abs/2111.03634} {arXiv:2111.03634 [astro-ph.HE]}
  \BibitemShut {NoStop}%
\bibitem [{\citenamefont {Abbott}\ \emph
  {et~al.}(2021{\natexlab{b}})\citenamefont {Abbott} \emph
  {et~al.}}]{LIGOScientific:2021aug}%
  \BibitemOpen
  \bibfield  {author} {\bibinfo {author} {\bibfnamefont {R.}~\bibnamefont
  {Abbott}} \emph {et~al.} (\bibinfo {collaboration} {LIGO Scientific, VIRGO,
  KAGRA}),\ }\href@noop {} {\  (\bibinfo {year} {2021}{\natexlab{b}})},\
  \Eprint {http://arxiv.org/abs/2111.03604} {arXiv:2111.03604 [astro-ph.CO]}
  \BibitemShut {NoStop}%
\bibitem [{\citenamefont {Abbott}\ \emph {et~al.}(2018)\citenamefont {Abbott}
  \emph {et~al.}}]{LIGOScientific:2018cki}%
  \BibitemOpen
  \bibfield  {author} {\bibinfo {author} {\bibfnamefont {B.~P.}\ \bibnamefont
  {Abbott}} \emph {et~al.} (\bibinfo {collaboration} {LIGO Scientific,
  Virgo}),\ }\href {\doibase 10.1103/PhysRevLett.121.161101} {\bibfield
  {journal} {\bibinfo  {journal} {Phys. Rev. Lett.}\ }\textbf {\bibinfo
  {volume} {121}},\ \bibinfo {pages} {161101} (\bibinfo {year} {2018})},\
  \Eprint {http://arxiv.org/abs/1805.11581} {arXiv:1805.11581 [gr-qc]}
  \BibitemShut {NoStop}%
\bibitem [{\citenamefont {Abbott}\ \emph
  {et~al.}(2019{\natexlab{a}})\citenamefont {Abbott} \emph
  {et~al.}}]{LIGOScientific:2018hze}%
  \BibitemOpen
  \bibfield  {author} {\bibinfo {author} {\bibfnamefont {B.~P.}\ \bibnamefont
  {Abbott}} \emph {et~al.} (\bibinfo {collaboration} {LIGO Scientific,
  Virgo}),\ }\href {\doibase 10.1103/PhysRevX.9.011001} {\bibfield  {journal}
  {\bibinfo  {journal} {Phys. Rev. X}\ }\textbf {\bibinfo {volume} {9}},\
  \bibinfo {pages} {011001} (\bibinfo {year} {2019}{\natexlab{a}})},\ \Eprint
  {http://arxiv.org/abs/1805.11579} {arXiv:1805.11579 [gr-qc]} \BibitemShut
  {NoStop}%
\bibitem [{\citenamefont {Capano}\ \emph {et~al.}(2020)\citenamefont {Capano},
  \citenamefont {Tews}, \citenamefont {Brown}, \citenamefont {Margalit},
  \citenamefont {De}, \citenamefont {Kumar}, \citenamefont {Brown},
  \citenamefont {Krishnan},\ and\ \citenamefont {Reddy}}]{Capano:2019eae}%
  \BibitemOpen
  \bibfield  {author} {\bibinfo {author} {\bibfnamefont {C.~D.}\ \bibnamefont
  {Capano}}, \bibinfo {author} {\bibfnamefont {I.}~\bibnamefont {Tews}},
  \bibinfo {author} {\bibfnamefont {S.~M.}\ \bibnamefont {Brown}}, \bibinfo
  {author} {\bibfnamefont {B.}~\bibnamefont {Margalit}}, \bibinfo {author}
  {\bibfnamefont {S.}~\bibnamefont {De}}, \bibinfo {author} {\bibfnamefont
  {S.}~\bibnamefont {Kumar}}, \bibinfo {author} {\bibfnamefont {D.~A.}\
  \bibnamefont {Brown}}, \bibinfo {author} {\bibfnamefont {B.}~\bibnamefont
  {Krishnan}}, \ and\ \bibinfo {author} {\bibfnamefont {S.}~\bibnamefont
  {Reddy}},\ }\href {\doibase 10.1038/s41550-020-1014-6} {\bibfield  {journal}
  {\bibinfo  {journal} {Nature Astron.}\ }\textbf {\bibinfo {volume} {4}},\
  \bibinfo {pages} {625} (\bibinfo {year} {2020})},\ \Eprint
  {http://arxiv.org/abs/1908.10352} {arXiv:1908.10352 [astro-ph.HE]}
  \BibitemShut {NoStop}%
\bibitem [{\citenamefont {Blanchet}(2014)}]{Blanchet:2013haa}%
  \BibitemOpen
  \bibfield  {author} {\bibinfo {author} {\bibfnamefont {L.}~\bibnamefont
  {Blanchet}},\ }\href {\doibase 10.12942/lrr-2014-2} {\bibfield  {journal}
  {\bibinfo  {journal} {Living Rev. Rel.}\ }\textbf {\bibinfo {volume} {17}},\
  \bibinfo {pages} {2} (\bibinfo {year} {2014})},\ \Eprint
  {http://arxiv.org/abs/1310.1528} {arXiv:1310.1528 [gr-qc]} \BibitemShut
  {NoStop}%
\bibitem [{\citenamefont {Campanelli}\ \emph {et~al.}(2006)\citenamefont
  {Campanelli}, \citenamefont {Lousto}, \citenamefont {Marronetti},\ and\
  \citenamefont {Zlochower}}]{Campanelli:2005dd}%
  \BibitemOpen
  \bibfield  {author} {\bibinfo {author} {\bibfnamefont {M.}~\bibnamefont
  {Campanelli}}, \bibinfo {author} {\bibfnamefont {C.~O.}\ \bibnamefont
  {Lousto}}, \bibinfo {author} {\bibfnamefont {P.}~\bibnamefont {Marronetti}},
  \ and\ \bibinfo {author} {\bibfnamefont {Y.}~\bibnamefont {Zlochower}},\
  }\href {\doibase 10.1103/PhysRevLett.96.111101} {\bibfield  {journal}
  {\bibinfo  {journal} {Phys. Rev. Lett.}\ }\textbf {\bibinfo {volume} {96}},\
  \bibinfo {pages} {111101} (\bibinfo {year} {2006})},\ \Eprint
  {http://arxiv.org/abs/gr-qc/0511048} {arXiv:gr-qc/0511048} \BibitemShut
  {NoStop}%
\bibitem [{\citenamefont {Pretorius}(2005)}]{Pretorius:2005gq}%
  \BibitemOpen
  \bibfield  {author} {\bibinfo {author} {\bibfnamefont {F.}~\bibnamefont
  {Pretorius}},\ }\href {\doibase 10.1103/PhysRevLett.95.121101} {\bibfield
  {journal} {\bibinfo  {journal} {Phys. Rev. Lett.}\ }\textbf {\bibinfo
  {volume} {95}},\ \bibinfo {pages} {121101} (\bibinfo {year} {2005})},\
  \Eprint {http://arxiv.org/abs/gr-qc/0507014} {arXiv:gr-qc/0507014}
  \BibitemShut {NoStop}%
\bibitem [{\citenamefont {Baker}\ \emph {et~al.}(2006)\citenamefont {Baker},
  \citenamefont {Centrella}, \citenamefont {Choi}, \citenamefont {Koppitz},\
  and\ \citenamefont {van Meter}}]{Baker:2005vv}%
  \BibitemOpen
  \bibfield  {author} {\bibinfo {author} {\bibfnamefont {J.~G.}\ \bibnamefont
  {Baker}}, \bibinfo {author} {\bibfnamefont {J.}~\bibnamefont {Centrella}},
  \bibinfo {author} {\bibfnamefont {D.-I.}\ \bibnamefont {Choi}}, \bibinfo
  {author} {\bibfnamefont {M.}~\bibnamefont {Koppitz}}, \ and\ \bibinfo
  {author} {\bibfnamefont {J.}~\bibnamefont {van Meter}},\ }\href {\doibase
  10.1103/PhysRevLett.96.111102} {\bibfield  {journal} {\bibinfo  {journal}
  {Phys. Rev. Lett.}\ }\textbf {\bibinfo {volume} {96}},\ \bibinfo {pages}
  {111102} (\bibinfo {year} {2006})},\ \Eprint
  {http://arxiv.org/abs/gr-qc/0511103} {arXiv:gr-qc/0511103} \BibitemShut
  {NoStop}%
\bibitem [{\citenamefont {Abbott}\ \emph
  {et~al.}(2016{\natexlab{b}})\citenamefont {Abbott} \emph
  {et~al.}}]{Abbott:2016xvh}%
  \BibitemOpen
  \bibfield  {author} {\bibinfo {author} {\bibfnamefont {B.~P.}\ \bibnamefont
  {Abbott}} \emph {et~al.},\ }\href {\doibase 10.1103/PhysRevD.93.112004}
  {\bibfield  {journal} {\bibinfo  {journal} {Phys. Rev. D}\ }\textbf {\bibinfo
  {volume} {93}},\ \bibinfo {pages} {112004} (\bibinfo {year}
  {2016}{\natexlab{b}})},\ \bibinfo {note} {[Addendum: Phys.Rev.D 97, 059901
  (2018)]},\ \Eprint {http://arxiv.org/abs/1604.00439} {arXiv:1604.00439
  [astro-ph.IM]} \BibitemShut {NoStop}%
\bibitem [{\citenamefont {Saleem}\ \emph {et~al.}(2022)\citenamefont {Saleem}
  \emph {et~al.}}]{Saleem:2021iwi}%
  \BibitemOpen
  \bibfield  {author} {\bibinfo {author} {\bibfnamefont {M.}~\bibnamefont
  {Saleem}} \emph {et~al.},\ }\href {\doibase 10.1088/1361-6382/ac3b99}
  {\bibfield  {journal} {\bibinfo  {journal} {Class. Quant. Grav.}\ }\textbf
  {\bibinfo {volume} {39}},\ \bibinfo {pages} {025004} (\bibinfo {year}
  {2022})},\ \Eprint {http://arxiv.org/abs/2105.01716} {arXiv:2105.01716
  [gr-qc]} \BibitemShut {NoStop}%
\bibitem [{\citenamefont {Punturo}\ \emph {et~al.}(2010)\citenamefont {Punturo}
  \emph {et~al.}}]{Punturo:2010zza}%
  \BibitemOpen
  \bibfield  {author} {\bibinfo {author} {\bibfnamefont {M.}~\bibnamefont
  {Punturo}} \emph {et~al.},\ }\href {\doibase 10.1088/0264-9381/27/8/084007}
  {\bibfield  {journal} {\bibinfo  {journal} {Class. Quant. Grav.}\ }\textbf
  {\bibinfo {volume} {27}},\ \bibinfo {pages} {084007} (\bibinfo {year}
  {2010})}\BibitemShut {NoStop}%
\bibitem [{\citenamefont {Hild}\ \emph {et~al.}(2011)\citenamefont {Hild} \emph
  {et~al.}}]{Hild:2010id}%
  \BibitemOpen
  \bibfield  {author} {\bibinfo {author} {\bibfnamefont {S.}~\bibnamefont
  {Hild}} \emph {et~al.},\ }\href {\doibase 10.1088/0264-9381/28/9/094013}
  {\bibfield  {journal} {\bibinfo  {journal} {Class. Quant. Grav.}\ }\textbf
  {\bibinfo {volume} {28}},\ \bibinfo {pages} {094013} (\bibinfo {year}
  {2011})},\ \Eprint {http://arxiv.org/abs/1012.0908} {arXiv:1012.0908 [gr-qc]}
  \BibitemShut {NoStop}%
\bibitem [{\citenamefont {Evans}\ \emph {et~al.}(2021)\citenamefont {Evans}
  \emph {et~al.}}]{Evans:2021gyd}%
  \BibitemOpen
  \bibfield  {author} {\bibinfo {author} {\bibfnamefont {M.}~\bibnamefont
  {Evans}} \emph {et~al.},\ }\href@noop {} {\  (\bibinfo {year} {2021})},\
  \Eprint {http://arxiv.org/abs/2109.09882} {arXiv:2109.09882 [astro-ph.IM]}
  \BibitemShut {NoStop}%
\bibitem [{\citenamefont {Srivastava}\ \emph {et~al.}(2022)\citenamefont
  {Srivastava}, \citenamefont {Davis}, \citenamefont {Kuns}, \citenamefont
  {Landry}, \citenamefont {Ballmer}, \citenamefont {Evans}, \citenamefont
  {Hall}, \citenamefont {Read},\ and\ \citenamefont
  {Sathyaprakash}}]{Srivastava:2022slt}%
  \BibitemOpen
  \bibfield  {author} {\bibinfo {author} {\bibfnamefont {V.}~\bibnamefont
  {Srivastava}}, \bibinfo {author} {\bibfnamefont {D.}~\bibnamefont {Davis}},
  \bibinfo {author} {\bibfnamefont {K.}~\bibnamefont {Kuns}}, \bibinfo {author}
  {\bibfnamefont {P.}~\bibnamefont {Landry}}, \bibinfo {author} {\bibfnamefont
  {S.}~\bibnamefont {Ballmer}}, \bibinfo {author} {\bibfnamefont
  {M.}~\bibnamefont {Evans}}, \bibinfo {author} {\bibfnamefont {E.~D.}\
  \bibnamefont {Hall}}, \bibinfo {author} {\bibfnamefont {J.}~\bibnamefont
  {Read}}, \ and\ \bibinfo {author} {\bibfnamefont {B.~S.}\ \bibnamefont
  {Sathyaprakash}},\ }\href {\doibase 10.3847/1538-4357/ac5f04} {\bibfield
  {journal} {\bibinfo  {journal} {Astrophys. J.}\ }\textbf {\bibinfo {volume}
  {931}},\ \bibinfo {pages} {22} (\bibinfo {year} {2022})},\ \Eprint
  {http://arxiv.org/abs/2201.10668} {arXiv:2201.10668 [gr-qc]} \BibitemShut
  {NoStop}%
\bibitem [{\citenamefont {Evans}\ \emph {et~al.}(2023)\citenamefont {Evans}
  \emph {et~al.}}]{Evans:2023euw}%
  \BibitemOpen
  \bibfield  {author} {\bibinfo {author} {\bibfnamefont {M.}~\bibnamefont
  {Evans}} \emph {et~al.},\ }\href@noop {} {\  (\bibinfo {year} {2023})},\
  \Eprint {http://arxiv.org/abs/2306.13745} {arXiv:2306.13745 [astro-ph.IM]}
  \BibitemShut {NoStop}%
\bibitem [{\citenamefont {Lindblom}\ \emph {et~al.}(2008)\citenamefont
  {Lindblom}, \citenamefont {Owen},\ and\ \citenamefont
  {Brown}}]{Lindblom:2008cm}%
  \BibitemOpen
  \bibfield  {author} {\bibinfo {author} {\bibfnamefont {L.}~\bibnamefont
  {Lindblom}}, \bibinfo {author} {\bibfnamefont {B.~J.}\ \bibnamefont {Owen}},
  \ and\ \bibinfo {author} {\bibfnamefont {D.~A.}\ \bibnamefont {Brown}},\
  }\href {\doibase 10.1103/PhysRevD.78.124020} {\bibfield  {journal} {\bibinfo
  {journal} {Phys. Rev. D}\ }\textbf {\bibinfo {volume} {78}},\ \bibinfo
  {pages} {124020} (\bibinfo {year} {2008})},\ \Eprint
  {http://arxiv.org/abs/0809.3844} {arXiv:0809.3844 [gr-qc]} \BibitemShut
  {NoStop}%
\bibitem [{\citenamefont {P\"urrer}\ and\ \citenamefont
  {Haster}(2020)}]{Purrer:2019jcp}%
  \BibitemOpen
  \bibfield  {author} {\bibinfo {author} {\bibfnamefont {M.}~\bibnamefont
  {P\"urrer}}\ and\ \bibinfo {author} {\bibfnamefont {C.-J.}\ \bibnamefont
  {Haster}},\ }\href {\doibase 10.1103/PhysRevResearch.2.023151} {\bibfield
  {journal} {\bibinfo  {journal} {Phys. Rev. Res.}\ }\textbf {\bibinfo {volume}
  {2}},\ \bibinfo {pages} {023151} (\bibinfo {year} {2020})},\ \Eprint
  {http://arxiv.org/abs/1912.10055} {arXiv:1912.10055 [gr-qc]} \BibitemShut
  {NoStop}%
\bibitem [{\citenamefont {Dhani}\ \emph {et~al.}(2024)\citenamefont {Dhani},
  \citenamefont {V\"olkel}, \citenamefont {Buonanno}, \citenamefont {Estelles},
  \citenamefont {Gair}, \citenamefont {Pfeiffer}, \citenamefont {Pompili},\
  and\ \citenamefont {Toubiana}}]{Dhani:2024jja}%
  \BibitemOpen
  \bibfield  {author} {\bibinfo {author} {\bibfnamefont {A.}~\bibnamefont
  {Dhani}}, \bibinfo {author} {\bibfnamefont {S.}~\bibnamefont {V\"olkel}},
  \bibinfo {author} {\bibfnamefont {A.}~\bibnamefont {Buonanno}}, \bibinfo
  {author} {\bibfnamefont {H.}~\bibnamefont {Estelles}}, \bibinfo {author}
  {\bibfnamefont {J.}~\bibnamefont {Gair}}, \bibinfo {author} {\bibfnamefont
  {H.~P.}\ \bibnamefont {Pfeiffer}}, \bibinfo {author} {\bibfnamefont
  {L.}~\bibnamefont {Pompili}}, \ and\ \bibinfo {author} {\bibfnamefont
  {A.}~\bibnamefont {Toubiana}},\ }\href@noop {} {\  (\bibinfo {year}
  {2024})},\ \Eprint {http://arxiv.org/abs/2404.05811} {arXiv:2404.05811
  [gr-qc]} \BibitemShut {NoStop}%
\bibitem [{\citenamefont {Kumar}\ \emph {et~al.}(2022)\citenamefont {Kumar},
  \citenamefont {Nitz},\ and\ \citenamefont {Forteza}}]{Kumar:2022tto}%
  \BibitemOpen
  \bibfield  {author} {\bibinfo {author} {\bibfnamefont {S.}~\bibnamefont
  {Kumar}}, \bibinfo {author} {\bibfnamefont {A.~H.}\ \bibnamefont {Nitz}}, \
  and\ \bibinfo {author} {\bibfnamefont {X.~J.}\ \bibnamefont {Forteza}},\
  }\href@noop {} {\  (\bibinfo {year} {2022})},\ \Eprint
  {http://arxiv.org/abs/2202.12762} {arXiv:2202.12762 [astro-ph.IM]}
  \BibitemShut {NoStop}%
\bibitem [{\citenamefont {Gupta}\ \emph {et~al.}(2024)\citenamefont {Gupta}
  \emph {et~al.}}]{Gupta:2024gun}%
  \BibitemOpen
  \bibfield  {author} {\bibinfo {author} {\bibfnamefont {A.}~\bibnamefont
  {Gupta}} \emph {et~al.},\ }\href {\doibase 10.21468/SciPostPhysCommRep.5} {\
  (\bibinfo {year} {2024}),\ 10.21468/SciPostPhysCommRep.5},\ \Eprint
  {http://arxiv.org/abs/2405.02197} {arXiv:2405.02197 [gr-qc]} \BibitemShut
  {NoStop}%
\bibitem [{\citenamefont {Chandramouli}\ \emph {et~al.}(2025)\citenamefont
  {Chandramouli}, \citenamefont {Prokup}, \citenamefont {Berti},\ and\
  \citenamefont {Yunes}}]{Chandramouli:2024vhw}%
  \BibitemOpen
  \bibfield  {author} {\bibinfo {author} {\bibfnamefont {R.~S.}\ \bibnamefont
  {Chandramouli}}, \bibinfo {author} {\bibfnamefont {K.}~\bibnamefont
  {Prokup}}, \bibinfo {author} {\bibfnamefont {E.}~\bibnamefont {Berti}}, \
  and\ \bibinfo {author} {\bibfnamefont {N.}~\bibnamefont {Yunes}},\ }\href
  {\doibase 10.1103/PhysRevD.111.044026} {\bibfield  {journal} {\bibinfo
  {journal} {Phys. Rev. D}\ }\textbf {\bibinfo {volume} {111}},\ \bibinfo
  {pages} {044026} (\bibinfo {year} {2025})},\ \Eprint
  {http://arxiv.org/abs/2410.06254} {arXiv:2410.06254 [gr-qc]} \BibitemShut
  {NoStop}%
\bibitem [{\citenamefont {Read}(2023)}]{Read:2023hkv}%
  \BibitemOpen
  \bibfield  {author} {\bibinfo {author} {\bibfnamefont {J.~S.}\ \bibnamefont
  {Read}},\ }\href {\doibase 10.1088/1361-6382/acd29b} {\bibfield  {journal}
  {\bibinfo  {journal} {Class. Quant. Grav.}\ }\textbf {\bibinfo {volume}
  {40}},\ \bibinfo {pages} {135002} (\bibinfo {year} {2023})},\ \Eprint
  {http://arxiv.org/abs/2301.06630} {arXiv:2301.06630 [gr-qc]} \BibitemShut
  {NoStop}%
\bibitem [{\citenamefont {Ashton}\ and\ \citenamefont
  {Khan}(2020)}]{Ashton:2019leq}%
  \BibitemOpen
  \bibfield  {author} {\bibinfo {author} {\bibfnamefont {G.}~\bibnamefont
  {Ashton}}\ and\ \bibinfo {author} {\bibfnamefont {S.}~\bibnamefont {Khan}},\
  }\href {\doibase 10.1103/PhysRevD.101.064037} {\bibfield  {journal} {\bibinfo
   {journal} {Phys. Rev. D}\ }\textbf {\bibinfo {volume} {101}},\ \bibinfo
  {pages} {064037} (\bibinfo {year} {2020})},\ \Eprint
  {http://arxiv.org/abs/1910.09138} {arXiv:1910.09138 [gr-qc]} \BibitemShut
  {NoStop}%
\bibitem [{\citenamefont {Jan}\ \emph {et~al.}(2020)\citenamefont {Jan},
  \citenamefont {Yelikar}, \citenamefont {Lange},\ and\ \citenamefont
  {O'Shaughnessy}}]{Jan:2020bdz}%
  \BibitemOpen
  \bibfield  {author} {\bibinfo {author} {\bibfnamefont {A.~Z.}\ \bibnamefont
  {Jan}}, \bibinfo {author} {\bibfnamefont {A.~B.}\ \bibnamefont {Yelikar}},
  \bibinfo {author} {\bibfnamefont {J.}~\bibnamefont {Lange}}, \ and\ \bibinfo
  {author} {\bibfnamefont {R.}~\bibnamefont {O'Shaughnessy}},\ }\href {\doibase
  10.1103/PhysRevD.102.124069} {\bibfield  {journal} {\bibinfo  {journal}
  {Phys. Rev. D}\ }\textbf {\bibinfo {volume} {102}},\ \bibinfo {pages}
  {124069} (\bibinfo {year} {2020})},\ \Eprint
  {http://arxiv.org/abs/2011.03571} {arXiv:2011.03571 [gr-qc]} \BibitemShut
  {NoStop}%
\bibitem [{\citenamefont {Ashton}\ and\ \citenamefont
  {Dietrich}(2022)}]{Ashton:2021cub}%
  \BibitemOpen
  \bibfield  {author} {\bibinfo {author} {\bibfnamefont {G.}~\bibnamefont
  {Ashton}}\ and\ \bibinfo {author} {\bibfnamefont {T.}~\bibnamefont
  {Dietrich}},\ }\href {\doibase 10.1038/s41550-022-01707-x} {\bibfield
  {journal} {\bibinfo  {journal} {Nature Astron.}\ }\textbf {\bibinfo {volume}
  {6}},\ \bibinfo {pages} {961} (\bibinfo {year} {2022})},\ \Eprint
  {http://arxiv.org/abs/2111.09214} {arXiv:2111.09214 [gr-qc]} \BibitemShut
  {NoStop}%
\bibitem [{\citenamefont {Hoy}(2022)}]{Hoy:2022tst}%
  \BibitemOpen
  \bibfield  {author} {\bibinfo {author} {\bibfnamefont {C.}~\bibnamefont
  {Hoy}},\ }\href {\doibase 10.1103/PhysRevD.106.083003} {\bibfield  {journal}
  {\bibinfo  {journal} {Phys. Rev. D}\ }\textbf {\bibinfo {volume} {106}},\
  \bibinfo {pages} {083003} (\bibinfo {year} {2022})},\ \Eprint
  {http://arxiv.org/abs/2208.00106} {arXiv:2208.00106 [gr-qc]} \BibitemShut
  {NoStop}%
\bibitem [{\citenamefont {Puecher}\ \emph {et~al.}(2024)\citenamefont
  {Puecher}, \citenamefont {Samajdar}, \citenamefont {Ashton}, \citenamefont
  {Van Den~Broeck},\ and\ \citenamefont {Dietrich}}]{Puecher:2023rxw}%
  \BibitemOpen
  \bibfield  {author} {\bibinfo {author} {\bibfnamefont {A.}~\bibnamefont
  {Puecher}}, \bibinfo {author} {\bibfnamefont {A.}~\bibnamefont {Samajdar}},
  \bibinfo {author} {\bibfnamefont {G.}~\bibnamefont {Ashton}}, \bibinfo
  {author} {\bibfnamefont {C.}~\bibnamefont {Van Den~Broeck}}, \ and\ \bibinfo
  {author} {\bibfnamefont {T.}~\bibnamefont {Dietrich}},\ }\href {\doibase
  10.1103/PhysRevD.109.023019} {\bibfield  {journal} {\bibinfo  {journal}
  {Phys. Rev. D}\ }\textbf {\bibinfo {volume} {109}},\ \bibinfo {pages}
  {023019} (\bibinfo {year} {2024})},\ \Eprint
  {http://arxiv.org/abs/2310.03555} {arXiv:2310.03555 [gr-qc]} \BibitemShut
  {NoStop}%
\bibitem [{\citenamefont {Samajdar}\ and\ \citenamefont
  {Dietrich}(2018)}]{PhysRevD.98.124030}%
  \BibitemOpen
  \bibfield  {author} {\bibinfo {author} {\bibfnamefont {A.}~\bibnamefont
  {Samajdar}}\ and\ \bibinfo {author} {\bibfnamefont {T.}~\bibnamefont
  {Dietrich}},\ }\href {\doibase 10.1103/PhysRevD.98.124030} {\bibfield
  {journal} {\bibinfo  {journal} {Phys. Rev. D}\ }\textbf {\bibinfo {volume}
  {98}},\ \bibinfo {pages} {124030} (\bibinfo {year} {2018})}\BibitemShut
  {NoStop}%
\bibitem [{\citenamefont {Samajdar}\ and\ \citenamefont
  {Dietrich}(2019)}]{PhysRevD.100.024046}%
  \BibitemOpen
  \bibfield  {author} {\bibinfo {author} {\bibfnamefont {A.}~\bibnamefont
  {Samajdar}}\ and\ \bibinfo {author} {\bibfnamefont {T.}~\bibnamefont
  {Dietrich}},\ }\href {\doibase 10.1103/PhysRevD.100.024046} {\bibfield
  {journal} {\bibinfo  {journal} {Phys. Rev. D}\ }\textbf {\bibinfo {volume}
  {100}},\ \bibinfo {pages} {024046} (\bibinfo {year} {2019})}\BibitemShut
  {NoStop}%
\bibitem [{\citenamefont {Hoy}\ \emph {et~al.}(2024)\citenamefont {Hoy},
  \citenamefont {Akcay}, \citenamefont {Mac~Uilliam},\ and\ \citenamefont
  {Thompson}}]{Hoy:2024vpc}%
  \BibitemOpen
  \bibfield  {author} {\bibinfo {author} {\bibfnamefont {C.}~\bibnamefont
  {Hoy}}, \bibinfo {author} {\bibfnamefont {S.}~\bibnamefont {Akcay}}, \bibinfo
  {author} {\bibfnamefont {J.}~\bibnamefont {Mac~Uilliam}}, \ and\ \bibinfo
  {author} {\bibfnamefont {J.~E.}\ \bibnamefont {Thompson}},\ }\href@noop {} {\
   (\bibinfo {year} {2024})},\ \Eprint {http://arxiv.org/abs/2409.19404}
  {arXiv:2409.19404 [gr-qc]} \BibitemShut {NoStop}%
\bibitem [{\citenamefont {Moore}\ and\ \citenamefont
  {Gair}(2014)}]{PhysRevLett.113.251101}%
  \BibitemOpen
  \bibfield  {author} {\bibinfo {author} {\bibfnamefont {C.~J.}\ \bibnamefont
  {Moore}}\ and\ \bibinfo {author} {\bibfnamefont {J.~R.}\ \bibnamefont
  {Gair}},\ }\href {\doibase 10.1103/PhysRevLett.113.251101} {\bibfield
  {journal} {\bibinfo  {journal} {Phys. Rev. Lett.}\ }\textbf {\bibinfo
  {volume} {113}},\ \bibinfo {pages} {251101} (\bibinfo {year}
  {2014})}\BibitemShut {NoStop}%
\bibitem [{\citenamefont {Pompili}\ \emph {et~al.}(2024)\citenamefont
  {Pompili}, \citenamefont {Buonanno},\ and\ \citenamefont
  {P\"urrer}}]{Pompili:2024yec}%
  \BibitemOpen
  \bibfield  {author} {\bibinfo {author} {\bibfnamefont {L.}~\bibnamefont
  {Pompili}}, \bibinfo {author} {\bibfnamefont {A.}~\bibnamefont {Buonanno}}, \
  and\ \bibinfo {author} {\bibfnamefont {M.}~\bibnamefont {P\"urrer}},\
  }\href@noop {} {\  (\bibinfo {year} {2024})},\ \Eprint
  {http://arxiv.org/abs/2410.16859} {arXiv:2410.16859 [gr-qc]} \BibitemShut
  {NoStop}%
\bibitem [{\citenamefont {Bachhar}\ \emph {et~al.}(2024)\citenamefont
  {Bachhar}, \citenamefont {P\"urrer},\ and\ \citenamefont
  {Green}}]{Bachhar:2024olc}%
  \BibitemOpen
  \bibfield  {author} {\bibinfo {author} {\bibfnamefont {R.}~\bibnamefont
  {Bachhar}}, \bibinfo {author} {\bibfnamefont {M.}~\bibnamefont {P\"urrer}}, \
  and\ \bibinfo {author} {\bibfnamefont {S.~R.}\ \bibnamefont {Green}},\
  }\href@noop {} {\  (\bibinfo {year} {2024})},\ \Eprint
  {http://arxiv.org/abs/2410.17168} {arXiv:2410.17168 [gr-qc]} \BibitemShut
  {NoStop}%
\bibitem [{\citenamefont {Khan}(2024)}]{PhysRevD.109.104045}%
  \BibitemOpen
  \bibfield  {author} {\bibinfo {author} {\bibfnamefont {S.}~\bibnamefont
  {Khan}},\ }\href {\doibase 10.1103/PhysRevD.109.104045} {\bibfield  {journal}
  {\bibinfo  {journal} {Phys. Rev. D}\ }\textbf {\bibinfo {volume} {109}},\
  \bibinfo {pages} {104045} (\bibinfo {year} {2024})}\BibitemShut {NoStop}%
\bibitem [{\citenamefont {Owen}\ \emph {et~al.}(2023)\citenamefont {Owen},
  \citenamefont {Haster}, \citenamefont {Perkins}, \citenamefont {Cornish},\
  and\ \citenamefont {Yunes}}]{Owen:2023mid}%
  \BibitemOpen
  \bibfield  {author} {\bibinfo {author} {\bibfnamefont {C.~B.}\ \bibnamefont
  {Owen}}, \bibinfo {author} {\bibfnamefont {C.-J.}\ \bibnamefont {Haster}},
  \bibinfo {author} {\bibfnamefont {S.}~\bibnamefont {Perkins}}, \bibinfo
  {author} {\bibfnamefont {N.~J.}\ \bibnamefont {Cornish}}, \ and\ \bibinfo
  {author} {\bibfnamefont {N.}~\bibnamefont {Yunes}},\ }\href {\doibase
  10.1103/PhysRevD.108.044018} {\bibfield  {journal} {\bibinfo  {journal}
  {Phys. Rev. D}\ }\textbf {\bibinfo {volume} {108}},\ \bibinfo {pages}
  {044018} (\bibinfo {year} {2023})},\ \Eprint
  {http://arxiv.org/abs/2301.11941} {arXiv:2301.11941 [gr-qc]} \BibitemShut
  {NoStop}%
\bibitem [{\citenamefont {Detweiler}\ and\ \citenamefont
  {Brown}(1997)}]{Detweiler:1996mq}%
  \BibitemOpen
  \bibfield  {author} {\bibinfo {author} {\bibfnamefont {S.~L.}\ \bibnamefont
  {Detweiler}}\ and\ \bibinfo {author} {\bibfnamefont {L.~H.}\ \bibnamefont
  {Brown}, \bibfnamefont {Jr.}},\ }\href {\doibase 10.1103/PhysRevD.56.826}
  {\bibfield  {journal} {\bibinfo  {journal} {Phys. Rev. D}\ }\textbf {\bibinfo
  {volume} {56}},\ \bibinfo {pages} {826} (\bibinfo {year} {1997})},\ \Eprint
  {http://arxiv.org/abs/gr-qc/9609010} {arXiv:gr-qc/9609010} \BibitemShut
  {NoStop}%
\bibitem [{\citenamefont {Bern}\ \emph {et~al.}(2019)\citenamefont {Bern},
  \citenamefont {Cheung}, \citenamefont {Roiban}, \citenamefont {Shen},
  \citenamefont {Solon},\ and\ \citenamefont {Zeng}}]{Bern:2019nnu}%
  \BibitemOpen
  \bibfield  {author} {\bibinfo {author} {\bibfnamefont {Z.}~\bibnamefont
  {Bern}}, \bibinfo {author} {\bibfnamefont {C.}~\bibnamefont {Cheung}},
  \bibinfo {author} {\bibfnamefont {R.}~\bibnamefont {Roiban}}, \bibinfo
  {author} {\bibfnamefont {C.-H.}\ \bibnamefont {Shen}}, \bibinfo {author}
  {\bibfnamefont {M.~P.}\ \bibnamefont {Solon}}, \ and\ \bibinfo {author}
  {\bibfnamefont {M.}~\bibnamefont {Zeng}},\ }\href {\doibase
  10.1103/PhysRevLett.122.201603} {\bibfield  {journal} {\bibinfo  {journal}
  {Phys. Rev. Lett.}\ }\textbf {\bibinfo {volume} {122}},\ \bibinfo {pages}
  {201603} (\bibinfo {year} {2019})},\ \Eprint
  {http://arxiv.org/abs/1901.04424} {arXiv:1901.04424 [hep-th]} \BibitemShut
  {NoStop}%
\bibitem [{\citenamefont {Vishveshwara}(1970)}]{Vishveshwara:1970zz}%
  \BibitemOpen
  \bibfield  {author} {\bibinfo {author} {\bibfnamefont {C.~V.}\ \bibnamefont
  {Vishveshwara}},\ }\href {\doibase 10.1038/227936a0} {\bibfield  {journal}
  {\bibinfo  {journal} {Nature}\ }\textbf {\bibinfo {volume} {227}},\ \bibinfo
  {pages} {936} (\bibinfo {year} {1970})}\BibitemShut {NoStop}%
\bibitem [{\citenamefont {Chandrasekhar}\ and\ \citenamefont
  {Detweiler}(1975)}]{Chandrasekhar:1975zza}%
  \BibitemOpen
  \bibfield  {author} {\bibinfo {author} {\bibfnamefont {S.}~\bibnamefont
  {Chandrasekhar}}\ and\ \bibinfo {author} {\bibfnamefont {S.~L.}\ \bibnamefont
  {Detweiler}},\ }\href {\doibase 10.1098/rspa.1975.0112} {\bibfield  {journal}
  {\bibinfo  {journal} {Proc. Roy. Soc. Lond. A}\ }\textbf {\bibinfo {volume}
  {344}},\ \bibinfo {pages} {441} (\bibinfo {year} {1975})}\BibitemShut
  {NoStop}%
\bibitem [{\citenamefont {Ohme}(2012)}]{Ohme:2011rm}%
  \BibitemOpen
  \bibfield  {author} {\bibinfo {author} {\bibfnamefont {F.}~\bibnamefont
  {Ohme}},\ }\href {\doibase 10.1088/0264-9381/29/12/124002} {\bibfield
  {journal} {\bibinfo  {journal} {Class. Quant. Grav.}\ }\textbf {\bibinfo
  {volume} {29}},\ \bibinfo {pages} {124002} (\bibinfo {year} {2012})},\
  \Eprint {http://arxiv.org/abs/1111.3737} {arXiv:1111.3737 [gr-qc]}
  \BibitemShut {NoStop}%
\bibitem [{\citenamefont {Ajith}\ \emph {et~al.}(2008)\citenamefont {Ajith}
  \emph {et~al.}}]{Ajith:2007kx}%
  \BibitemOpen
  \bibfield  {author} {\bibinfo {author} {\bibfnamefont {P.}~\bibnamefont
  {Ajith}} \emph {et~al.},\ }\href {\doibase 10.1103/PhysRevD.77.104017}
  {\bibfield  {journal} {\bibinfo  {journal} {Phys. Rev. D}\ }\textbf {\bibinfo
  {volume} {77}},\ \bibinfo {pages} {104017} (\bibinfo {year} {2008})},\
  \bibinfo {note} {[Erratum: Phys.Rev.D 79, 129901 (2009)]},\ \Eprint
  {http://arxiv.org/abs/0710.2335} {arXiv:0710.2335 [gr-qc]} \BibitemShut
  {NoStop}%
\bibitem [{\citenamefont {Ajith}\ \emph {et~al.}(2011)\citenamefont {Ajith}
  \emph {et~al.}}]{Ajith:2009bn}%
  \BibitemOpen
  \bibfield  {author} {\bibinfo {author} {\bibfnamefont {P.}~\bibnamefont
  {Ajith}} \emph {et~al.},\ }\href {\doibase 10.1103/PhysRevLett.106.241101}
  {\bibfield  {journal} {\bibinfo  {journal} {Phys. Rev. Lett.}\ }\textbf
  {\bibinfo {volume} {106}},\ \bibinfo {pages} {241101} (\bibinfo {year}
  {2011})},\ \Eprint {http://arxiv.org/abs/0909.2867} {arXiv:0909.2867 [gr-qc]}
  \BibitemShut {NoStop}%
\bibitem [{\citenamefont {Santamaria}\ \emph {et~al.}(2010)\citenamefont
  {Santamaria} \emph {et~al.}}]{Santamaria:2010yb}%
  \BibitemOpen
  \bibfield  {author} {\bibinfo {author} {\bibfnamefont {L.}~\bibnamefont
  {Santamaria}} \emph {et~al.},\ }\href {\doibase 10.1103/PhysRevD.82.064016}
  {\bibfield  {journal} {\bibinfo  {journal} {Phys. Rev. D}\ }\textbf {\bibinfo
  {volume} {82}},\ \bibinfo {pages} {064016} (\bibinfo {year} {2010})},\
  \Eprint {http://arxiv.org/abs/1005.3306} {arXiv:1005.3306 [gr-qc]}
  \BibitemShut {NoStop}%
\bibitem [{\citenamefont {Khan}\ \emph {et~al.}(2016)\citenamefont {Khan},
  \citenamefont {Husa}, \citenamefont {Hannam}, \citenamefont {Ohme},
  \citenamefont {P\"urrer}, \citenamefont {Jim\'enez~Forteza},\ and\
  \citenamefont {Boh\'e}}]{Khan:2015jqa}%
  \BibitemOpen
  \bibfield  {author} {\bibinfo {author} {\bibfnamefont {S.}~\bibnamefont
  {Khan}}, \bibinfo {author} {\bibfnamefont {S.}~\bibnamefont {Husa}}, \bibinfo
  {author} {\bibfnamefont {M.}~\bibnamefont {Hannam}}, \bibinfo {author}
  {\bibfnamefont {F.}~\bibnamefont {Ohme}}, \bibinfo {author} {\bibfnamefont
  {M.}~\bibnamefont {P\"urrer}}, \bibinfo {author} {\bibfnamefont
  {X.}~\bibnamefont {Jim\'enez~Forteza}}, \ and\ \bibinfo {author}
  {\bibfnamefont {A.}~\bibnamefont {Boh\'e}},\ }\href {\doibase
  10.1103/PhysRevD.93.044007} {\bibfield  {journal} {\bibinfo  {journal} {Phys.
  Rev. D}\ }\textbf {\bibinfo {volume} {93}},\ \bibinfo {pages} {044007}
  (\bibinfo {year} {2016})},\ \Eprint {http://arxiv.org/abs/1508.07253}
  {arXiv:1508.07253 [gr-qc]} \BibitemShut {NoStop}%
\bibitem [{\citenamefont {Khan}\ \emph {et~al.}(2019)\citenamefont {Khan},
  \citenamefont {Chatziioannou}, \citenamefont {Hannam},\ and\ \citenamefont
  {Ohme}}]{Khan:2018fmp}%
  \BibitemOpen
  \bibfield  {author} {\bibinfo {author} {\bibfnamefont {S.}~\bibnamefont
  {Khan}}, \bibinfo {author} {\bibfnamefont {K.}~\bibnamefont {Chatziioannou}},
  \bibinfo {author} {\bibfnamefont {M.}~\bibnamefont {Hannam}}, \ and\ \bibinfo
  {author} {\bibfnamefont {F.}~\bibnamefont {Ohme}},\ }\href {\doibase
  10.1103/PhysRevD.100.024059} {\bibfield  {journal} {\bibinfo  {journal}
  {Phys. Rev. D}\ }\textbf {\bibinfo {volume} {100}},\ \bibinfo {pages}
  {024059} (\bibinfo {year} {2019})},\ \Eprint
  {http://arxiv.org/abs/1809.10113} {arXiv:1809.10113 [gr-qc]} \BibitemShut
  {NoStop}%
\bibitem [{\citenamefont {Khan}\ \emph {et~al.}(2020)\citenamefont {Khan},
  \citenamefont {Ohme}, \citenamefont {Chatziioannou},\ and\ \citenamefont
  {Hannam}}]{Khan:2019kot}%
  \BibitemOpen
  \bibfield  {author} {\bibinfo {author} {\bibfnamefont {S.}~\bibnamefont
  {Khan}}, \bibinfo {author} {\bibfnamefont {F.}~\bibnamefont {Ohme}}, \bibinfo
  {author} {\bibfnamefont {K.}~\bibnamefont {Chatziioannou}}, \ and\ \bibinfo
  {author} {\bibfnamefont {M.}~\bibnamefont {Hannam}},\ }\href {\doibase
  10.1103/PhysRevD.101.024056} {\bibfield  {journal} {\bibinfo  {journal}
  {Phys. Rev. D}\ }\textbf {\bibinfo {volume} {101}},\ \bibinfo {pages}
  {024056} (\bibinfo {year} {2020})},\ \Eprint
  {http://arxiv.org/abs/1911.06050} {arXiv:1911.06050 [gr-qc]} \BibitemShut
  {NoStop}%
\bibitem [{\citenamefont {Husa}\ \emph {et~al.}(2016)\citenamefont {Husa},
  \citenamefont {Khan}, \citenamefont {Hannam}, \citenamefont {P\"urrer},
  \citenamefont {Ohme}, \citenamefont {Jim\'enez~Forteza},\ and\ \citenamefont
  {Boh\'e}}]{Husa:2015iqa}%
  \BibitemOpen
  \bibfield  {author} {\bibinfo {author} {\bibfnamefont {S.}~\bibnamefont
  {Husa}}, \bibinfo {author} {\bibfnamefont {S.}~\bibnamefont {Khan}}, \bibinfo
  {author} {\bibfnamefont {M.}~\bibnamefont {Hannam}}, \bibinfo {author}
  {\bibfnamefont {M.}~\bibnamefont {P\"urrer}}, \bibinfo {author}
  {\bibfnamefont {F.}~\bibnamefont {Ohme}}, \bibinfo {author} {\bibfnamefont
  {X.}~\bibnamefont {Jim\'enez~Forteza}}, \ and\ \bibinfo {author}
  {\bibfnamefont {A.}~\bibnamefont {Boh\'e}},\ }\href {\doibase
  10.1103/PhysRevD.93.044006} {\bibfield  {journal} {\bibinfo  {journal} {Phys.
  Rev. D}\ }\textbf {\bibinfo {volume} {93}},\ \bibinfo {pages} {044006}
  (\bibinfo {year} {2016})},\ \Eprint {http://arxiv.org/abs/1508.07250}
  {arXiv:1508.07250 [gr-qc]} \BibitemShut {NoStop}%
\bibitem [{\citenamefont {London}\ \emph {et~al.}(2018)\citenamefont {London},
  \citenamefont {Khan}, \citenamefont {Fauchon-Jones}, \citenamefont
  {Garc\'\i{}a}, \citenamefont {Hannam}, \citenamefont {Husa}, \citenamefont
  {Jim\'enez-Forteza}, \citenamefont {Kalaghatgi}, \citenamefont {Ohme},\ and\
  \citenamefont {Pannarale}}]{London:2017bcn}%
  \BibitemOpen
  \bibfield  {author} {\bibinfo {author} {\bibfnamefont {L.}~\bibnamefont
  {London}}, \bibinfo {author} {\bibfnamefont {S.}~\bibnamefont {Khan}},
  \bibinfo {author} {\bibfnamefont {E.}~\bibnamefont {Fauchon-Jones}}, \bibinfo
  {author} {\bibfnamefont {C.}~\bibnamefont {Garc\'\i{}a}}, \bibinfo {author}
  {\bibfnamefont {M.}~\bibnamefont {Hannam}}, \bibinfo {author} {\bibfnamefont
  {S.}~\bibnamefont {Husa}}, \bibinfo {author} {\bibfnamefont {X.}~\bibnamefont
  {Jim\'enez-Forteza}}, \bibinfo {author} {\bibfnamefont {C.}~\bibnamefont
  {Kalaghatgi}}, \bibinfo {author} {\bibfnamefont {F.}~\bibnamefont {Ohme}}, \
  and\ \bibinfo {author} {\bibfnamefont {F.}~\bibnamefont {Pannarale}},\ }\href
  {\doibase 10.1103/PhysRevLett.120.161102} {\bibfield  {journal} {\bibinfo
  {journal} {Phys. Rev. Lett.}\ }\textbf {\bibinfo {volume} {120}},\ \bibinfo
  {pages} {161102} (\bibinfo {year} {2018})},\ \Eprint
  {http://arxiv.org/abs/1708.00404} {arXiv:1708.00404 [gr-qc]} \BibitemShut
  {NoStop}%
\bibitem [{\citenamefont {Hannam}\ \emph {et~al.}(2014)\citenamefont {Hannam},
  \citenamefont {Schmidt}, \citenamefont {Boh\'e}, \citenamefont {Haegel},
  \citenamefont {Husa}, \citenamefont {Ohme}, \citenamefont {Pratten},\ and\
  \citenamefont {P\"urrer}}]{Hannam:2013oca}%
  \BibitemOpen
  \bibfield  {author} {\bibinfo {author} {\bibfnamefont {M.}~\bibnamefont
  {Hannam}}, \bibinfo {author} {\bibfnamefont {P.}~\bibnamefont {Schmidt}},
  \bibinfo {author} {\bibfnamefont {A.}~\bibnamefont {Boh\'e}}, \bibinfo
  {author} {\bibfnamefont {L.}~\bibnamefont {Haegel}}, \bibinfo {author}
  {\bibfnamefont {S.}~\bibnamefont {Husa}}, \bibinfo {author} {\bibfnamefont
  {F.}~\bibnamefont {Ohme}}, \bibinfo {author} {\bibfnamefont {G.}~\bibnamefont
  {Pratten}}, \ and\ \bibinfo {author} {\bibfnamefont {M.}~\bibnamefont
  {P\"urrer}},\ }\href {\doibase 10.1103/PhysRevLett.113.151101} {\bibfield
  {journal} {\bibinfo  {journal} {Phys. Rev. Lett.}\ }\textbf {\bibinfo
  {volume} {113}},\ \bibinfo {pages} {151101} (\bibinfo {year} {2014})},\
  \Eprint {http://arxiv.org/abs/1308.3271} {arXiv:1308.3271 [gr-qc]}
  \BibitemShut {NoStop}%
\bibitem [{\citenamefont {Garc\'\i{}a-Quir\'os}\ \emph
  {et~al.}(2020)\citenamefont {Garc\'\i{}a-Quir\'os}, \citenamefont {Colleoni},
  \citenamefont {Husa}, \citenamefont {Estell\'es}, \citenamefont {Pratten},
  \citenamefont {Ramos-Buades}, \citenamefont {Mateu-Lucena},\ and\
  \citenamefont {Jaume}}]{Garcia-Quiros:2020qpx}%
  \BibitemOpen
  \bibfield  {author} {\bibinfo {author} {\bibfnamefont {C.}~\bibnamefont
  {Garc\'\i{}a-Quir\'os}}, \bibinfo {author} {\bibfnamefont {M.}~\bibnamefont
  {Colleoni}}, \bibinfo {author} {\bibfnamefont {S.}~\bibnamefont {Husa}},
  \bibinfo {author} {\bibfnamefont {H.}~\bibnamefont {Estell\'es}}, \bibinfo
  {author} {\bibfnamefont {G.}~\bibnamefont {Pratten}}, \bibinfo {author}
  {\bibfnamefont {A.}~\bibnamefont {Ramos-Buades}}, \bibinfo {author}
  {\bibfnamefont {M.}~\bibnamefont {Mateu-Lucena}}, \ and\ \bibinfo {author}
  {\bibfnamefont {R.}~\bibnamefont {Jaume}},\ }\href {\doibase
  10.1103/PhysRevD.102.064002} {\bibfield  {journal} {\bibinfo  {journal}
  {Phys. Rev. D}\ }\textbf {\bibinfo {volume} {102}},\ \bibinfo {pages}
  {064002} (\bibinfo {year} {2020})},\ \Eprint
  {http://arxiv.org/abs/2001.10914} {arXiv:2001.10914 [gr-qc]} \BibitemShut
  {NoStop}%
\bibitem [{\citenamefont {Pratten}\ \emph {et~al.}(2021)\citenamefont {Pratten}
  \emph {et~al.}}]{Pratten:2020ceb}%
  \BibitemOpen
  \bibfield  {author} {\bibinfo {author} {\bibfnamefont {G.}~\bibnamefont
  {Pratten}} \emph {et~al.},\ }\href {\doibase 10.1103/PhysRevD.103.104056}
  {\bibfield  {journal} {\bibinfo  {journal} {Phys. Rev. D}\ }\textbf {\bibinfo
  {volume} {103}},\ \bibinfo {pages} {104056} (\bibinfo {year} {2021})},\
  \Eprint {http://arxiv.org/abs/2004.06503} {arXiv:2004.06503 [gr-qc]}
  \BibitemShut {NoStop}%
\bibitem [{\citenamefont {Pratten}\ \emph {et~al.}(2020)\citenamefont
  {Pratten}, \citenamefont {Husa}, \citenamefont {Garcia-Quiros}, \citenamefont
  {Colleoni}, \citenamefont {Ramos-Buades}, \citenamefont {Estelles},\ and\
  \citenamefont {Jaume}}]{Pratten:2020fqn}%
  \BibitemOpen
  \bibfield  {author} {\bibinfo {author} {\bibfnamefont {G.}~\bibnamefont
  {Pratten}}, \bibinfo {author} {\bibfnamefont {S.}~\bibnamefont {Husa}},
  \bibinfo {author} {\bibfnamefont {C.}~\bibnamefont {Garcia-Quiros}}, \bibinfo
  {author} {\bibfnamefont {M.}~\bibnamefont {Colleoni}}, \bibinfo {author}
  {\bibfnamefont {A.}~\bibnamefont {Ramos-Buades}}, \bibinfo {author}
  {\bibfnamefont {H.}~\bibnamefont {Estelles}}, \ and\ \bibinfo {author}
  {\bibfnamefont {R.}~\bibnamefont {Jaume}},\ }\href {\doibase
  10.1103/PhysRevD.102.064001} {\bibfield  {journal} {\bibinfo  {journal}
  {Phys. Rev. D}\ }\textbf {\bibinfo {volume} {102}},\ \bibinfo {pages}
  {064001} (\bibinfo {year} {2020})},\ \Eprint
  {http://arxiv.org/abs/2001.11412} {arXiv:2001.11412 [gr-qc]} \BibitemShut
  {NoStop}%
\bibitem [{\citenamefont {Estell\'es}\ \emph {et~al.}(2021)\citenamefont
  {Estell\'es}, \citenamefont {Ramos-Buades}, \citenamefont {Husa},
  \citenamefont {Garc\'\i{}a-Quir\'os}, \citenamefont {Colleoni}, \citenamefont
  {Haegel},\ and\ \citenamefont {Jaume}}]{Estelles:2020osj}%
  \BibitemOpen
  \bibfield  {author} {\bibinfo {author} {\bibfnamefont {H.}~\bibnamefont
  {Estell\'es}}, \bibinfo {author} {\bibfnamefont {A.}~\bibnamefont
  {Ramos-Buades}}, \bibinfo {author} {\bibfnamefont {S.}~\bibnamefont {Husa}},
  \bibinfo {author} {\bibfnamefont {C.}~\bibnamefont {Garc\'\i{}a-Quir\'os}},
  \bibinfo {author} {\bibfnamefont {M.}~\bibnamefont {Colleoni}}, \bibinfo
  {author} {\bibfnamefont {L.}~\bibnamefont {Haegel}}, \ and\ \bibinfo {author}
  {\bibfnamefont {R.}~\bibnamefont {Jaume}},\ }\href {\doibase
  10.1103/PhysRevD.103.124060} {\bibfield  {journal} {\bibinfo  {journal}
  {Phys. Rev. D}\ }\textbf {\bibinfo {volume} {103}},\ \bibinfo {pages}
  {124060} (\bibinfo {year} {2021})},\ \Eprint
  {http://arxiv.org/abs/2004.08302} {arXiv:2004.08302 [gr-qc]} \BibitemShut
  {NoStop}%
\bibitem [{\citenamefont {Estell\'es}\ \emph {et~al.}(2022)\citenamefont
  {Estell\'es}, \citenamefont {Colleoni}, \citenamefont {Garc\'\i{}a-Quir\'os},
  \citenamefont {Husa}, \citenamefont {Keitel}, \citenamefont {Mateu-Lucena},
  \citenamefont {Planas},\ and\ \citenamefont
  {Ramos-Buades}}]{Estelles:2021gvs}%
  \BibitemOpen
  \bibfield  {author} {\bibinfo {author} {\bibfnamefont {H.}~\bibnamefont
  {Estell\'es}}, \bibinfo {author} {\bibfnamefont {M.}~\bibnamefont
  {Colleoni}}, \bibinfo {author} {\bibfnamefont {C.}~\bibnamefont
  {Garc\'\i{}a-Quir\'os}}, \bibinfo {author} {\bibfnamefont {S.}~\bibnamefont
  {Husa}}, \bibinfo {author} {\bibfnamefont {D.}~\bibnamefont {Keitel}},
  \bibinfo {author} {\bibfnamefont {M.}~\bibnamefont {Mateu-Lucena}}, \bibinfo
  {author} {\bibfnamefont {M.~d.~L.}\ \bibnamefont {Planas}}, \ and\ \bibinfo
  {author} {\bibfnamefont {A.}~\bibnamefont {Ramos-Buades}},\ }\href {\doibase
  10.1103/PhysRevD.105.084040} {\bibfield  {journal} {\bibinfo  {journal}
  {Phys. Rev. D}\ }\textbf {\bibinfo {volume} {105}},\ \bibinfo {pages}
  {084040} (\bibinfo {year} {2022})},\ \Eprint
  {http://arxiv.org/abs/2105.05872} {arXiv:2105.05872 [gr-qc]} \BibitemShut
  {NoStop}%
\bibitem [{\citenamefont {Ghosh}\ \emph {et~al.}(2024)\citenamefont {Ghosh},
  \citenamefont {Kolitsidou},\ and\ \citenamefont {Hannam}}]{Ghosh:2023mhc}%
  \BibitemOpen
  \bibfield  {author} {\bibinfo {author} {\bibfnamefont {S.}~\bibnamefont
  {Ghosh}}, \bibinfo {author} {\bibfnamefont {P.}~\bibnamefont {Kolitsidou}}, \
  and\ \bibinfo {author} {\bibfnamefont {M.}~\bibnamefont {Hannam}},\ }\href
  {\doibase 10.1103/PhysRevD.109.024061} {\bibfield  {journal} {\bibinfo
  {journal} {Phys. Rev. D}\ }\textbf {\bibinfo {volume} {109}},\ \bibinfo
  {pages} {024061} (\bibinfo {year} {2024})},\ \Eprint
  {http://arxiv.org/abs/2310.16980} {arXiv:2310.16980 [gr-qc]} \BibitemShut
  {NoStop}%
\bibitem [{\citenamefont {Thompson}\ \emph {et~al.}(2024)\citenamefont
  {Thompson}, \citenamefont {Hamilton}, \citenamefont {London}, \citenamefont
  {Ghosh}, \citenamefont {Kolitsidou}, \citenamefont {Hoy},\ and\ \citenamefont
  {Hannam}}]{Thompson:2023ase}%
  \BibitemOpen
  \bibfield  {author} {\bibinfo {author} {\bibfnamefont {J.~E.}\ \bibnamefont
  {Thompson}}, \bibinfo {author} {\bibfnamefont {E.}~\bibnamefont {Hamilton}},
  \bibinfo {author} {\bibfnamefont {L.}~\bibnamefont {London}}, \bibinfo
  {author} {\bibfnamefont {S.}~\bibnamefont {Ghosh}}, \bibinfo {author}
  {\bibfnamefont {P.}~\bibnamefont {Kolitsidou}}, \bibinfo {author}
  {\bibfnamefont {C.}~\bibnamefont {Hoy}}, \ and\ \bibinfo {author}
  {\bibfnamefont {M.}~\bibnamefont {Hannam}},\ }\href {\doibase
  10.1103/PhysRevD.109.063012} {\bibfield  {journal} {\bibinfo  {journal}
  {Phys. Rev. D}\ }\textbf {\bibinfo {volume} {109}},\ \bibinfo {pages}
  {063012} (\bibinfo {year} {2024})},\ \Eprint
  {http://arxiv.org/abs/2312.10025} {arXiv:2312.10025 [gr-qc]} \BibitemShut
  {NoStop}%
\bibitem [{\citenamefont {Buonanno}\ and\ \citenamefont
  {Damour}(1999)}]{Buonanno:1998gg}%
  \BibitemOpen
  \bibfield  {author} {\bibinfo {author} {\bibfnamefont {A.}~\bibnamefont
  {Buonanno}}\ and\ \bibinfo {author} {\bibfnamefont {T.}~\bibnamefont
  {Damour}},\ }\href {\doibase 10.1103/PhysRevD.59.084006} {\bibfield
  {journal} {\bibinfo  {journal} {Phys. Rev. D}\ }\textbf {\bibinfo {volume}
  {59}},\ \bibinfo {pages} {084006} (\bibinfo {year} {1999})},\ \Eprint
  {http://arxiv.org/abs/gr-qc/9811091} {arXiv:gr-qc/9811091} \BibitemShut
  {NoStop}%
\bibitem [{\citenamefont {Pan}\ \emph {et~al.}(2011)\citenamefont {Pan},
  \citenamefont {Buonanno}, \citenamefont {Boyle}, \citenamefont {Buchman},
  \citenamefont {Kidder}, \citenamefont {Pfeiffer},\ and\ \citenamefont
  {Scheel}}]{Pan:2011gk}%
  \BibitemOpen
  \bibfield  {author} {\bibinfo {author} {\bibfnamefont {Y.}~\bibnamefont
  {Pan}}, \bibinfo {author} {\bibfnamefont {A.}~\bibnamefont {Buonanno}},
  \bibinfo {author} {\bibfnamefont {M.}~\bibnamefont {Boyle}}, \bibinfo
  {author} {\bibfnamefont {L.~T.}\ \bibnamefont {Buchman}}, \bibinfo {author}
  {\bibfnamefont {L.~E.}\ \bibnamefont {Kidder}}, \bibinfo {author}
  {\bibfnamefont {H.~P.}\ \bibnamefont {Pfeiffer}}, \ and\ \bibinfo {author}
  {\bibfnamefont {M.~A.}\ \bibnamefont {Scheel}},\ }\href {\doibase
  10.1103/PhysRevD.84.124052} {\bibfield  {journal} {\bibinfo  {journal} {Phys.
  Rev. D}\ }\textbf {\bibinfo {volume} {84}},\ \bibinfo {pages} {124052}
  (\bibinfo {year} {2011})},\ \Eprint {http://arxiv.org/abs/1106.1021}
  {arXiv:1106.1021 [gr-qc]} \BibitemShut {NoStop}%
\bibitem [{\citenamefont {Pan}\ \emph {et~al.}(2014)\citenamefont {Pan},
  \citenamefont {Buonanno}, \citenamefont {Taracchini}, \citenamefont {Kidder},
  \citenamefont {Mrou\'e}, \citenamefont {Pfeiffer}, \citenamefont {Scheel},\
  and\ \citenamefont {Szil\'agyi}}]{Pan:2013rra}%
  \BibitemOpen
  \bibfield  {author} {\bibinfo {author} {\bibfnamefont {Y.}~\bibnamefont
  {Pan}}, \bibinfo {author} {\bibfnamefont {A.}~\bibnamefont {Buonanno}},
  \bibinfo {author} {\bibfnamefont {A.}~\bibnamefont {Taracchini}}, \bibinfo
  {author} {\bibfnamefont {L.~E.}\ \bibnamefont {Kidder}}, \bibinfo {author}
  {\bibfnamefont {A.~H.}\ \bibnamefont {Mrou\'e}}, \bibinfo {author}
  {\bibfnamefont {H.~P.}\ \bibnamefont {Pfeiffer}}, \bibinfo {author}
  {\bibfnamefont {M.~A.}\ \bibnamefont {Scheel}}, \ and\ \bibinfo {author}
  {\bibfnamefont {B.}~\bibnamefont {Szil\'agyi}},\ }\href {\doibase
  10.1103/PhysRevD.89.084006} {\bibfield  {journal} {\bibinfo  {journal} {Phys.
  Rev. D}\ }\textbf {\bibinfo {volume} {89}},\ \bibinfo {pages} {084006}
  (\bibinfo {year} {2014})},\ \Eprint {http://arxiv.org/abs/1307.6232}
  {arXiv:1307.6232 [gr-qc]} \BibitemShut {NoStop}%
\bibitem [{\citenamefont {Taracchini}\ \emph {et~al.}(2014)\citenamefont
  {Taracchini} \emph {et~al.}}]{Taracchini:2013rva}%
  \BibitemOpen
  \bibfield  {author} {\bibinfo {author} {\bibfnamefont {A.}~\bibnamefont
  {Taracchini}} \emph {et~al.},\ }\href {\doibase 10.1103/PhysRevD.89.061502}
  {\bibfield  {journal} {\bibinfo  {journal} {Phys. Rev. D}\ }\textbf {\bibinfo
  {volume} {89}},\ \bibinfo {pages} {061502} (\bibinfo {year} {2014})},\
  \Eprint {http://arxiv.org/abs/1311.2544} {arXiv:1311.2544 [gr-qc]}
  \BibitemShut {NoStop}%
\bibitem [{\citenamefont {Boh\'e}\ \emph {et~al.}(2017)\citenamefont {Boh\'e}
  \emph {et~al.}}]{Bohe:2016gbl}%
  \BibitemOpen
  \bibfield  {author} {\bibinfo {author} {\bibfnamefont {A.}~\bibnamefont
  {Boh\'e}} \emph {et~al.},\ }\href {\doibase 10.1103/PhysRevD.95.044028}
  {\bibfield  {journal} {\bibinfo  {journal} {Phys. Rev. D}\ }\textbf {\bibinfo
  {volume} {95}},\ \bibinfo {pages} {044028} (\bibinfo {year} {2017})},\
  \Eprint {http://arxiv.org/abs/1611.03703} {arXiv:1611.03703 [gr-qc]}
  \BibitemShut {NoStop}%
\bibitem [{\citenamefont {Cotesta}\ \emph {et~al.}(2018)\citenamefont
  {Cotesta}, \citenamefont {Buonanno}, \citenamefont {Boh\'e}, \citenamefont
  {Taracchini}, \citenamefont {Hinder},\ and\ \citenamefont
  {Ossokine}}]{Cotesta:2018fcv}%
  \BibitemOpen
  \bibfield  {author} {\bibinfo {author} {\bibfnamefont {R.}~\bibnamefont
  {Cotesta}}, \bibinfo {author} {\bibfnamefont {A.}~\bibnamefont {Buonanno}},
  \bibinfo {author} {\bibfnamefont {A.}~\bibnamefont {Boh\'e}}, \bibinfo
  {author} {\bibfnamefont {A.}~\bibnamefont {Taracchini}}, \bibinfo {author}
  {\bibfnamefont {I.}~\bibnamefont {Hinder}}, \ and\ \bibinfo {author}
  {\bibfnamefont {S.}~\bibnamefont {Ossokine}},\ }\href {\doibase
  10.1103/PhysRevD.98.084028} {\bibfield  {journal} {\bibinfo  {journal} {Phys.
  Rev. D}\ }\textbf {\bibinfo {volume} {98}},\ \bibinfo {pages} {084028}
  (\bibinfo {year} {2018})},\ \Eprint {http://arxiv.org/abs/1803.10701}
  {arXiv:1803.10701 [gr-qc]} \BibitemShut {NoStop}%
\bibitem [{\citenamefont {Ossokine}\ \emph {et~al.}(2020)\citenamefont
  {Ossokine} \emph {et~al.}}]{Ossokine:2020kjp}%
  \BibitemOpen
  \bibfield  {author} {\bibinfo {author} {\bibfnamefont {S.}~\bibnamefont
  {Ossokine}} \emph {et~al.},\ }\href {\doibase 10.1103/PhysRevD.102.044055}
  {\bibfield  {journal} {\bibinfo  {journal} {Phys. Rev. D}\ }\textbf {\bibinfo
  {volume} {102}},\ \bibinfo {pages} {044055} (\bibinfo {year} {2020})},\
  \Eprint {http://arxiv.org/abs/2004.09442} {arXiv:2004.09442 [gr-qc]}
  \BibitemShut {NoStop}%
\bibitem [{\citenamefont {Pompili}\ \emph {et~al.}(2023)\citenamefont {Pompili}
  \emph {et~al.}}]{Pompili:2023tna}%
  \BibitemOpen
  \bibfield  {author} {\bibinfo {author} {\bibfnamefont {L.}~\bibnamefont
  {Pompili}} \emph {et~al.},\ }\href {\doibase 10.1103/PhysRevD.108.124035}
  {\bibfield  {journal} {\bibinfo  {journal} {Phys. Rev. D}\ }\textbf {\bibinfo
  {volume} {108}},\ \bibinfo {pages} {124035} (\bibinfo {year} {2023})},\
  \Eprint {http://arxiv.org/abs/2303.18039} {arXiv:2303.18039 [gr-qc]}
  \BibitemShut {NoStop}%
\bibitem [{\citenamefont {Gamboa}\ \emph {et~al.}(2024)\citenamefont {Gamboa}
  \emph {et~al.}}]{Gamboa:2024hli}%
  \BibitemOpen
  \bibfield  {author} {\bibinfo {author} {\bibfnamefont {A.}~\bibnamefont
  {Gamboa}} \emph {et~al.},\ }\href@noop {} {\  (\bibinfo {year} {2024})},\
  \Eprint {http://arxiv.org/abs/2412.12823} {arXiv:2412.12823 [gr-qc]}
  \BibitemShut {NoStop}%
\bibitem [{\citenamefont {Damour}\ \emph {et~al.}(2013)\citenamefont {Damour},
  \citenamefont {Nagar},\ and\ \citenamefont {Bernuzzi}}]{Damour:2012ky}%
  \BibitemOpen
  \bibfield  {author} {\bibinfo {author} {\bibfnamefont {T.}~\bibnamefont
  {Damour}}, \bibinfo {author} {\bibfnamefont {A.}~\bibnamefont {Nagar}}, \
  and\ \bibinfo {author} {\bibfnamefont {S.}~\bibnamefont {Bernuzzi}},\ }\href
  {\doibase 10.1103/PhysRevD.87.084035} {\bibfield  {journal} {\bibinfo
  {journal} {Phys. Rev. D}\ }\textbf {\bibinfo {volume} {87}},\ \bibinfo
  {pages} {084035} (\bibinfo {year} {2013})},\ \Eprint
  {http://arxiv.org/abs/1212.4357} {arXiv:1212.4357 [gr-qc]} \BibitemShut
  {NoStop}%
\bibitem [{\citenamefont {Gamba}\ \emph {et~al.}(2021)\citenamefont {Gamba},
  \citenamefont {Bernuzzi},\ and\ \citenamefont {Nagar}}]{Gamba:2020ljo}%
  \BibitemOpen
  \bibfield  {author} {\bibinfo {author} {\bibfnamefont {R.}~\bibnamefont
  {Gamba}}, \bibinfo {author} {\bibfnamefont {S.}~\bibnamefont {Bernuzzi}}, \
  and\ \bibinfo {author} {\bibfnamefont {A.}~\bibnamefont {Nagar}},\ }\href
  {\doibase 10.1103/PhysRevD.104.084058} {\bibfield  {journal} {\bibinfo
  {journal} {Phys. Rev. D}\ }\textbf {\bibinfo {volume} {104}},\ \bibinfo
  {pages} {084058} (\bibinfo {year} {2021})},\ \Eprint
  {http://arxiv.org/abs/2012.00027} {arXiv:2012.00027 [gr-qc]} \BibitemShut
  {NoStop}%
\bibitem [{\citenamefont {Nagar}\ \emph {et~al.}(2024)\citenamefont {Nagar},
  \citenamefont {Bernuzzi}, \citenamefont {Chiaramello}, \citenamefont
  {Fantini}, \citenamefont {Gamba}, \citenamefont {Panzeri},\ and\
  \citenamefont {Rettegno}}]{Nagar:2024oyk}%
  \BibitemOpen
  \bibfield  {author} {\bibinfo {author} {\bibfnamefont {A.}~\bibnamefont
  {Nagar}}, \bibinfo {author} {\bibfnamefont {S.}~\bibnamefont {Bernuzzi}},
  \bibinfo {author} {\bibfnamefont {D.}~\bibnamefont {Chiaramello}}, \bibinfo
  {author} {\bibfnamefont {V.}~\bibnamefont {Fantini}}, \bibinfo {author}
  {\bibfnamefont {R.}~\bibnamefont {Gamba}}, \bibinfo {author} {\bibfnamefont
  {M.}~\bibnamefont {Panzeri}}, \ and\ \bibinfo {author} {\bibfnamefont
  {P.}~\bibnamefont {Rettegno}},\ }\href@noop {} {\  (\bibinfo {year}
  {2024})},\ \Eprint {http://arxiv.org/abs/2407.04762} {arXiv:2407.04762
  [gr-qc]} \BibitemShut {NoStop}%
\bibitem [{\citenamefont {Field}\ \emph {et~al.}(2014)\citenamefont {Field},
  \citenamefont {Galley}, \citenamefont {Hesthaven}, \citenamefont {Kaye},\
  and\ \citenamefont {Tiglio}}]{Field:2013cfa}%
  \BibitemOpen
  \bibfield  {author} {\bibinfo {author} {\bibfnamefont {S.~E.}\ \bibnamefont
  {Field}}, \bibinfo {author} {\bibfnamefont {C.~R.}\ \bibnamefont {Galley}},
  \bibinfo {author} {\bibfnamefont {J.~S.}\ \bibnamefont {Hesthaven}}, \bibinfo
  {author} {\bibfnamefont {J.}~\bibnamefont {Kaye}}, \ and\ \bibinfo {author}
  {\bibfnamefont {M.}~\bibnamefont {Tiglio}},\ }\href {\doibase
  10.1103/PhysRevX.4.031006} {\bibfield  {journal} {\bibinfo  {journal} {Phys.
  Rev. X}\ }\textbf {\bibinfo {volume} {4}},\ \bibinfo {pages} {031006}
  (\bibinfo {year} {2014})},\ \Eprint {http://arxiv.org/abs/1308.3565}
  {arXiv:1308.3565 [gr-qc]} \BibitemShut {NoStop}%
\bibitem [{\citenamefont {Blackman}\ \emph
  {et~al.}(2017{\natexlab{a}})\citenamefont {Blackman}, \citenamefont {Field},
  \citenamefont {Scheel}, \citenamefont {Galley}, \citenamefont {Hemberger},
  \citenamefont {Schmidt},\ and\ \citenamefont {Smith}}]{Blackman:2017dfb}%
  \BibitemOpen
  \bibfield  {author} {\bibinfo {author} {\bibfnamefont {J.}~\bibnamefont
  {Blackman}}, \bibinfo {author} {\bibfnamefont {S.~E.}\ \bibnamefont {Field}},
  \bibinfo {author} {\bibfnamefont {M.~A.}\ \bibnamefont {Scheel}}, \bibinfo
  {author} {\bibfnamefont {C.~R.}\ \bibnamefont {Galley}}, \bibinfo {author}
  {\bibfnamefont {D.~A.}\ \bibnamefont {Hemberger}}, \bibinfo {author}
  {\bibfnamefont {P.}~\bibnamefont {Schmidt}}, \ and\ \bibinfo {author}
  {\bibfnamefont {R.}~\bibnamefont {Smith}},\ }\href {\doibase
  10.1103/PhysRevD.95.104023} {\bibfield  {journal} {\bibinfo  {journal} {Phys.
  Rev. D}\ }\textbf {\bibinfo {volume} {95}},\ \bibinfo {pages} {104023}
  (\bibinfo {year} {2017}{\natexlab{a}})},\ \Eprint
  {http://arxiv.org/abs/1701.00550} {arXiv:1701.00550 [gr-qc]} \BibitemShut
  {NoStop}%
\bibitem [{\citenamefont {Blackman}\ \emph
  {et~al.}(2017{\natexlab{b}})\citenamefont {Blackman}, \citenamefont {Field},
  \citenamefont {Scheel}, \citenamefont {Galley}, \citenamefont {Ott},
  \citenamefont {Boyle}, \citenamefont {Kidder}, \citenamefont {Pfeiffer},\
  and\ \citenamefont {Szil\'agyi}}]{Blackman:2017pcm}%
  \BibitemOpen
  \bibfield  {author} {\bibinfo {author} {\bibfnamefont {J.}~\bibnamefont
  {Blackman}}, \bibinfo {author} {\bibfnamefont {S.~E.}\ \bibnamefont {Field}},
  \bibinfo {author} {\bibfnamefont {M.~A.}\ \bibnamefont {Scheel}}, \bibinfo
  {author} {\bibfnamefont {C.~R.}\ \bibnamefont {Galley}}, \bibinfo {author}
  {\bibfnamefont {C.~D.}\ \bibnamefont {Ott}}, \bibinfo {author} {\bibfnamefont
  {M.}~\bibnamefont {Boyle}}, \bibinfo {author} {\bibfnamefont {L.~E.}\
  \bibnamefont {Kidder}}, \bibinfo {author} {\bibfnamefont {H.~P.}\
  \bibnamefont {Pfeiffer}}, \ and\ \bibinfo {author} {\bibfnamefont
  {B.}~\bibnamefont {Szil\'agyi}},\ }\href {\doibase
  10.1103/PhysRevD.96.024058} {\bibfield  {journal} {\bibinfo  {journal} {Phys.
  Rev. D}\ }\textbf {\bibinfo {volume} {96}},\ \bibinfo {pages} {024058}
  (\bibinfo {year} {2017}{\natexlab{b}})},\ \Eprint
  {http://arxiv.org/abs/1705.07089} {arXiv:1705.07089 [gr-qc]} \BibitemShut
  {NoStop}%
\bibitem [{\citenamefont {Varma}\ \emph
  {et~al.}(2019{\natexlab{a}})\citenamefont {Varma}, \citenamefont {Field},
  \citenamefont {Scheel}, \citenamefont {Blackman}, \citenamefont {Kidder},\
  and\ \citenamefont {Pfeiffer}}]{Varma:2018mmi}%
  \BibitemOpen
  \bibfield  {author} {\bibinfo {author} {\bibfnamefont {V.}~\bibnamefont
  {Varma}}, \bibinfo {author} {\bibfnamefont {S.~E.}\ \bibnamefont {Field}},
  \bibinfo {author} {\bibfnamefont {M.~A.}\ \bibnamefont {Scheel}}, \bibinfo
  {author} {\bibfnamefont {J.}~\bibnamefont {Blackman}}, \bibinfo {author}
  {\bibfnamefont {L.~E.}\ \bibnamefont {Kidder}}, \ and\ \bibinfo {author}
  {\bibfnamefont {H.~P.}\ \bibnamefont {Pfeiffer}},\ }\href {\doibase
  10.1103/PhysRevD.99.064045} {\bibfield  {journal} {\bibinfo  {journal} {Phys.
  Rev. D}\ }\textbf {\bibinfo {volume} {99}},\ \bibinfo {pages} {064045}
  (\bibinfo {year} {2019}{\natexlab{a}})},\ \Eprint
  {http://arxiv.org/abs/1812.07865} {arXiv:1812.07865 [gr-qc]} \BibitemShut
  {NoStop}%
\bibitem [{\citenamefont {Varma}\ \emph
  {et~al.}(2019{\natexlab{b}})\citenamefont {Varma}, \citenamefont {Field},
  \citenamefont {Scheel}, \citenamefont {Blackman}, \citenamefont {Gerosa},
  \citenamefont {Stein}, \citenamefont {Kidder},\ and\ \citenamefont
  {Pfeiffer}}]{Varma:2019csw}%
  \BibitemOpen
  \bibfield  {author} {\bibinfo {author} {\bibfnamefont {V.}~\bibnamefont
  {Varma}}, \bibinfo {author} {\bibfnamefont {S.~E.}\ \bibnamefont {Field}},
  \bibinfo {author} {\bibfnamefont {M.~A.}\ \bibnamefont {Scheel}}, \bibinfo
  {author} {\bibfnamefont {J.}~\bibnamefont {Blackman}}, \bibinfo {author}
  {\bibfnamefont {D.}~\bibnamefont {Gerosa}}, \bibinfo {author} {\bibfnamefont
  {L.~C.}\ \bibnamefont {Stein}}, \bibinfo {author} {\bibfnamefont {L.~E.}\
  \bibnamefont {Kidder}}, \ and\ \bibinfo {author} {\bibfnamefont {H.~P.}\
  \bibnamefont {Pfeiffer}},\ }\href {\doibase 10.1103/PhysRevResearch.1.033015}
  {\bibfield  {journal} {\bibinfo  {journal} {Phys. Rev. Research.}\ }\textbf
  {\bibinfo {volume} {1}},\ \bibinfo {pages} {033015} (\bibinfo {year}
  {2019}{\natexlab{b}})},\ \Eprint {http://arxiv.org/abs/1905.09300}
  {arXiv:1905.09300 [gr-qc]} \BibitemShut {NoStop}%
\bibitem [{\citenamefont {Abbott}\ \emph
  {et~al.}(2020{\natexlab{a}})\citenamefont {Abbott} \emph
  {et~al.}}]{LIGOScientific:2019hgc}%
  \BibitemOpen
  \bibfield  {author} {\bibinfo {author} {\bibfnamefont {B.~P.}\ \bibnamefont
  {Abbott}} \emph {et~al.} (\bibinfo {collaboration} {LIGO Scientific,
  Virgo}),\ }\href {\doibase 10.1088/1361-6382/ab685e} {\bibfield  {journal}
  {\bibinfo  {journal} {Class. Quant. Grav.}\ }\textbf {\bibinfo {volume}
  {37}},\ \bibinfo {pages} {055002} (\bibinfo {year} {2020}{\natexlab{a}})},\
  \Eprint {http://arxiv.org/abs/1908.11170} {arXiv:1908.11170 [gr-qc]}
  \BibitemShut {NoStop}%
\bibitem [{\citenamefont {Prix}(2007)}]{Prix:2007ks}%
  \BibitemOpen
  \bibfield  {author} {\bibinfo {author} {\bibfnamefont {R.}~\bibnamefont
  {Prix}},\ }\href {\doibase 10.1088/0264-9381/24/19/S11} {\bibfield  {journal}
  {\bibinfo  {journal} {Class. Quant. Grav.}\ }\textbf {\bibinfo {volume}
  {24}},\ \bibinfo {pages} {S481} (\bibinfo {year} {2007})},\ \Eprint
  {http://arxiv.org/abs/0707.0428} {arXiv:0707.0428 [gr-qc]} \BibitemShut
  {NoStop}%
\bibitem [{\citenamefont {Usman}\ \emph {et~al.}(2016)\citenamefont {Usman}
  \emph {et~al.}}]{Usman:2015kfa}%
  \BibitemOpen
  \bibfield  {author} {\bibinfo {author} {\bibfnamefont {S.~A.}\ \bibnamefont
  {Usman}} \emph {et~al.},\ }\href {\doibase 10.1088/0264-9381/33/21/215004}
  {\bibfield  {journal} {\bibinfo  {journal} {Class. Quant. Grav.}\ }\textbf
  {\bibinfo {volume} {33}},\ \bibinfo {pages} {215004} (\bibinfo {year}
  {2016})},\ \Eprint {http://arxiv.org/abs/1508.02357} {arXiv:1508.02357
  [gr-qc]} \BibitemShut {NoStop}%
\bibitem [{\citenamefont {Allen}(2021)}]{Allen:2021yuy}%
  \BibitemOpen
  \bibfield  {author} {\bibinfo {author} {\bibfnamefont {B.}~\bibnamefont
  {Allen}},\ }\href {\doibase 10.1103/PhysRevD.104.042005} {\bibfield
  {journal} {\bibinfo  {journal} {Phys. Rev. D}\ }\textbf {\bibinfo {volume}
  {104}},\ \bibinfo {pages} {042005} (\bibinfo {year} {2021})},\ \Eprint
  {http://arxiv.org/abs/2102.11254} {arXiv:2102.11254 [astro-ph.IM]}
  \BibitemShut {NoStop}%
\bibitem [{\citenamefont {Veitch}\ \emph {et~al.}(2015)\citenamefont {Veitch}
  \emph {et~al.}}]{Veitch:2014wba}%
  \BibitemOpen
  \bibfield  {author} {\bibinfo {author} {\bibfnamefont {J.}~\bibnamefont
  {Veitch}} \emph {et~al.},\ }\href {\doibase 10.1103/PhysRevD.91.042003}
  {\bibfield  {journal} {\bibinfo  {journal} {Phys. Rev. D}\ }\textbf {\bibinfo
  {volume} {91}},\ \bibinfo {pages} {042003} (\bibinfo {year} {2015})},\
  \Eprint {http://arxiv.org/abs/1409.7215} {arXiv:1409.7215 [gr-qc]}
  \BibitemShut {NoStop}%
\bibitem [{\citenamefont {Biwer}\ \emph {et~al.}(2019)\citenamefont {Biwer},
  \citenamefont {Capano}, \citenamefont {De}, \citenamefont {Cabero},
  \citenamefont {Brown}, \citenamefont {Nitz},\ and\ \citenamefont
  {Raymond}}]{Biwer:2018osg}%
  \BibitemOpen
  \bibfield  {author} {\bibinfo {author} {\bibfnamefont {C.~M.}\ \bibnamefont
  {Biwer}}, \bibinfo {author} {\bibfnamefont {C.~D.}\ \bibnamefont {Capano}},
  \bibinfo {author} {\bibfnamefont {S.}~\bibnamefont {De}}, \bibinfo {author}
  {\bibfnamefont {M.}~\bibnamefont {Cabero}}, \bibinfo {author} {\bibfnamefont
  {D.~A.}\ \bibnamefont {Brown}}, \bibinfo {author} {\bibfnamefont {A.~H.}\
  \bibnamefont {Nitz}}, \ and\ \bibinfo {author} {\bibfnamefont
  {V.}~\bibnamefont {Raymond}},\ }\href {\doibase 10.1088/1538-3873/aaef0b}
  {\bibfield  {journal} {\bibinfo  {journal} {Publ. Astron. Soc. Pac.}\
  }\textbf {\bibinfo {volume} {131}},\ \bibinfo {pages} {024503} (\bibinfo
  {year} {2019})},\ \Eprint {http://arxiv.org/abs/1807.10312} {arXiv:1807.10312
  [astro-ph.IM]} \BibitemShut {NoStop}%
\bibitem [{\citenamefont {Mozzon}\ \emph {et~al.}(2020)\citenamefont {Mozzon},
  \citenamefont {Nuttall}, \citenamefont {Lundgren}, \citenamefont {Dent},
  \citenamefont {Kumar},\ and\ \citenamefont {Nitz}}]{Mozzon2020}%
  \BibitemOpen
  \bibfield  {author} {\bibinfo {author} {\bibfnamefont {S.}~\bibnamefont
  {Mozzon}}, \bibinfo {author} {\bibfnamefont {L.~K.}\ \bibnamefont {Nuttall}},
  \bibinfo {author} {\bibfnamefont {A.}~\bibnamefont {Lundgren}}, \bibinfo
  {author} {\bibfnamefont {T.}~\bibnamefont {Dent}}, \bibinfo {author}
  {\bibfnamefont {S.}~\bibnamefont {Kumar}}, \ and\ \bibinfo {author}
  {\bibfnamefont {A.~H.}\ \bibnamefont {Nitz}},\ }\href {\doibase
  10.1088/1361-6382/abac6c} {\bibfield  {journal} {\bibinfo  {journal}
  {Classical and Quantum Gravity}\ }\textbf {\bibinfo {volume} {37}},\ \bibinfo
  {pages} {215014} (\bibinfo {year} {2020})}\BibitemShut {NoStop}%
\bibitem [{\citenamefont {Edy}\ \emph {et~al.}(2021)\citenamefont {Edy},
  \citenamefont {Lundgren},\ and\ \citenamefont {Nuttall}}]{Edy2021}%
  \BibitemOpen
  \bibfield  {author} {\bibinfo {author} {\bibfnamefont {O.}~\bibnamefont
  {Edy}}, \bibinfo {author} {\bibfnamefont {A.}~\bibnamefont {Lundgren}}, \
  and\ \bibinfo {author} {\bibfnamefont {L.~K.}\ \bibnamefont {Nuttall}},\
  }\href {\doibase 10.1103/physrevd.103.124061} {\bibfield  {journal} {\bibinfo
   {journal} {Physical Review D}\ }\textbf {\bibinfo {volume} {103}} (\bibinfo
  {year} {2021}),\ 10.1103/physrevd.103.124061}\BibitemShut {NoStop}%
\bibitem [{lig()}]{ligoLIGOT2100313v3LIGO}%
  \BibitemOpen
  \href@noop {} {\enquote {\bibinfo {title} {{L}{I}{G}{O}-{T}2100313-v3:
  {L}{I}{G}{O} and {V}irgo {C}alibration {U}ncertainty ({O}1, {O}2 and {O}3)
  --- dcc.ligo.org},}\ }\bibinfo {howpublished}
  {\url{https://dcc.ligo.org/T2100313/public}},\ \bibinfo {note} {[Accessed
  10-02-2025]}\BibitemShut {NoStop}%
\bibitem [{\citenamefont {{Cahillane}}\ \emph {et~al.}(2017)\citenamefont
  {{Cahillane}}, \citenamefont {{Betzwieser}}, \citenamefont {{Brown}},
  \citenamefont {{Goetz}}, \citenamefont {{Hall}}, \citenamefont {{Izumi}},
  \citenamefont {{Kandhasamy}}, \citenamefont {{Karki}}, \citenamefont
  {{Kissel}}, \citenamefont {{Mendell}}, \citenamefont {{Savage}},
  \citenamefont {{Tuyenbayev}}, \citenamefont {{Urban}}, \citenamefont
  {{Viets}}, \citenamefont {{Wade}},\ and\ \citenamefont
  {{Weinstein}}}]{2017PhRvD..96j2001C}%
  \BibitemOpen
  \bibfield  {author} {\bibinfo {author} {\bibfnamefont {C.}~\bibnamefont
  {{Cahillane}}}, \bibinfo {author} {\bibfnamefont {J.}~\bibnamefont
  {{Betzwieser}}}, \bibinfo {author} {\bibfnamefont {D.~A.}\ \bibnamefont
  {{Brown}}}, \bibinfo {author} {\bibfnamefont {E.}~\bibnamefont {{Goetz}}},
  \bibinfo {author} {\bibfnamefont {E.~D.}\ \bibnamefont {{Hall}}}, \bibinfo
  {author} {\bibfnamefont {K.}~\bibnamefont {{Izumi}}}, \bibinfo {author}
  {\bibfnamefont {S.}~\bibnamefont {{Kandhasamy}}}, \bibinfo {author}
  {\bibfnamefont {S.}~\bibnamefont {{Karki}}}, \bibinfo {author} {\bibfnamefont
  {J.~S.}\ \bibnamefont {{Kissel}}}, \bibinfo {author} {\bibfnamefont
  {G.}~\bibnamefont {{Mendell}}}, \bibinfo {author} {\bibfnamefont {R.~L.}\
  \bibnamefont {{Savage}}}, \bibinfo {author} {\bibfnamefont {D.}~\bibnamefont
  {{Tuyenbayev}}}, \bibinfo {author} {\bibfnamefont {A.}~\bibnamefont
  {{Urban}}}, \bibinfo {author} {\bibfnamefont {A.}~\bibnamefont {{Viets}}},
  \bibinfo {author} {\bibfnamefont {M.}~\bibnamefont {{Wade}}}, \ and\ \bibinfo
  {author} {\bibfnamefont {A.~J.}\ \bibnamefont {{Weinstein}}},\ }\href
  {\doibase 10.1103/PhysRevD.96.102001} {\bibfield  {journal} {\bibinfo
  {journal} {\prd}\ }\textbf {\bibinfo {volume} {96}},\ \bibinfo {eid} {102001}
  (\bibinfo {year} {2017})},\ \Eprint {http://arxiv.org/abs/1708.03023}
  {arXiv:1708.03023 [astro-ph.IM]} \BibitemShut {NoStop}%
\bibitem [{\citenamefont {Sun}\ \emph {et~al.}(2020)\citenamefont {Sun} \emph
  {et~al.}}]{Sun:2020wke}%
  \BibitemOpen
  \bibfield  {author} {\bibinfo {author} {\bibfnamefont {L.}~\bibnamefont
  {Sun}} \emph {et~al.},\ }\href {\doibase 10.1088/1361-6382/abb14e} {\bibfield
   {journal} {\bibinfo  {journal} {Class. Quant. Grav.}\ }\textbf {\bibinfo
  {volume} {37}},\ \bibinfo {pages} {225008} (\bibinfo {year} {2020})},\
  \Eprint {http://arxiv.org/abs/2005.02531} {arXiv:2005.02531 [astro-ph.IM]}
  \BibitemShut {NoStop}%
\bibitem [{\citenamefont {Sun}\ \emph {et~al.}(2021)\citenamefont {Sun} \emph
  {et~al.}}]{Sun:2021qcg}%
  \BibitemOpen
  \bibfield  {author} {\bibinfo {author} {\bibfnamefont {L.}~\bibnamefont
  {Sun}} \emph {et~al.},\ }\href@noop {} {\  (\bibinfo {year} {2021})},\
  \Eprint {http://arxiv.org/abs/2107.00129} {arXiv:2107.00129 [astro-ph.IM]}
  \BibitemShut {NoStop}%
\bibitem [{\citenamefont {Acernese}\ \emph {et~al.}(2022)\citenamefont
  {Acernese} \emph {et~al.}}]{VIRGO:2021kfv}%
  \BibitemOpen
  \bibfield  {author} {\bibinfo {author} {\bibfnamefont {F.}~\bibnamefont
  {Acernese}} \emph {et~al.} (\bibinfo {collaboration} {VIRGO}),\ }\href
  {\doibase 10.1088/1361-6382/ac3c8e} {\bibfield  {journal} {\bibinfo
  {journal} {Class. Quant. Grav.}\ }\textbf {\bibinfo {volume} {39}},\ \bibinfo
  {pages} {045006} (\bibinfo {year} {2022})},\ \Eprint
  {http://arxiv.org/abs/2107.03294} {arXiv:2107.03294 [gr-qc]} \BibitemShut
  {NoStop}%
\bibitem [{\citenamefont {Acernese}\ \emph {et~al.}(2018)\citenamefont
  {Acernese} \emph {et~al.}}]{Virgo:2018gxa}%
  \BibitemOpen
  \bibfield  {author} {\bibinfo {author} {\bibfnamefont {F.}~\bibnamefont
  {Acernese}} \emph {et~al.} (\bibinfo {collaboration} {Virgo}),\ }\href
  {\doibase 10.1088/1361-6382/aadf1a} {\bibfield  {journal} {\bibinfo
  {journal} {Class. Quant. Grav.}\ }\textbf {\bibinfo {volume} {35}},\ \bibinfo
  {pages} {205004} (\bibinfo {year} {2018})},\ \Eprint
  {http://arxiv.org/abs/1807.03275} {arXiv:1807.03275 [gr-qc]} \BibitemShut
  {NoStop}%
\bibitem [{\citenamefont {Farr}\ \emph {et~al.}(2015)\citenamefont {Farr},
  \citenamefont {Farr},\ and\ \citenamefont
  {Littenberg}}]{SplineCalMarg-T1400682}%
  \BibitemOpen
  \bibfield  {author} {\bibinfo {author} {\bibfnamefont {W.~M.}\ \bibnamefont
  {Farr}}, \bibinfo {author} {\bibfnamefont {B.}~\bibnamefont {Farr}}, \ and\
  \bibinfo {author} {\bibfnamefont {T.}~\bibnamefont {Littenberg}},\ }\href
  {https://dcc.ligo.org/P1500262/} {\emph {\bibinfo {title} {Modelling
  Calibration Errors In CBC Waveforms}}},\ \bibinfo {type} {Tech. Rep.}\
  \bibinfo {number} {{LIGO}-T1400682}\ (\bibinfo  {institution} {{LIGO}
  Project},\ \bibinfo {year} {2015})\BibitemShut {NoStop}%
\bibitem [{\citenamefont {Cahillane}\ \emph {et~al.}(2017)\citenamefont
  {Cahillane} \emph {et~al.}}]{LIGOScientific:2017aaj}%
  \BibitemOpen
  \bibfield  {author} {\bibinfo {author} {\bibfnamefont {C.}~\bibnamefont
  {Cahillane}} \emph {et~al.} (\bibinfo {collaboration} {LIGO Scientific}),\
  }\href {\doibase 10.1103/PhysRevD.96.102001} {\bibfield  {journal} {\bibinfo
  {journal} {Phys. Rev. D}\ }\textbf {\bibinfo {volume} {96}},\ \bibinfo
  {pages} {102001} (\bibinfo {year} {2017})},\ \Eprint
  {http://arxiv.org/abs/1708.03023} {arXiv:1708.03023 [astro-ph.IM]}
  \BibitemShut {NoStop}%
\bibitem [{\citenamefont {Ashton}\ \emph {et~al.}(2019)\citenamefont {Ashton}
  \emph {et~al.}}]{Ashton:2018jfp}%
  \BibitemOpen
  \bibfield  {author} {\bibinfo {author} {\bibfnamefont {G.}~\bibnamefont
  {Ashton}} \emph {et~al.},\ }\href {\doibase 10.3847/1538-4365/ab06fc}
  {\bibfield  {journal} {\bibinfo  {journal} {Astrophys. J. Suppl.}\ }\textbf
  {\bibinfo {volume} {241}},\ \bibinfo {pages} {27} (\bibinfo {year} {2019})},\
  \Eprint {http://arxiv.org/abs/1811.02042} {arXiv:1811.02042 [astro-ph.IM]}
  \BibitemShut {NoStop}%
\bibitem [{\citenamefont {Edelman}\ \emph {et~al.}(2021)\citenamefont {Edelman}
  \emph {et~al.}}]{Edelman:2020aqj}%
  \BibitemOpen
  \bibfield  {author} {\bibinfo {author} {\bibfnamefont {B.}~\bibnamefont
  {Edelman}} \emph {et~al.},\ }\href {\doibase 10.1103/PhysRevD.103.042004}
  {\bibfield  {journal} {\bibinfo  {journal} {Phys. Rev. D}\ }\textbf {\bibinfo
  {volume} {103}},\ \bibinfo {pages} {042004} (\bibinfo {year} {2021})},\
  \Eprint {http://arxiv.org/abs/2008.06436} {arXiv:2008.06436 [gr-qc]}
  \BibitemShut {NoStop}%
\bibitem [{\citenamefont {Harris}\ \emph {et~al.}(2020)\citenamefont {Harris},
  \citenamefont {Millman}, \citenamefont {van~der Walt}, \citenamefont
  {Gommers}, \citenamefont {Virtanen}, \citenamefont {Cournapeau},
  \citenamefont {Wieser}, \citenamefont {Taylor}, \citenamefont {Berg},
  \citenamefont {Smith}, \citenamefont {Kern}, \citenamefont {Picus},
  \citenamefont {Hoyer}, \citenamefont {van Kerkwijk}, \citenamefont {Brett},
  \citenamefont {Haldane}, \citenamefont {del R{\'{i}}o}, \citenamefont
  {Wiebe}, \citenamefont {Peterson}, \citenamefont {G{\'{e}}rard-Marchant},
  \citenamefont {Sheppard}, \citenamefont {Reddy}, \citenamefont {Weckesser},
  \citenamefont {Abbasi}, \citenamefont {Gohlke},\ and\ \citenamefont
  {Oliphant}}]{harris2020array}%
  \BibitemOpen
  \bibfield  {author} {\bibinfo {author} {\bibfnamefont {C.~R.}\ \bibnamefont
  {Harris}}, \bibinfo {author} {\bibfnamefont {K.~J.}\ \bibnamefont {Millman}},
  \bibinfo {author} {\bibfnamefont {S.~J.}\ \bibnamefont {van~der Walt}},
  \bibinfo {author} {\bibfnamefont {R.}~\bibnamefont {Gommers}}, \bibinfo
  {author} {\bibfnamefont {P.}~\bibnamefont {Virtanen}}, \bibinfo {author}
  {\bibfnamefont {D.}~\bibnamefont {Cournapeau}}, \bibinfo {author}
  {\bibfnamefont {E.}~\bibnamefont {Wieser}}, \bibinfo {author} {\bibfnamefont
  {J.}~\bibnamefont {Taylor}}, \bibinfo {author} {\bibfnamefont
  {S.}~\bibnamefont {Berg}}, \bibinfo {author} {\bibfnamefont {N.~J.}\
  \bibnamefont {Smith}}, \bibinfo {author} {\bibfnamefont {R.}~\bibnamefont
  {Kern}}, \bibinfo {author} {\bibfnamefont {M.}~\bibnamefont {Picus}},
  \bibinfo {author} {\bibfnamefont {S.}~\bibnamefont {Hoyer}}, \bibinfo
  {author} {\bibfnamefont {M.~H.}\ \bibnamefont {van Kerkwijk}}, \bibinfo
  {author} {\bibfnamefont {M.}~\bibnamefont {Brett}}, \bibinfo {author}
  {\bibfnamefont {A.}~\bibnamefont {Haldane}}, \bibinfo {author} {\bibfnamefont
  {J.~F.}\ \bibnamefont {del R{\'{i}}o}}, \bibinfo {author} {\bibfnamefont
  {M.}~\bibnamefont {Wiebe}}, \bibinfo {author} {\bibfnamefont
  {P.}~\bibnamefont {Peterson}}, \bibinfo {author} {\bibfnamefont
  {P.}~\bibnamefont {G{\'{e}}rard-Marchant}}, \bibinfo {author} {\bibfnamefont
  {K.}~\bibnamefont {Sheppard}}, \bibinfo {author} {\bibfnamefont
  {T.}~\bibnamefont {Reddy}}, \bibinfo {author} {\bibfnamefont
  {W.}~\bibnamefont {Weckesser}}, \bibinfo {author} {\bibfnamefont
  {H.}~\bibnamefont {Abbasi}}, \bibinfo {author} {\bibfnamefont
  {C.}~\bibnamefont {Gohlke}}, \ and\ \bibinfo {author} {\bibfnamefont {T.~E.}\
  \bibnamefont {Oliphant}},\ }\href {\doibase 10.1038/s41586-020-2649-2}
  {\bibfield  {journal} {\bibinfo  {journal} {Nature}\ }\textbf {\bibinfo
  {volume} {585}},\ \bibinfo {pages} {357} (\bibinfo {year}
  {2020})}\BibitemShut {NoStop}%
\bibitem [{\citenamefont {Virtanen}\ \emph {et~al.}(2020)\citenamefont
  {Virtanen}, \citenamefont {Gommers}, \citenamefont {Oliphant}, \citenamefont
  {Haberland}, \citenamefont {Reddy}, \citenamefont {Cournapeau}, \citenamefont
  {Burovski}, \citenamefont {Peterson}, \citenamefont {Weckesser},
  \citenamefont {Bright}, \citenamefont {{van der Walt}}, \citenamefont
  {Brett}, \citenamefont {Wilson}, \citenamefont {Millman}, \citenamefont
  {Mayorov}, \citenamefont {Nelson}, \citenamefont {Jones}, \citenamefont
  {Kern}, \citenamefont {Larson}, \citenamefont {Carey}, \citenamefont {Polat},
  \citenamefont {Feng}, \citenamefont {Moore}, \citenamefont {{VanderPlas}},
  \citenamefont {Laxalde}, \citenamefont {Perktold}, \citenamefont {Cimrman},
  \citenamefont {Henriksen}, \citenamefont {Quintero}, \citenamefont {Harris},
  \citenamefont {Archibald}, \citenamefont {Ribeiro}, \citenamefont
  {Pedregosa}, \citenamefont {{van Mulbregt}},\ and\ \citenamefont {{SciPy 1.0
  Contributors}}}]{2020SciPy-NMeth}%
  \BibitemOpen
  \bibfield  {author} {\bibinfo {author} {\bibfnamefont {P.}~\bibnamefont
  {Virtanen}}, \bibinfo {author} {\bibfnamefont {R.}~\bibnamefont {Gommers}},
  \bibinfo {author} {\bibfnamefont {T.~E.}\ \bibnamefont {Oliphant}}, \bibinfo
  {author} {\bibfnamefont {M.}~\bibnamefont {Haberland}}, \bibinfo {author}
  {\bibfnamefont {T.}~\bibnamefont {Reddy}}, \bibinfo {author} {\bibfnamefont
  {D.}~\bibnamefont {Cournapeau}}, \bibinfo {author} {\bibfnamefont
  {E.}~\bibnamefont {Burovski}}, \bibinfo {author} {\bibfnamefont
  {P.}~\bibnamefont {Peterson}}, \bibinfo {author} {\bibfnamefont
  {W.}~\bibnamefont {Weckesser}}, \bibinfo {author} {\bibfnamefont
  {J.}~\bibnamefont {Bright}}, \bibinfo {author} {\bibfnamefont {S.~J.}\
  \bibnamefont {{van der Walt}}}, \bibinfo {author} {\bibfnamefont
  {M.}~\bibnamefont {Brett}}, \bibinfo {author} {\bibfnamefont
  {J.}~\bibnamefont {Wilson}}, \bibinfo {author} {\bibfnamefont {K.~J.}\
  \bibnamefont {Millman}}, \bibinfo {author} {\bibfnamefont {N.}~\bibnamefont
  {Mayorov}}, \bibinfo {author} {\bibfnamefont {A.~R.~J.}\ \bibnamefont
  {Nelson}}, \bibinfo {author} {\bibfnamefont {E.}~\bibnamefont {Jones}},
  \bibinfo {author} {\bibfnamefont {R.}~\bibnamefont {Kern}}, \bibinfo {author}
  {\bibfnamefont {E.}~\bibnamefont {Larson}}, \bibinfo {author} {\bibfnamefont
  {C.~J.}\ \bibnamefont {Carey}}, \bibinfo {author} {\bibfnamefont
  {{\.I}.}~\bibnamefont {Polat}}, \bibinfo {author} {\bibfnamefont
  {Y.}~\bibnamefont {Feng}}, \bibinfo {author} {\bibfnamefont {E.~W.}\
  \bibnamefont {Moore}}, \bibinfo {author} {\bibfnamefont {J.}~\bibnamefont
  {{VanderPlas}}}, \bibinfo {author} {\bibfnamefont {D.}~\bibnamefont
  {Laxalde}}, \bibinfo {author} {\bibfnamefont {J.}~\bibnamefont {Perktold}},
  \bibinfo {author} {\bibfnamefont {R.}~\bibnamefont {Cimrman}}, \bibinfo
  {author} {\bibfnamefont {I.}~\bibnamefont {Henriksen}}, \bibinfo {author}
  {\bibfnamefont {E.~A.}\ \bibnamefont {Quintero}}, \bibinfo {author}
  {\bibfnamefont {C.~R.}\ \bibnamefont {Harris}}, \bibinfo {author}
  {\bibfnamefont {A.~M.}\ \bibnamefont {Archibald}}, \bibinfo {author}
  {\bibfnamefont {A.~H.}\ \bibnamefont {Ribeiro}}, \bibinfo {author}
  {\bibfnamefont {F.}~\bibnamefont {Pedregosa}}, \bibinfo {author}
  {\bibfnamefont {P.}~\bibnamefont {{van Mulbregt}}}, \ and\ \bibinfo {author}
  {\bibnamefont {{SciPy 1.0 Contributors}}},\ }\href {\doibase
  10.1038/s41592-019-0686-2} {\bibfield  {journal} {\bibinfo  {journal} {Nature
  Methods}\ }\textbf {\bibinfo {volume} {17}},\ \bibinfo {pages} {261}
  (\bibinfo {year} {2020})}\BibitemShut {NoStop}%
\bibitem [{\citenamefont {Abbott}\ \emph
  {et~al.}(2020{\natexlab{b}})\citenamefont {Abbott} \emph
  {et~al.}}]{LIGOScientific:2020stg}%
  \BibitemOpen
  \bibfield  {author} {\bibinfo {author} {\bibfnamefont {R.}~\bibnamefont
  {Abbott}} \emph {et~al.} (\bibinfo {collaboration} {LIGO Scientific,
  Virgo}),\ }\href {\doibase 10.1103/PhysRevD.102.043015} {\bibfield  {journal}
  {\bibinfo  {journal} {Phys. Rev. D}\ }\textbf {\bibinfo {volume} {102}},\
  \bibinfo {pages} {043015} (\bibinfo {year} {2020}{\natexlab{b}})},\ \Eprint
  {http://arxiv.org/abs/2004.08342} {arXiv:2004.08342 [astro-ph.HE]}
  \BibitemShut {NoStop}%
\bibitem [{\citenamefont {Dupletsa}\ \emph {et~al.}(2024)\citenamefont
  {Dupletsa}, \citenamefont {Harms}, \citenamefont {Ng}, \citenamefont
  {Tissino}, \citenamefont {Santoliquido},\ and\ \citenamefont
  {Cozzumbo}}]{dupletsa2024}%
  \BibitemOpen
  \bibfield  {author} {\bibinfo {author} {\bibfnamefont {U.}~\bibnamefont
  {Dupletsa}}, \bibinfo {author} {\bibfnamefont {J.}~\bibnamefont {Harms}},
  \bibinfo {author} {\bibfnamefont {K.~K.~Y.}\ \bibnamefont {Ng}}, \bibinfo
  {author} {\bibfnamefont {J.}~\bibnamefont {Tissino}}, \bibinfo {author}
  {\bibfnamefont {F.}~\bibnamefont {Santoliquido}}, \ and\ \bibinfo {author}
  {\bibfnamefont {A.}~\bibnamefont {Cozzumbo}},\ }\href
  {https://arxiv.org/abs/2404.16103} {\enquote {\bibinfo {title} {Validating
  prior-informed fisher-matrix analyses against gwtc data},}\ } (\bibinfo
  {year} {2024}),\ \Eprint {http://arxiv.org/abs/2404.16103} {arXiv:2404.16103
  [gr-qc]} \BibitemShut {NoStop}%
\bibitem [{\citenamefont {Iacovelli}\ \emph
  {et~al.}(2022{\natexlab{a}})\citenamefont {Iacovelli}, \citenamefont
  {Mancarella}, \citenamefont {Foffa},\ and\ \citenamefont
  {Maggiore}}]{Iacovelli:2022bbs}%
  \BibitemOpen
  \bibfield  {author} {\bibinfo {author} {\bibfnamefont {F.}~\bibnamefont
  {Iacovelli}}, \bibinfo {author} {\bibfnamefont {M.}~\bibnamefont
  {Mancarella}}, \bibinfo {author} {\bibfnamefont {S.}~\bibnamefont {Foffa}}, \
  and\ \bibinfo {author} {\bibfnamefont {M.}~\bibnamefont {Maggiore}},\ }\href
  {\doibase 10.3847/1538-4357/ac9cd4} {\bibfield  {journal} {\bibinfo
  {journal} {Astrophys. J.}\ }\textbf {\bibinfo {volume} {941}},\ \bibinfo
  {pages} {208} (\bibinfo {year} {2022}{\natexlab{a}})},\ \Eprint
  {http://arxiv.org/abs/2207.02771} {arXiv:2207.02771 [gr-qc]} \BibitemShut
  {NoStop}%
\bibitem [{\citenamefont {Iacovelli}\ \emph
  {et~al.}(2022{\natexlab{b}})\citenamefont {Iacovelli}, \citenamefont
  {Mancarella}, \citenamefont {Foffa},\ and\ \citenamefont
  {Maggiore}}]{Iacovelli:2022mbg}%
  \BibitemOpen
  \bibfield  {author} {\bibinfo {author} {\bibfnamefont {F.}~\bibnamefont
  {Iacovelli}}, \bibinfo {author} {\bibfnamefont {M.}~\bibnamefont
  {Mancarella}}, \bibinfo {author} {\bibfnamefont {S.}~\bibnamefont {Foffa}}, \
  and\ \bibinfo {author} {\bibfnamefont {M.}~\bibnamefont {Maggiore}},\ }\href
  {\doibase 10.3847/1538-4365/ac9129} {\bibfield  {journal} {\bibinfo
  {journal} {Astrophys. J. Supp.}\ }\textbf {\bibinfo {volume} {263}},\
  \bibinfo {pages} {2} (\bibinfo {year} {2022}{\natexlab{b}})},\ \Eprint
  {http://arxiv.org/abs/2207.06910} {arXiv:2207.06910 [astro-ph.IM]}
  \BibitemShut {NoStop}%
\bibitem [{\citenamefont {Li}\ \emph {et~al.}(2022)\citenamefont {Li},
  \citenamefont {Heng}, \citenamefont {Chan}, \citenamefont {Messenger},\ and\
  \citenamefont {Fan}}]{Li:2021mbo}%
  \BibitemOpen
  \bibfield  {author} {\bibinfo {author} {\bibfnamefont {Y.}~\bibnamefont
  {Li}}, \bibinfo {author} {\bibfnamefont {I.~S.}\ \bibnamefont {Heng}},
  \bibinfo {author} {\bibfnamefont {M.~L.}\ \bibnamefont {Chan}}, \bibinfo
  {author} {\bibfnamefont {C.}~\bibnamefont {Messenger}}, \ and\ \bibinfo
  {author} {\bibfnamefont {X.}~\bibnamefont {Fan}},\ }\href {\doibase
  10.1103/PhysRevD.105.043010} {\bibfield  {journal} {\bibinfo  {journal}
  {Phys. Rev. D}\ }\textbf {\bibinfo {volume} {105}},\ \bibinfo {pages}
  {043010} (\bibinfo {year} {2022})},\ \Eprint
  {http://arxiv.org/abs/2109.07389} {arXiv:2109.07389 [astro-ph.IM]}
  \BibitemShut {NoStop}%
\bibitem [{\citenamefont {Abbott}\ \emph
  {et~al.}(2019{\natexlab{b}})\citenamefont {Abbott} \emph {et~al.}}]{gwtc1}%
  \BibitemOpen
  \bibfield  {author} {\bibinfo {author} {\bibfnamefont {B.~P.}\ \bibnamefont
  {Abbott}} \emph {et~al.},\ }\href {\doibase 10.1103/physrevx.9.031040}
  {\bibfield  {journal} {\bibinfo  {journal} {Physical Review X}\ }\textbf
  {\bibinfo {volume} {9}} (\bibinfo {year} {2019}{\natexlab{b}}),\
  10.1103/physrevx.9.031040}\BibitemShut {NoStop}%
\bibitem [{\citenamefont {Abbott}\ \emph
  {et~al.}(2021{\natexlab{c}})\citenamefont {Abbott} \emph {et~al.}}]{gwtc2}%
  \BibitemOpen
  \bibfield  {author} {\bibinfo {author} {\bibfnamefont {R.}~\bibnamefont
  {Abbott}} \emph {et~al.},\ }\href {\doibase 10.1103/physrevx.11.021053}
  {\bibfield  {journal} {\bibinfo  {journal} {Physical Review X}\ }\textbf
  {\bibinfo {volume} {11}} (\bibinfo {year} {2021}{\natexlab{c}}),\
  10.1103/physrevx.11.021053}\BibitemShut {NoStop}%
\bibitem [{\citenamefont {Abbott}\ \emph {et~al.}(2024)\citenamefont {Abbott}
  \emph {et~al.}}]{LIGOScientific:2021usb}%
  \BibitemOpen
  \bibfield  {author} {\bibinfo {author} {\bibfnamefont {R.}~\bibnamefont
  {Abbott}} \emph {et~al.} (\bibinfo {collaboration} {LIGO Scientific,
  VIRGO}),\ }\href {\doibase 10.1103/PhysRevD.109.022001} {\bibfield  {journal}
  {\bibinfo  {journal} {Phys. Rev. D}\ }\textbf {\bibinfo {volume} {109}},\
  \bibinfo {pages} {022001} (\bibinfo {year} {2024})},\ \Eprint
  {http://arxiv.org/abs/2108.01045} {arXiv:2108.01045 [gr-qc]} \BibitemShut
  {NoStop}%
\bibitem [{\citenamefont {Vallisneri}(2008)}]{Vallisneri_2008}%
  \BibitemOpen
  \bibfield  {author} {\bibinfo {author} {\bibfnamefont {M.}~\bibnamefont
  {Vallisneri}},\ }\href {\doibase 10.1103/physrevd.77.042001} {\bibfield
  {journal} {\bibinfo  {journal} {Physical Review D}\ }\textbf {\bibinfo
  {volume} {77}} (\bibinfo {year} {2008}),\
  10.1103/physrevd.77.042001}\BibitemShut {NoStop}%
\bibitem [{\citenamefont {Speagle}(2020)}]{speagle:2019}%
  \BibitemOpen
  \bibfield  {author} {\bibinfo {author} {\bibfnamefont {J.~S.}\ \bibnamefont
  {Speagle}},\ }\href {\doibase 10.1093/mnras/staa278} {\bibfield  {journal}
  {\bibinfo  {journal} {Monthly Notices of the Royal Astronomical Society}\
  }\textbf {\bibinfo {volume} {493}},\ \bibinfo {pages} {3132} (\bibinfo {year}
  {2020})},\ \Eprint
  {http://arxiv.org/abs/https://academic.oup.com/mnras/article-pdf/493/3/3132/32890730/staa278.pdf}
  {https://academic.oup.com/mnras/article-pdf/493/3/3132/32890730/staa278.pdf}
  \BibitemShut {NoStop}%
\bibitem [{\citenamefont {Kumar}\ \emph
  {et~al.}(2025{\natexlab{a}})\citenamefont {Kumar}, \citenamefont {Melching},\
  and\ \citenamefont {Ohme}}]{GitHubPlugin}%
  \BibitemOpen
  \bibfield  {author} {\bibinfo {author} {\bibfnamefont {S.}~\bibnamefont
  {Kumar}}, \bibinfo {author} {\bibfnamefont {M.}~\bibnamefont {Melching}}, \
  and\ \bibinfo {author} {\bibfnamefont {F.}~\bibnamefont {Ohme}},\ }\href@noop
  {} {\enquote {\bibinfo {title} {{PyCBC} plugin for waveform errors},}\
  }\bibinfo {howpublished}
  {\url{https://github.com/gwastro/pycbc_wferrors_plugin}} (\bibinfo {year}
  {2025}{\natexlab{a}})\BibitemShut {NoStop}%
\bibitem [{\citenamefont {{LIGO Scientific Collaboration}}(2018)}]{lalsuite}%
  \BibitemOpen
  \bibfield  {author} {\bibinfo {author} {\bibnamefont {{LIGO Scientific
  Collaboration}}},\ }\href {\doibase 10.7935/GT1W-FZ16} {\enquote {\bibinfo
  {title} {{LIGO} {A}lgorithm {L}ibrary - {LALS}uite},}\ }\bibinfo
  {howpublished} {free software (GPL)} (\bibinfo {year} {2018})\BibitemShut
  {NoStop}%
\bibitem [{\citenamefont {Paul}\ and\ \citenamefont
  {Mishra}(2023)}]{Paul:2022xfy}%
  \BibitemOpen
  \bibfield  {author} {\bibinfo {author} {\bibfnamefont {K.}~\bibnamefont
  {Paul}}\ and\ \bibinfo {author} {\bibfnamefont {C.~K.}\ \bibnamefont
  {Mishra}},\ }\href {\doibase 10.1103/PhysRevD.108.024023} {\bibfield
  {journal} {\bibinfo  {journal} {Phys. Rev. D}\ }\textbf {\bibinfo {volume}
  {108}},\ \bibinfo {pages} {024023} (\bibinfo {year} {2023})},\ \Eprint
  {http://arxiv.org/abs/2211.04155} {arXiv:2211.04155 [gr-qc]} \BibitemShut
  {NoStop}%
\bibitem [{\citenamefont {Henry}\ and\ \citenamefont
  {Khalil}(2023)}]{Henry:2023tka}%
  \BibitemOpen
  \bibfield  {author} {\bibinfo {author} {\bibfnamefont {Q.}~\bibnamefont
  {Henry}}\ and\ \bibinfo {author} {\bibfnamefont {M.}~\bibnamefont {Khalil}},\
  }\href {\doibase 10.1103/PhysRevD.108.104016} {\bibfield  {journal} {\bibinfo
   {journal} {Phys. Rev. D}\ }\textbf {\bibinfo {volume} {108}},\ \bibinfo
  {pages} {104016} (\bibinfo {year} {2023})},\ \Eprint
  {http://arxiv.org/abs/2308.13606} {arXiv:2308.13606 [gr-qc]} \BibitemShut
  {NoStop}%
\bibitem [{\citenamefont {Ramos-Buades}\ \emph {et~al.}(2022)\citenamefont
  {Ramos-Buades}, \citenamefont {Buonanno}, \citenamefont {Khalil},\ and\
  \citenamefont {Ossokine}}]{Ramos-Buades:2021adz}%
  \BibitemOpen
  \bibfield  {author} {\bibinfo {author} {\bibfnamefont {A.}~\bibnamefont
  {Ramos-Buades}}, \bibinfo {author} {\bibfnamefont {A.}~\bibnamefont
  {Buonanno}}, \bibinfo {author} {\bibfnamefont {M.}~\bibnamefont {Khalil}}, \
  and\ \bibinfo {author} {\bibfnamefont {S.}~\bibnamefont {Ossokine}},\ }\href
  {\doibase 10.1103/PhysRevD.105.044035} {\bibfield  {journal} {\bibinfo
  {journal} {Phys. Rev. D}\ }\textbf {\bibinfo {volume} {105}},\ \bibinfo
  {pages} {044035} (\bibinfo {year} {2022})},\ \Eprint
  {http://arxiv.org/abs/2112.06952} {arXiv:2112.06952 [gr-qc]} \BibitemShut
  {NoStop}%
\bibitem [{\citenamefont {Paul}\ \emph {et~al.}(2024)\citenamefont {Paul},
  \citenamefont {Maurya}, \citenamefont {Henry}, \citenamefont {Sharma},
  \citenamefont {Satheesh}, \citenamefont {Divyajyoti}, \citenamefont {Kumar},\
  and\ \citenamefont {Mishra}}]{Paul:2024ujx}%
  \BibitemOpen
  \bibfield  {author} {\bibinfo {author} {\bibfnamefont {K.}~\bibnamefont
  {Paul}}, \bibinfo {author} {\bibfnamefont {A.}~\bibnamefont {Maurya}},
  \bibinfo {author} {\bibfnamefont {Q.}~\bibnamefont {Henry}}, \bibinfo
  {author} {\bibfnamefont {K.}~\bibnamefont {Sharma}}, \bibinfo {author}
  {\bibfnamefont {P.}~\bibnamefont {Satheesh}}, \bibinfo {author} {\bibnamefont
  {Divyajyoti}}, \bibinfo {author} {\bibfnamefont {P.}~\bibnamefont {Kumar}}, \
  and\ \bibinfo {author} {\bibfnamefont {C.~K.}\ \bibnamefont {Mishra}},\
  }\href@noop {} {\  (\bibinfo {year} {2024})},\ \Eprint
  {http://arxiv.org/abs/2409.13866} {arXiv:2409.13866 [gr-qc]} \BibitemShut
  {NoStop}%
\bibitem [{\citenamefont {Hu}\ and\ \citenamefont {Veitch}(2022)}]{Hu:2022rjq}%
  \BibitemOpen
  \bibfield  {author} {\bibinfo {author} {\bibfnamefont {Q.}~\bibnamefont
  {Hu}}\ and\ \bibinfo {author} {\bibfnamefont {J.}~\bibnamefont {Veitch}},\
  }\href {\doibase 10.1103/PhysRevD.106.044042} {\bibfield  {journal} {\bibinfo
   {journal} {Phys. Rev. D}\ }\textbf {\bibinfo {volume} {106}},\ \bibinfo
  {pages} {044042} (\bibinfo {year} {2022})},\ \Eprint
  {http://arxiv.org/abs/2205.08448} {arXiv:2205.08448 [gr-qc]} \BibitemShut
  {NoStop}%
\bibitem [{\citenamefont {Hammond}\ \emph {et~al.}(2021)\citenamefont
  {Hammond}, \citenamefont {Hawke},\ and\ \citenamefont
  {Andersson}}]{Hammond:2021vtv}%
  \BibitemOpen
  \bibfield  {author} {\bibinfo {author} {\bibfnamefont {P.}~\bibnamefont
  {Hammond}}, \bibinfo {author} {\bibfnamefont {I.}~\bibnamefont {Hawke}}, \
  and\ \bibinfo {author} {\bibfnamefont {N.}~\bibnamefont {Andersson}},\ }\href
  {\doibase 10.1103/PhysRevD.104.103006} {\bibfield  {journal} {\bibinfo
  {journal} {Phys. Rev. D}\ }\textbf {\bibinfo {volume} {104}},\ \bibinfo
  {pages} {103006} (\bibinfo {year} {2021})},\ \Eprint
  {http://arxiv.org/abs/2108.08649} {arXiv:2108.08649 [astro-ph.HE]}
  \BibitemShut {NoStop}%
\bibitem [{\citenamefont {Gittins}\ \emph {et~al.}(2025)\citenamefont
  {Gittins}, \citenamefont {Matur}, \citenamefont {Andersson},\ and\
  \citenamefont {Hawke}}]{Gittins:2024jui}%
  \BibitemOpen
  \bibfield  {author} {\bibinfo {author} {\bibfnamefont {F.}~\bibnamefont
  {Gittins}}, \bibinfo {author} {\bibfnamefont {R.}~\bibnamefont {Matur}},
  \bibinfo {author} {\bibfnamefont {N.}~\bibnamefont {Andersson}}, \ and\
  \bibinfo {author} {\bibfnamefont {I.}~\bibnamefont {Hawke}},\ }\href
  {\doibase 10.1103/PhysRevD.111.023049} {\bibfield  {journal} {\bibinfo
  {journal} {Phys. Rev. D}\ }\textbf {\bibinfo {volume} {111}},\ \bibinfo
  {pages} {023049} (\bibinfo {year} {2025})},\ \Eprint
  {http://arxiv.org/abs/2409.13468} {arXiv:2409.13468 [gr-qc]} \BibitemShut
  {NoStop}%
\bibitem [{\citenamefont {Kumar}\ \emph
  {et~al.}(2025{\natexlab{b}})\citenamefont {Kumar}, \citenamefont {Melching},\
  and\ \citenamefont {Ohme}}]{kumar_2025_14975178}%
  \BibitemOpen
  \bibfield  {author} {\bibinfo {author} {\bibfnamefont {S.}~\bibnamefont
  {Kumar}}, \bibinfo {author} {\bibfnamefont {M.}~\bibnamefont {Melching}}, \
  and\ \bibinfo {author} {\bibfnamefont {F.}~\bibnamefont {Ohme}},\ }\href
  {\doibase 10.5281/zenodo.14975178} {\enquote {\bibinfo {title} {Raw data and
  configuration files for analysis in arxiv:2502.17400, accounting for waveform
  systematic errors},}\ } (\bibinfo {year} {2025}{\natexlab{b}})\BibitemShut
  {NoStop}%
\bibitem [{\citenamefont {Cutler}\ and\ \citenamefont
  {Vallisneri}(2007)}]{Cutler_2007}%
  \BibitemOpen
  \bibfield  {author} {\bibinfo {author} {\bibfnamefont {C.}~\bibnamefont
  {Cutler}}\ and\ \bibinfo {author} {\bibfnamefont {M.}~\bibnamefont
  {Vallisneri}},\ }\href {\doibase 10.1103/physrevd.76.104018} {\bibfield
  {journal} {\bibinfo  {journal} {Physical Review D}\ }\textbf {\bibinfo
  {volume} {76}} (\bibinfo {year} {2007}),\
  10.1103/physrevd.76.104018}\BibitemShut {NoStop}%
\bibitem [{\citenamefont {Balasubramanian}\ \emph {et~al.}(1996)\citenamefont
  {Balasubramanian}, \citenamefont {Sathyaprakash},\ and\ \citenamefont
  {Dhurandhar}}]{Balasub_1996}%
  \BibitemOpen
  \bibfield  {author} {\bibinfo {author} {\bibfnamefont {R.}~\bibnamefont
  {Balasubramanian}}, \bibinfo {author} {\bibfnamefont {B.~S.}\ \bibnamefont
  {Sathyaprakash}}, \ and\ \bibinfo {author} {\bibfnamefont {S.~V.}\
  \bibnamefont {Dhurandhar}},\ }\href {\doibase 10.1103/PhysRevD.53.3033}
  {\bibfield  {journal} {\bibinfo  {journal} {Phys. Rev. D}\ }\textbf {\bibinfo
  {volume} {53}},\ \bibinfo {pages} {3033} (\bibinfo {year}
  {1996})}\BibitemShut {NoStop}%
\bibitem [{\citenamefont {Allen}(2005)}]{Allen_2005}%
  \BibitemOpen
  \bibfield  {author} {\bibinfo {author} {\bibfnamefont {B.}~\bibnamefont
  {Allen}},\ }\href {\doibase 10.1103/physrevd.71.062001} {\bibfield  {journal}
  {\bibinfo  {journal} {Physical Review D}\ }\textbf {\bibinfo {volume} {71}}
  (\bibinfo {year} {2005}),\ 10.1103/physrevd.71.062001}\BibitemShut {NoStop}%
\end{thebibliography}%
\end{document}